\documentclass[12pt]{article}

\usepackage{jheppub, amsmath,amssymb,amsfonts,amsxtra, mathrsfs, makeidx,graphics,graphicx,amsthm,epsfig, bm,longtable,float, color,mathtools,xfrac,footnote,rotating, lscape, makecell, environ,mathtools, empheq}

\DeclareMathOperator\Gh{\Gamma_h}
\def\Gh[#1] {
	\Gamma_h \left(#1\right)
	}
\DeclareMathOperator\Ge{\Gamma_e}
\def\Ge[#1] {
	\Gamma_e \left(#1\right)
	}

\title{
\begin{center}
Rank-two tensors and deconfinement in 3d $\mathcal{N}=2$ $SU(N)$ gauge theories 
\end{center}
}

\author[a]{Antonio Amariti,}	
\author[b]{Fabio Mantegazza}
\author[c,d]{Simone Rota}

\affiliation[a]{INFN, Sezione di Milano, Via Celoria 16, I-20133 Milano, Italy}
\affiliation[b]{Deutsches Elektronen-Synchrotron DESY, Notkestr. 85, 22607 Hamburg, Germany}
\affiliation[c]{SISSA, Via Bonomea 265, 34136 Trieste, Italy}
\affiliation[d]{INFN, Sezione di Trieste, Via Valerio 2, 34127 Trieste, Italy}

\emailAdd{antonio.amariti@mi.infn.it}
\emailAdd{fabio.mantegazza@desy.de}
\emailAdd{srota@sissa.it}

\preprint{DESY-25-071}

\abstract{
In this paper we study the IR dynamics of $SU(N)$ gauge theories with four supercharges in 3d in presence of symmetric or antisymmetric tensor. Using the tensor deconfinement technique we provide some proofs of results previously claimed in the literature about confining dualities for $SU(N)$ with two antisymmetric tensors.
Furthermore we study 3d confining dualities with symmetric tensors and linear monopole superpotentials.
These confining dualities have been obtained by applying the duplication formula for the hyperbolic Gamma functions on effective dualities on $\mathbb{R}^{1,2} \times S^1$ with $SU(N)$ gauge groups and antisymmetric tensors. We conclude the analysis providing an alternative derivation, again using the tensor deconfinement technique.
}

\begin{document}
\maketitle
\flushbottom
\allowdisplaybreaks

\section{Introduction}

The classification of S-confining gauge theories in supersymmetric gauge theories in 4d with four supercharges  has been shown to be complete for gauge theories with a single gauge group and vanishing superpotential \cite{Csaki:1996sm,Csaki:1996zb}.

A similar classification in 3d theories with the same amount of supersymmetry is not available yet, and indeed there are more confining gauge theories than in the 4d parent cases. There are various reasons behind this fact, i.e. the existence of a Coulomb branch, the absence of global anomalies for the flavor symmetries, the existence of topological symmetries and the possibility of introducing a Chern-Simons action for the gauge group. 
We refer the reader to \cite{Aharony:1997bx,deBoer:1997kr,Intriligator:2013lca} for general review of the basic aspects of 3d supersymmetric theories with four supercharges.

Despite the larger landscape of 3d confining gauge theories, there is a uniform way to construct IR dualities and their confining  limits both in 4d and in 3d by using the technique of deconfining two index tensor matter fields, along the lines of the seminal papers of \cite{Berkooz:1995km,Luty:1996cg,Pouliot:1995me}.

Such procedure of obtaining dualities from dualities has been largely applied in the last few years in various dimensions \cite{Pasquetti:2019uop,Benvenuti:2020wpc,Etxebarria:2021lmq,Benvenuti:2021nwt,Bottini:2022vpy,Bajeot:2022lah,Bajeot:2022wmu,Amariti:2022wae,Amariti:2023wts,Amariti:2024sde,Amariti:2024gco,Benvenuti:2024glr,Hwang:2024hhy}. An important result obtained in \cite{Benvenuti:2020wpc} has shown that the models classified by \cite{Csaki:1996zb} for special unitary and symplectic gauge groups and two index tensors can all be obtained 
from two basic S-confining dualities, corresponding to the limiting case of Seiberg  \cite{Seiberg:1994pq} and Intriligator-Pouliot \cite{Intriligator:1995ne} duality.
In 3d this approach has allowed various collaborations to obtain new confining dualities, that do not have necessarily a 4d origin in terms of a parent s-confining duality reduced on the circle, with the aid of the  prescription of  \cite{Aharony:2013dha}.

Furthermore other confining dualities have been claimed in the 3d literature by a deep inspection of the moduli space. For example special unitary gauge groups with antisymmetric tensors have been fully explored in \cite{Nii:2019ebv}.  Some of the models discussed in this reference do not have an origin from any known 4d s-confining duality, This is for example the case $SU(N)$ gauge group with 2 antisymmetric tensors, $n_f$ fundamentals and $n_a$ antifundamentals, with $n_f+n_a=4$. 

This duality has been studied from tensor deconfinement for $n_{f}=4$ and $n_f=3$ in \cite{Amariti:2024gco}, with the analysis for $n_f<3$ is so far missing in the literature. 
The motivation of \cite{Amariti:2024gco} to study such cases is related to the fact that proving the duality through tensor deconfinement automatically gives the expected integral identity for the squashed 3-sphere
partition functions of \cite{Hama:2011ea}. 
Then, by an opportune \emph{freezing} of at least three mass parameters for the fundamentals and by applying 
the duplication formula for the hyperbolic Gamma function on such identities, one obtains new identities associated to $SU(N)$ gauge theories with symmetric tensors.  
Then these dualities, claimed from localization, have been further studied, showing that they can be obtained from tensor deconfinement as well.

In this paper we aim then  to extend the analysis started in \cite{Amariti:2024gco} by proving the dualities of \cite{Nii:2019ebv} with two antisymmetric tensors for $n_f<3$. 
We show that the cases with $n_f<3$ behave similarly to the case $n_f=3$ discussed in \cite{Amariti:2024gco}, i.e. after (i) deconfining the two antisymmetric tensors, using 3d symplectic SQCD theories, (ii) dualizing the $SU(n)$ gauge node and (iii) confining the antisymmetric tensors back, we arrive to a model with two conjugate antisymmetric tensors and four antifundamentals.
Up to an overall conjugation this corresponds to the model with $n_f=4$, that has been already studied using the  tensor deconfinement technique in \cite{Amariti:2024gco}. The details of the derivation are strongly dependent from the parity of the rank of the gauge node and from the explicit values of $n_f$ 
and $n_a$. For this reason we  analyze each case separately, giving a complete classification.
The second part of the paper is devoted to the analysis of   some of the dualities with symmetric tensors claimed in \cite{Amariti:2024gco}. We restrict our interest to cases that originate from the application of the duplication formula,
for the hyperbolic Gamma functions,
to 3d dualities with an antisymmetric tensor or an antisymmetric flavor and a 4d origin. The reason is that in such cases it is possible to consider effective confining dualities on the $S^1$, i.e. in presence of Kaluza-Klein (KK) monopole superpotential along the discussion of \cite{Aharony:2013dha}.
Translating this dimensional reduction into the reduction of the relative supersymmetric index \cite{Romelsberger:2005eg,Kinney:2005ej} to the squashed three sphere partition function \cite{Hama:2011ea}, 
in these cases one can opportunely \emph{freeze} three or four mass parameters and then apply the  duplication formula for the hyperbolic Gamma function on these effective confining dualities. 
This procedure generates new identities, at the level of the partition function, for theories with symmetric tensors. Such identities are different from the ones found in  \cite{Amariti:2024gco}, having a larger amount of flavors and a constraint among the fugacities (denoted as balancing condition in the mathematical literature).
Translating the results from localization to the associated field theories we see that such effective confining dualities have non-trivial monopole deformations, that indeed enforces the balancing condition on the identities. However such monopole deformations do not correspond to any KK monopole deformations of any 4d model in theses cases, and are more similar to other types of  linear monopole superpotential, like the ones first proposed in \cite{Benini:2017dud}.
Nevertheless such effective confining dualities can be studied in 3d through tensors deconfinement and furthermore it can be shown that, through opportune real mass deformations, that remove the constraints from the monopole superpotentials,  they flow to the confining dualities with symmetric tensors already studied in \cite{Amariti:2024gco}.

\section{3d $SU(N)$  with two antisymmetric tensors}
\label{sec2}

In this section we study 3d $\mathcal{N}=2$ theories with an $SU(N)$ gauge group and two antisymmetric.
Such models have been claimed in \cite{Nii:2019ebv} to be confining if there are $F \leq 4$ fundamentals and $4-F$ antifundamentals and vanishing superpotential.
The cases with $F=4$ and $F=3$ have been studied in \cite{Amariti:2024gco} from the point of view of tensor deconfinement.
Here we focus on the other three cases. Our analysis is performed distinguishing in each case the parity of $N$, i.e. $N=2n$ and $N=2n+1$. By deconfining the two antisymmetric tensors we show that in each case we can dualize the 
special unitary gauge group and confine the two antisymmetric tensors back. In each case we obtain a model with an $SU(N'<N)$ gauge group, two conjugate antisymmetric tensors and four antifundamentals. This model, up to conjugation, is confining and the proof from tensor deconfinement has been discussed in \cite{Amariti:2024gco}. We have reviewed such  confining duality in Appendix \ref{app2AS4fund}.

%
%
%
%
%
%
%
%
%
%
%
%
%
%
%
%%%%%%%%%%%%%%%%%%%%%%%%%%%%%%%%%%%
%%%%%%%%%%%%%%%%%%%%%%%%%%%%%%%%%%%
%%%%%%%%%%%%%%%%%%%%%%%%%%%%%%%%%%%
\subsection{$SU(2n)$ with 2 fundamentals and 2 antifundamentals}
\label{Sub_AA22ev}
%%%%%%%%%%%%%%%%%%%%%%%%%%%%%%%%%%%
%%%%%%%%%%%%%%%%%%%%%%%%%%%%%%%%%%%
%%%%%%%%%%%%%%%%%%%%%%%%%%%%%%%%%%%

\begin{figure}
\begin{center}
  \includegraphics[width=14cm]{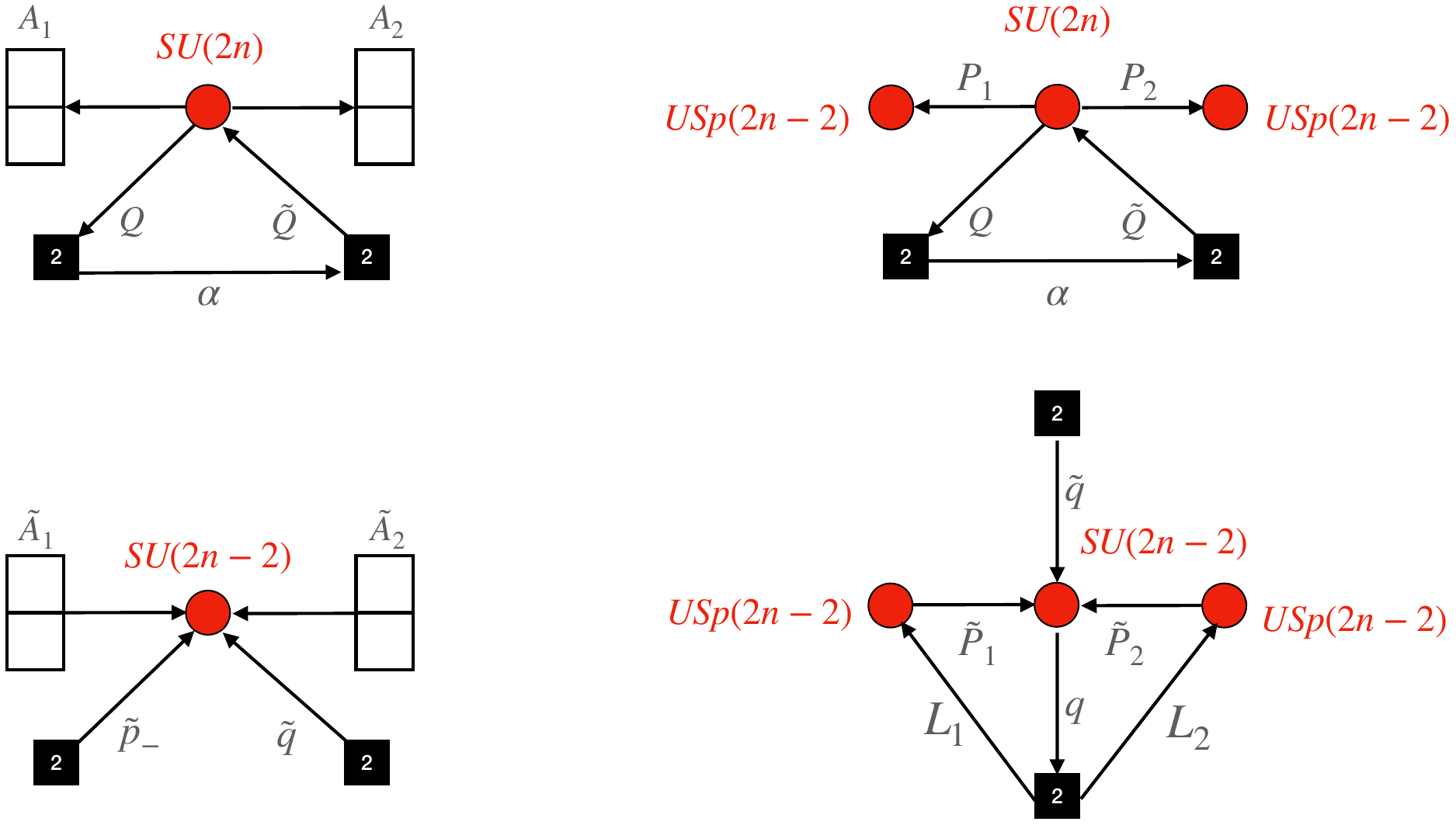}
  \end{center}
  \caption{In this figure we have summarized the various dual step implemented in the text of Subsection \ref{Sub_AA22ev}. On the top-left corner we provide the quiver for the starting $SU(2n)$ theory, with two antisymmetric tensors, 2 fundamentals and 2 antifundamentals. On the top-right corner we show the auxiliary quiver where the antisymmetric tensors are traded with two $USp(2n-2)$ gauge group with new bifundamentals $P_{1,2}$. On the bottom-right corner we present the quiver of the theory after having dualized the central $SU(2n)$ gauge node to $SU(2n-2)$. In the last bottom-left quiver we present the theory after the confinement of the two $USp(2n-2)$ gauge nodes. We refer to the text for the details regarding the superpotential, the duality mappings and the singlets appearing at each step. }
    \label{2AS22ev}
\end{figure}

We start our discussion by considering an $SU(2n)$ gauge group with two antisymmetric tensors $A_{1,2}$ and two fundamental flavors $Q$ and $\tilde Q$.
This theory was studied in \cite{Nii:2019ebv}, where a confining duality was proposed in terms of the singlets
$M=Q \tilde Q$, $\tilde B = A \tilde Q^2$, $T^{(n)} = A^n$, $T^{(n-1)} = A^{n-1} Q^2$, $P^{(1)} = A^{n-1} (A \tilde Q) Q$,
 $P^{(2)} = A^{n-1} (A \tilde Q)^2$ and $R=A^{n-2} (A \tilde Q)^2 Q^2$.
 There is also a dressed monopole
 \footnote{We refer the reader to \cite{Amariti:2024gco} for a long review on such states.
 See also \cite{Intriligator:2013lca,Csaki:2014cwa,Amariti:2015kha,Nii:2019ebv} for discussions. }
  $Y^{dressed} = Y^{bare}_{SU(2n-2)}A^{2n-4}$
 and the superpotential of the dual WZ model in this case is
 \begin{equation}
 \label{WKeita1}
 W = Y^d (M^2 {T^{(n)}}^2 + M P^{(1)} T^{(n)} + T^{(n)} R + T^{(n-1)} P^{(2)} + {P^{(1)}}^2 + \tilde B T^{(n-1)} T^{(n)}).
 \end{equation}

Here we want to obtain proof of such confining duality in terms of others (elementary) dualities along the lines of tensor deconfinement. Actually, instead of considering a vanishing superpotential, here we flip the operators Pf$\,A_{1,2}$, the operators $A_{1,2} \tilde Q^2$ and the meson $Q \tilde Q$. The flippers are denoted as $\rho_{1,2}$, $\beta_{1,2}$ and $\alpha$ respectively.
In this way the superpotential associated to the first quiver in Figure \ref{2AS22ev} is
\begin{equation}
\label{W2AS22evfirst}
W = \sum_{i=1}^2 (\rho_i \, \text{Pf} \, A_i+ \beta_i A_i \tilde Q^2) + \alpha Q \tilde Q.
\end{equation}
We then deconfine the two tensors $A_{1,2}$ using two $USp(2n-2)$ gauge groups. The two new bifundamentals emerging from the deconfinement are denoted as $P_{1,2}$ in Figure \ref{2AS22ev} and the
superpotential of this phase is
\begin{equation}
\label{W2AS22evsecond}
W =  \sum_{i=1}^2   \beta_i P_i^2 \tilde Q^2 + \alpha Q \tilde Q.
\end{equation}
The $SU(2n)$ gauge node has $4n-2$ fundamentals and $2$ antifundamentals. The dual theory was proposed in 
 \cite{Nii:2018bgf} and it was reviewed in appendix {\bf D.1.1} of \cite{Amariti:2024gco}.  We refer the reader to such reference for details. 
The dual gauge group becomes $SU(2n-2)$  
and the superpotential is 
\begin{equation}
\label{W2AS22evthirdmassless}
W =\sum_{i=1}^2   ( L_i \tilde P_i q+\beta_i L_i^2),
\end{equation}
where we have integrated out the massive meson $M_{Q \tilde Q}$ and the flipper $\alpha$.

At this point we observe that the two $USp(2n-2)$ gauge groups are confining and each one gives rise to a
conjugate antisymmetric, $\tilde A_{1}$ and $\tilde A_{2}$ respectively and a field in the antifundamental representation of $SU(2n-2)$,  $ \tilde p_1 = L_1 \tilde P_1$ and  $ \tilde p_2 = L_2 \tilde P_2$ respectively.
Integrating out the massive fields the final superpotential is
\begin{equation}
\label{W2AS22evfourth}
W = \sum_{i=1}^2   \tilde \rho_i \, \epsilon_{2n-2} \cdot (\tilde A_i^{n-2} \tilde p_{-}^2),
\end{equation}
where $\tilde p_- = \tilde p_1 - \tilde p_2$.

Next we can use the confinement of $SU(2n-2)$ with two antisymmetric and four fundamentals, proposed in \cite{Nii:2019ebv}
and studied from tensor deconfinement in \cite{Amariti:2024gco}.
Such confining duality has been reviewed in appendix \ref{app2AS4fund} for completeness.
Observe that in the case at hand here case the $SU(4) \times SU(2)$ non abelian flavor symmetry is partially broken by the superpotential (\ref{W2AS22evfourth}). Furthermore the model described in the fourth quiver in Figure \ref{2AS22ev} differs from the one in appendix  \ref{app2AS4fund}  for an overall conjugation.
Keeping in mind these differences here we split the indices of the $SU(4)$ flavor symmetry into the indices of two non-abelian $SU(2)$ flavor symmetries, denoting them as $SU(2)_a \times SU(2)_b$, in addition to an abelian $U(1)$.
Moreover we have the relation $m=n-1$ mapping the ranks of the gauge groups.
Using these rules the $SU(2n-2)$ model is confining in terms of the following singlets 
\begin{equation}
\label{singletsfirstrel}
\begin{array}{lclc}
t^{(n-1)}_j &=& \Tilde{A}_1^{j} \Tilde{A}_2^{n-1-j}, \quad \quad & j=0,...,n-1 \\
 t^{(n-2,aa)}_j &=& \Tilde{A}_1^{j} \Tilde{A}_2^{n-2-j} p_{-}^2, \quad \quad & j=0,...,n-2 \\
 t^{(n-2,ab)}_j &=& \Tilde{A}_1^{j} \Tilde{A}_2^{n-2-j} p_{-} \Tilde{q}, \quad  \quad& j=0,...,n-2 \\
 t^{(n-2,bb)}_j &=& \Tilde{A}_1^{j} \Tilde{A}_2^{n-2-j} \Tilde{q}^2, \quad  \quad& j=0,...,n-2 \\
t^{(n-3)}_j &=& \Tilde{A}_1^{j} \Tilde{A}_2^{n-3-j} p_{-}^2 \Tilde{q}^2, \quad & j=0,...,n-3 \\
\end{array}
\end{equation}

These singlets  are mapped to the singlets of the original $SU(2n)$ theory with $2 \square$ and $2 \Bar{\square}$ as 
\begin{equation}
\label{singletssecondrel}
    \begin{array}{lcllcllcl}
         t^{(n-1)}_j &\rightarrow& T^{(n-1)}_{n-1-j}, \quad \quad &t^{(n-2,aa)}_j &\rightarrow& R_{n-3-j}, \quad \quad&t^{(n-2,ab)}_j &\rightarrow& P^{(1)}_{n-1-j}, \\
         t^{(n-2,bb)}_j &\rightarrow& T^{(n)}_{n-1-j}, \quad \quad &t^{(n-3)}_j &\rightarrow& P^{(2)}_{n-3-j}, \quad \quad &y^d_j &\rightarrow& Y^d_{2n-4-j}.
    \end{array}
\end{equation}
The final superpotential is given by (\ref{Wconf4e}) in addition to the deformation (\ref{W2AS22evfourth}). Using the singlets in formula (\ref{singletsfirstrel}) this gives the following confining superpotential
\begin{equation}
\label{intermediofinale}
    W = y^d_k ( t^{(n-1)}_j t^{n-3}_{2n-4-k-j} + t^{(n-2,aa)}_\ell t^{(n-2,bb)}_{2n-4-k-\ell} + t^{(n-2,ab)}_s t^{(n-2,ab)}_{2n-4-k-s}) + \Tilde{\rho}_1 t^{(n-2,aa)}_0 + \Tilde{\rho}_2 t^{(n-2,aa)}_{n-2},
\end{equation}
where the sums over  $k=0,...,2n-4$, $j= 0,..., n-1 $, $\ell=1,...n-3$ and $s = 0,..., n-2$ are understood.
The deformation (\ref{W2AS22evfourth}) is mapped in the last two terms in (\ref{intermediofinale}) where the monopoles $\Tilde{\rho}_1$ give mass to  $t^{(m-2,aa)}_0$ and $t^{(m-2,aa)}_{n-2}$, which then are not mapped to any of the $R$ fields. Using  the mapping (\ref{singletssecondrel}), the final superpotential is 
\begin{equation}
\label{lastour}
    \begin{split}
    W=&  Y^d_{2n-4-k}  ( T^{(n-1)}_{n-1-j} P^{(2)}_{k+j+1-n} + R_{n-3-\ell} T^{(n)}_{k+\ell+3-n} + P^{(1)}_{n-1-s} P^{(1)}_{k+s+3-n}).
    \end{split}
\end{equation}
We conclude the analysis observing that  the superpotential (\ref{lastour})  is exactly the one that is obtained by adding the deformation (\ref{W2AS22evfirst}) to the  superpotential (\ref{WKeita1}).

We can also reproduce the proof of the confining duality given above by studying the matching of the three sphere partition function.
The relation that we want to prove in this case is

\begin{eqnarray}
&&\label{2fund2antifund2ASeven}
Z_{SU(2n)} (\vec \mu;\vec \nu;\cdot;\cdot;\vec \tau;\cdot)
=
\prod_{a,b =1}^2 \Gamma_h(\mu_a +\nu_b)
 \prod_{\ell=1}^2  
 \Gamma_h(\tau_\ell + \nu_1 +\nu_2)
\prod_{j=0}^{n} \Gamma_h(j \tau_1 + (n-j) \tau_2)
\nonumber \\
&&
\prod _{j=0}^{n-1} \Gamma_h (\tau _1 (n-1-j)+j \tau _2+\mu _1+\mu _2)
\prod _{j=0}^{n-4} \Gamma_h \left(\tau _1 (n-j-2)+(j+2) \tau _2+\sum_{a=1}^2 (\mu _a+\nu _a)\right)
\nonumber \\
&&
\prod _{j=0}^{n-3} \Gamma_h \left(\tau _1 (n-1-j)+(j+2) \tau _2+\nu _1+\nu _2\right)
\prod _{j=0}^{n-2} \prod_{a,b=1}^2\Gamma_h \left(\tau _1 (n-1-j)+(j+1) \tau _2+\mu _a+\nu _b \right)
\nonumber \\
&&
\prod _{j=0}^{2 n-4} \Gamma_h \left(2 \omega-\tau _1 (2 n-j-2)-(j+2) \tau _2
-\sum_{a=1}^{2} (\mu _a+\nu _a) \right).
\end{eqnarray}

In order to prove such relation we follow the deconfinement and 
duality steps discussed from the field theory approach above.
We start by reading the mass parameters of the flippers  $\alpha,\beta$
and $\rho$ in the first quiver in Figure \ref{2AS22ev}. They are
\begin{equation}
m_{\alpha^{(a,b)}} = 2\omega - \mu_a -\nu_b,\quad 
m_{\rho_{\ell}} = 2\omega-n \tau_{\ell},\quad
m_{\beta_{\ell}} = 2\omega - \tau_{\ell} - \nu_1 -\nu_2,
\end{equation}
with $\ell=1,2$ and $a,b=1,2$.
Then we deconfine the antisymmetric tensors, by reversing the identities associated to Aharony duality for symplectic SQCD, that have been discussed in a physical language in \cite{Willett:2011gp,Benini:2011mf}. The mass parameters for the new bifundamentals $P_{\ell}$ are
\begin{equation}
m_{P_{\ell}} = \frac{\tau_{\ell}}{2}.
\end{equation}
The duality step requires to define an auxiliary quantity
\begin{equation}
X =\frac{(n-1) (\tau_1+\tau_2)+\mu_1+\mu_2}{2n-2}=\frac{\tau_1+\tau_2}{2}+\frac{\mu_1+\mu_2}{2(n-1)},
\end{equation}
such that the masses of the fields in the third quiver in Figure \ref{2AS22ev} are
\begin{equation}
m_{q^{(a)}} = 2 \omega-X - \nu_{a},\quad
m_{\tilde q^{(a)}} = X - \mu_{a},\quad 
m_{\tilde P_\ell} = X - \frac{\tau_\ell}{2},\quad
m_{L_{\ell}^{(a)}}=\frac{\tau_{\ell}}{2} +\nu_{a}.
\end{equation}
There is a further field with mass parameter $m_{M_{Q \tilde Q}^{(a,b)}} = \mu_a +\nu_b$, but it disappears together with the flipper $\alpha$, thanks to the inversion relation 
\begin{equation}
\label{inversionHyp}
\Gamma_h(x) \Gamma_h(2\omega-x)=1,
\end{equation}
for the hyperbolic Gamma functions.
Then the confinement of the two symplectic nodes gives rise to the mesonic singlets $L_\ell^{(1)} \cdot L_\ell^{(2)} $ and to the singlets $\tilde \rho_{\ell}$, identified with the fundamental monopoles of the $USp(2n-2)$ gauge groups. The real masses for these singlets are 
\begin{equation}
m_{L_{\ell}^{(1)} \cdot L_{\ell}^{(2)} } = \tau_{\ell}+\nu_{1}+\nu_{2}
,\quad
m_{\tilde \rho_{1,2}}=2 \omega-(n-1)\tau_{1,2}+\tau_{2,1}-\sum_{a=1}^{2} (\mu_a+\nu_a).
\end{equation}
The singlets obtained from $L_{\ell}^{(1)} \cdot L_{\ell}^{(2)} $  disappear, due to the holomorphic mass
with the  flippers $\beta_{\ell} $ that can be read from (\ref{W2AS22evthirdmassless}). At the level of the partition function this translates to 
an application of the  inversion relation (\ref{inversionHyp}), i.e. 
$\Gamma_h(m_{\beta_{\ell}}) \Gamma_h(m_{L_{\ell}^{(1)}} \cdot m_{L_{\ell}^{(2)}}   ) =1$ for both $\ell=1$ and $\ell=2$.
The mass parameters of the charged matter fields are 
\begin{equation}
m_{\tilde A_{\ell}} = 2 X - \tau_{\ell}, \quad m_{\tilde p_-^{(a)}}= X+\nu_a
,\quad
m_{\tilde q ^{(a)}} =X-\mu_{a},
\end{equation}
where we can re-organize the four antifundamentals as
\begin{eqnarray}
\hat m = \{m_{\tilde q ^{(1)}},m_{\tilde q ^{(2)}}, m_{\tilde p_-^{(1)}},m_{\tilde p_-^{(2)}} \}.
\end{eqnarray}
In this way we have arrived to an identity between the partition function of the first and the last quivers in Figure \ref{2AS22ev}, that reads
\begin{eqnarray}
\label{id14even2n22}
&&
\prod_{\ell=1}^{2} \Gamma_h(m_{\rho_{\ell}},m_{\beta_{\ell}})
\prod_{a,b=1}^{2}  \Gamma_h(m_{\alpha^{(a,b)}} )
Z_{SU(2n)} (\vec \mu;\vec \nu;\cdot;\cdot;\vec \tau;\cdot)
\nonumber \\
=
&& 
\prod_{\ell=1}^{2}
\Gamma_h(m_{\tilde \rho_\ell})
Z_{SU(2n-2)} (\cdot ;\vec {\hat m};\cdot;\cdot;\cdot ;2X-\vec \tau).
\end{eqnarray}
The integral $Z_{SU(2n-2)} (\cdot ;\vec {\hat m};\cdot;\cdot;\cdot ;2X-\vec \tau)$ can be evaluated using (\ref{4fund2ASeven}), up to an overall conjugation that does not modify the result.
Once such integral is plugged in (\ref{id14even2n22}) we apply (\ref{inversionHyp}), moving the Gamma function of the flippers $\rho, \beta$ and $\alpha$ on the RHS and eliminating the singlets $\tilde \rho$.  Finally we obtained the expected relation (\ref{2fund2antifund2ASeven}).
%
%
%
%
%
%
%
%
%
%
%
%
%
%
%
%%%%%%%%%%%%%%%%%%%%%%%%%%%%%%%%%%%
%%%%%%%%%%%%%%%%%%%%%%%%%%%%%%%%%%%
%%%%%%%%%%%%%%%%%%%%%%%%%%%%%%%%%%%
\subsection{$SU(2n+1)$ with 2 fundamentals and 2 antifundamentals}
\label{Sub_AA22odd}
%%%%%%%%%%%%%%%%%%%%%%%%%%%%%%%%%%%
%%%%%%%%%%%%%%%%%%%%%%%%%%%%%%%%%%%
%%%%%%%%%%%%%%%%%%%%%%%%%%%%%%%%%%%
\begin{figure}
\begin{center}
  \includegraphics[width=14cm]{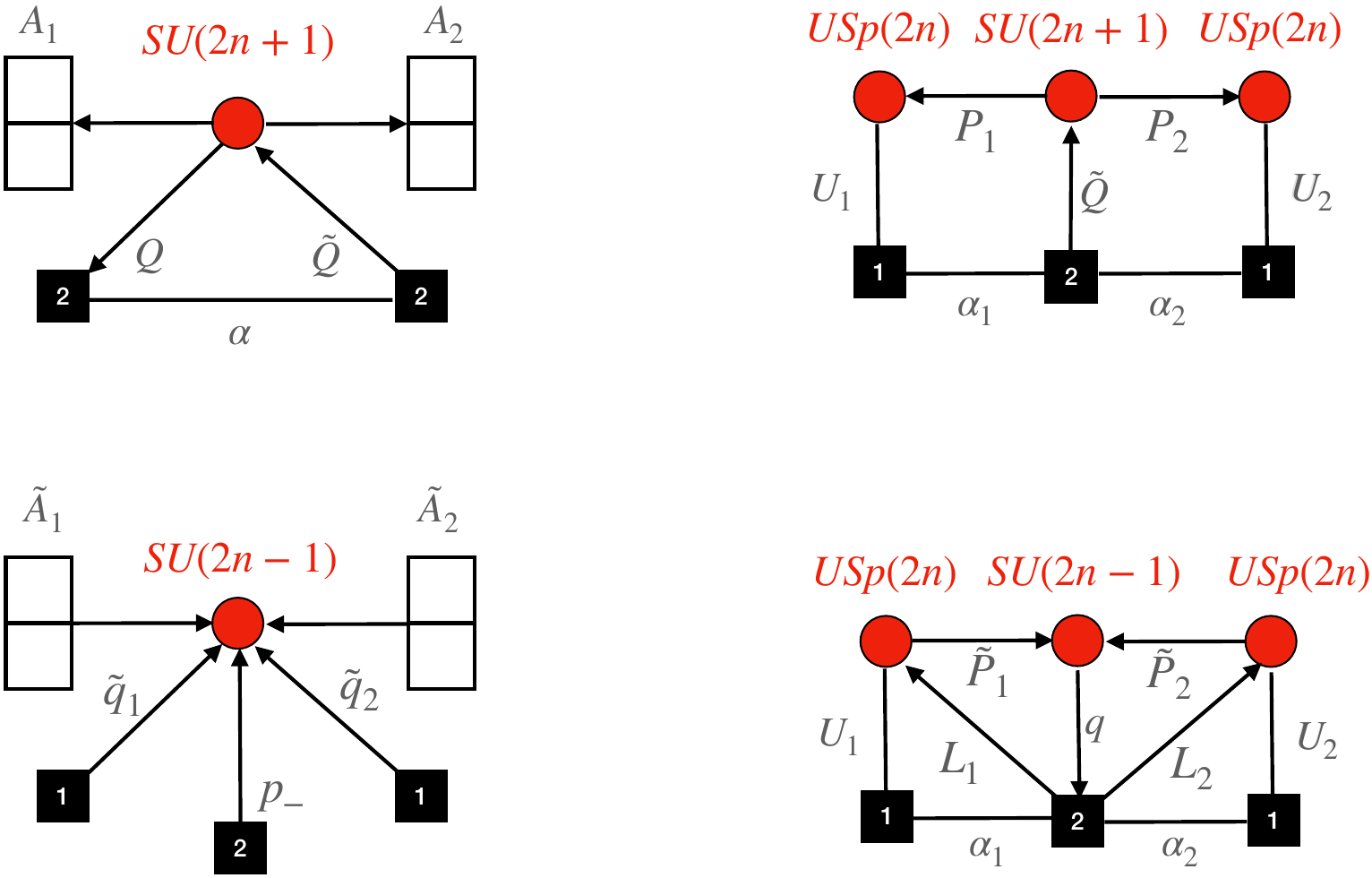}
  \end{center}
  \caption{ In this figure we have summarized the various dual step implemented in the text of Subsection \ref{Sub_AA22odd}. On the top-left corner we provide the quiver for the starting $SU(2n+1)$ theory, with two antisymmetric tensors, 2 fundamentals and 2 antifundamentals. On the top-right corner we show the auxiliary quiver where the antisymmetric tensors are traded with two $USp(2n)$ gauge group with new bifundamentals $P_{1,2}$ and new charged fields $U_{1,2}$. On the bottom-right corner we present the quiver of the theory after having dualized the central $SU(2n+1)$ gauge node to $SU(2n-1)$. In the last bottom-left quiver we present the theory after the confinement of the two $USp(2n)$ gauge nodes. We refer to the text for the details regarding the superpotential, the duality mappings and the singlets appearing at each step.  }
    \label{2AS22odd}
\end{figure}

We proceed by considering an $SU(2n+1)$ gauge group with two antisymmetric tensors $A_{1,2}$ and two fundamental flavors $Q$ and $\tilde Q$.
This theory was studied in \cite{Nii:2019ebv}, where a confining duality was proposed in terms of the singlets
$M=Q \tilde Q$, $\tilde B = A \tilde Q^2$, $T^{(n)} = A^n$, $T^{(n-1)} = A^{n-1} (A \tilde Q)^2 Q$, $P^{(n)} = A^{n} (A \tilde Q)$,
 $P^{(n-1)} = A^{n-1} (A \tilde Q) Q^2$.
 There is also a dressed monopole $Y^{dressed} = Y^{bare}_{SU(2n-1)}A^{2n-3}$
 and the superpotential of the dual WZ model in this case is
 \begin{equation}
 \label{WKeita2}
 W = Y^d (M {T^{(n)}} P^{(n)} +\tilde B {T^{(n)}}^2+T^{(n)}  T^{(n-1)} + P^{(n)}  P^{(n-1)} ).
 \end{equation}

Here we want to obtain proof of such confining duality in terms of others (elementary) dualities along the lines of tensor deconfinement. 
Actually, instead of considering a vanishing superpotential, here we flip the operators Pf$\,A_{1,2}$, the operators $A_{1,2} \tilde Q^2$ and the meson $Q \tilde Q$. The flippers are denoted as $\rho_{1,2}$, $\beta_{1,2}$ and $\alpha$ respectively.
In this way the superpotential associated to the first quiver in Figure \ref{2AS22odd} is

\begin{equation}
\label{W2AS22oddfirst}
W = \sum_{i=1}^2 (\rho_i \, A_i^n Q_i +\beta_i A_i \tilde Q^2) + \alpha Q \tilde Q.
\end{equation}
We then deconfine the two tensors $A_{1,2}$ using two $USp(2n)$ gauge groups. The two new bifundamentals emerging from the deconfinement are denoted as $P_{1,2}$ in Figure \ref{2AS22odd}. There are also two new $USp(2n)_{1,2}$ charged fields denoted as $U_{1,2}$, associated to the original fundamentals $Q_{1,2}$ through the relations $Q_{1,2} = U_{1,2} P_{1,2}$.
The singlet $\alpha$ splits into $\alpha_{1}$ and   $\alpha_2$ as well
and the superpotential of this phase is
\begin{equation}
\label{W2AS22oddsecond}
W = \sum_{i=1}^2 (\alpha_i U_i P_i \tilde Q +\beta_i P_i^2 \tilde Q^2).
\end{equation}

The $SU(2n+1)$ gauge node has $4n$ fundamentals and $2$ antifundamentals. The dual theory was proposed in 
 \cite{Nii:2018bgf} and it was reviewed in appendix {\bf D.1.1} of \cite{Amariti:2024gco}. It is an $SU(2n-1)$ gauge theory  
and its superpotential is 
\begin{equation}
\label{W2AS22oddthirdmassless}
W = \sum_{i=1}^2 ( L_i \tilde P_i q + U_i L_i \alpha_i +\beta_i L_i^2), 
\end{equation}
where $L_{1,2} = P_{1,2} \tilde Q$ are the mesons  of this duality.

At this point we observe that the two $USp(2n)$ gauge groups are confining and each one gives rise to a conjugate antisymmetric $\tilde A_{1,2} = \tilde P_{1,2}^2$, an antifundamental $\tilde q_{1,2} = U_{1,2} P_{1,2}$, another antifundamental
$p_{1,2} = L_{1,2} \tilde P_{1,2}$ and a singlet $\ell_{1,2} = L_{1,2}^2$.
Integrating out the massive fields the final superpotential is
\begin{equation}
\label{W2AS22oddfourth}
W = \sum_{i=1}^2 \tilde{\rho}_i \tilde{A}_i^{n-2} \tilde{q}_i  \tilde{p}_-^2, 
\end{equation}
where $\tilde p_- = \tilde p_1 - \tilde p_2$ and $\tilde \rho_{1,2}$ correspond to the fundamental  monopoles of $USp(2n)_{1,2}$.

Next we can use the confinement of $SU(2n-1)$ with two antisymmetric and four fundamentals, reviewed in appendix \ref{app2AS4fund} for completeness.
Observe that here the $SU(4) \times SU(2)$ non abelian flavor symmetry is partially broken by the superpotential (\ref{W2AS22oddfourth}). Furthermore the model described in the fourth quiver in Figure \ref{2AS22odd} differs from the one in appendix  \ref{app2AS4fund}  for an overall conjugation.
Keeping in mind these differences here we split the indices of the $SU(4)$ flavor symmetry into the indices of two abelian $U(1)$ and a non-abelian $SU(2)$ flavor symmetry, denoting them as $U(1)_a \times U(1)_b \times SU(2)_c$.
Moreover we have the relation $m=n-1$ mapping the ranks of the gauge groups.
Using these rules the $SU(2n-1)$ model is confining in terms of the following singlets 
\begin{equation}
\label{singletsfirstrelodd}
\begin{array}{lclc}
t^{(n-1,a)}_j &=& \Tilde{A}_1^{j} \Tilde{A}_2^{n-1-j} \tilde{q}_1, \quad \quad &\\
t^{(n-1,b)}_j &=& \Tilde{A}_1^{j} \Tilde{A}_2^{n-1-j} \tilde{q}_2, \quad \quad & j=0,...,n-1 \\
t^{(n-1,c)}_j &=& \Tilde{A}_1^{j} \Tilde{A}_2^{n-1-j} \tilde p_{-}, \quad \quad & \\
 t^{(n-1,abc)}_\ell &=& \Tilde{A}_1^{\ell} \Tilde{A}_2^{n-2-\ell} \tilde{q}_1 \tilde{q}_2 \tilde p_{-}, \quad \quad &  \\
 t^{(n-1,acc)}_\ell &=& \Tilde{A}_1^{\ell} \Tilde{A}_2^{n-2-\ell} \tilde{q}_1 \tilde p_{-}^2, \quad \quad & \ell=0,...,n-2 \\
 t^{(n-1,bcc)}_\ell &=& \Tilde{A}_1^{\ell} \Tilde{A}_2^{n-2-\ell} \tilde{q}_2 \tilde p_{-}^2, \quad \quad & \\
\end{array}
\end{equation}
and are mapped to the singlets of the $SU(2n+1)$ theory with $2 \square$ and $2 \Bar{\square}$ as 
\begin{equation}
\label{singletssecondrel2}
    \begin{array}{lllllllll}
         t^{(n-1,a)}_j &\rightarrow& T^{(n)}_{n-1-j}, \quad &\quad\quad
           t^{(n-1,b)}_j &\rightarrow& T^{(n)}_{n-j}, \quad &\quad\quad
           t^{(n-1,c)}_j &\rightarrow& P^{(n)}_{n-1-j}, \\ 
           t^{(n-2,abc)}_j &\rightarrow& P^{(n-1)}_{n-2-j}, \quad&\quad\quad
          t^{(n-2,acc)}_j &\rightarrow& T^{(n-1)}_{n-3-j}, \quad &\quad\quad
            t^{(n-2,bcc)}_j &\rightarrow& T^{(n-1)}_{n-2-j}.
    \end{array}
\end{equation}
The final superpotential is given by (\ref{Wconf4o}) in addition to the deformation (\ref{W2AS22oddfourth}). Using the singlets in formula (\ref{singletsfirstrelodd}) this gives the following confining superpotential
\begin{equation}
\label{intermediofinale2}
\begin{split}
W = & y^d_k (t^{(n-1,a)}_j t^{(n-2, acc)}_{2n-k-j-4} + t^{(n-1,a)}_j t^{(n-2, bcc)}_{2n-k-j-3} + t^{(n-1,b)}_j t^{(n-2, acc)}_{2n-k-j-3} + t^{(n-1,b)}_j t^{(n-2, bcc)}_{2n-k-j-2} +  \\  & t^{(n-1,c)}_j t^{(n-2, abc)}_{2n-k-j-3}) + \tilde{\rho}_1 t^{(n-2,acc)}_{n-2} + \tilde{\rho}_2 t^{(n-2,bcc)}_{0},
\end{split}
\end{equation}
where the sums over $k=0,...,2n-3$ and $j=0,...,n-1$ are understood.
The deformation (\ref{W2AS22oddfourth}) is mapped in the last two terms in (\ref{intermediofinale2})
where the monopoles $\Tilde{\rho}_1$ give mass to  $t^{(n-2,acc)}_{n-2}$ and $t^{(n-2,bcc)}_{0}$, which then are not mapped to any of the singlets. Using  the mapping (\ref{singletssecondrel2}), the final superpotential is 
\begin{equation}
\label{lastourbis}
    \begin{split}
    W =& \sum_{\{ k,j\}}  Y^d_{2n-3-k}  ( T^{(n,a)}_{n-1-j} T^{(n-1,acc)}_{k+j+1-n} + T^{(n,a)}_{n-1-j} T^{(n-1,bcc)}_{k+j+1-n} +T^{(n,b)}_{n-j} T^{(n-1,acc)}_{k+j-n} +\\
    & + T^{(n,b)}_{n-j} T^{(n-1,bcc)}_{k+j-n} + P^{(n)}_{n-1-j} P^{(n-1)}_{k+j+1-n}  ).
    \end{split}
\end{equation}
We can also reproduce the proof of the confining duality given above by studying the matching of the three sphere partition function.
The relation that we want to prove in this case is
\begin{eqnarray}
&&\label{2fund2antifund2ASodd}
Z_{SU(2n+1)} (\vec \mu;\vec \nu;\cdot;\cdot;\vec \tau;\cdot)
=
\prod_{a,b =1}^2 \Gamma_h(\mu_a +\nu_b)
 \prod_{\ell=1}^2 
 \Gamma_h(\tau_\ell + \nu_1 +\nu_2)
\nonumber \\
&&
\prod _{j=0}^{n-2} \prod_{a=1}^2
\Gamma_h \left((n-1-j)\tau _1 +(j+1) \tau _2+\mu _1+\mu _2+\nu _a\right)
\nonumber \\
&&
\prod _{j=0}^{n-3} \prod_{a=1}^2
\Gamma_h \left((n-1-j)\tau _1 +(j+2) \tau _2+\mu _a+\nu _1+\nu _2\right)
\nonumber \\
&&
\prod _{j=0}^{2 n-3} \Gamma_h \left(
2 \omega -
(2 n-1-j) \tau _1 -(j+2) \tau _2
-\sum_{a=1}^2 (\mu_a +\nu_a)\right)
\nonumber \\
&&
\prod_{a=1}^2 \left(
\prod _{j=0}^{n}   \Gamma_h \left(\tau _1 (n-j)+j \tau _2+\mu _a \right)
\cdot
\prod _{j=0}^{n-1}  \Gamma_h \left(\tau _1 (n-j)+(j+1) \tau _2+\nu _a\right)
\right).
\nonumber \\
\end{eqnarray}
In order to prove such relation we follow the deconfinement and 
duality steps discussed from the field theory approach above.
We start by reading the mass parameters of the flippers  $\alpha,\beta$
and $\rho$ in the first quiver in Figure \ref{2AS22odd}. They are
\begin{equation}
m_{\rho_{\ell}} = 2\omega-n \tau_{\ell},\quad
m_{\beta_{\ell}} = 2\omega - \tau_{\ell} - \nu_1 -\nu_2
,\quad
m_{\alpha^{(a,b)}} = 2\omega - \mu_a -\nu_b,
\end{equation}
with $\ell=1,2$ and $a,b=1,2$.
Then we deconfine the antisymmetric tensors and the mass parameter for the new bifundamentals $P_{1,2}$ and for the fields $U_{1,2}$ are
\begin{equation}
m_{P_{\ell}} = \frac{\tau_{\ell}}{2}, \quad
m_{U_\ell^{(a)}} = \mu_{a} - \frac{\tau_{\ell}}{2}.
\end{equation}
The duality step requires to define an auxiliary quantity
\begin{equation}
X =\frac{n (\tau_1+\tau_2)}{2n-1},
\end{equation}
such that the masses of the fields in the third quiver in Figure \ref{2AS22odd} are
\begin{equation}
m_{\tilde P_{\ell}} = X -\frac{\tau_{\ell}}{2},\quad
m_{q}^{(a)} = 2\omega - X - \nu_a,\quad
m_{L_\ell^{(a)}} = \frac{\tau_{\ell}}{2}+\nu_{a}.
\end{equation}
Then the confinement of the two symplectic nodes gives rise to the singlets 
\begin{eqnarray}
&&m_{L_{\ell}^{(1)} L_{\ell}^{(2)} } = \tau_{\ell}+\nu_{1}+\nu_{2}, \quad
m_{U_{a}^{(a)} L_{a}^{(b)}} = \mu_{a}+\nu_{b},  \\
&&m_{\tilde \rho_{1,2}} = 2\omega-n \tau_{2,1}- \tau_{1,2}-\nu_{1}-\nu_{2}-\mu_{1,2},  \nonumber 
\end{eqnarray}
while the terms in the first line disappear together with the flippers $\alpha$ and $\beta$, because of the inversion relation, the  singlets in the second line have to be considered.
The mass parameters of the charged matter fields are 
\begin{equation}
m_{\tilde A_{1,2}} = 2 X - \tau_{1,2}, \quad m_{\tilde p_-^{(a)}}= X+\nu_a
,\quad
m_{\tilde q^{(a)}} = X - \tau_{a}-\mu_{a},
\end{equation}
where we can reorganize the four antifundamentals as
\begin{eqnarray}
\hat m = \{m_{\tilde q^{(1)}} ,m_{\tilde q^{(2)}}, m_{\tilde p_-^{(1)}}, m_{\tilde p_-^{(2)}}\}.
\end{eqnarray}
In this way we have arrived to an identity between the partition function of the first and the last quivers in Figure \ref{2AS22odd}, that reads
\begin{eqnarray}
\label{id14odd2n22}
&&
\prod_{\ell=1}^{2} \Gamma_h(m_{\rho_{\ell}},m_{\beta_{\ell}})
\prod_{a,b=1}^{2}  \Gamma_h(m_{\alpha^{(a,b)}} )
Z_{SU(2n+1)} (\vec \mu;\vec \nu;\cdot;\cdot;\vec \tau;\cdot)
\nonumber \\
=
&& 
\prod_{\ell=1}^{2}
\Gamma_h(m_{\tilde \rho_\ell})
Z_{SU(2n-1)} (\cdot ;\vec {\hat m};\cdot;\cdot;\cdot ;2X-\vec \tau).
\end{eqnarray}
The integral $Z_{SU(2n-1)} (\cdot ;\vec {\hat m};\cdot;\cdot;\cdot ;2X-\vec \tau)$ can be evaluated using (\ref{4fund2ASodd}), up to an overall conjugation that does not modify the result.
Once such integral is plugged in (\ref{id14odd2n22}), we apply the inversion relation to move the Gamma function of the flippers $\rho, \beta$ and $\alpha$ on the RHS and eliminate the singlets $\tilde \rho$, finally obtaining (\ref{2fund2antifund2ASodd}).

%
%
%
%
%
%
%
%
%
%%%%%%%%%%%%%%%%%%%%%%%%%%%%%%%%%%
%%%%%%%%%%%%%%%%%%%%%%%%%%%%%%%%%%
%%%%%%%%%%%%%%%%%%%%%%%%%%%%%%%%%%
\subsection{$SU(2n)$ with 1 fundamental and 3 antifundamentals}
\label{Sub_AA13ev}
%%%%%%%%%%%%%%%%%%%%%%%%%%%%%%%%%%
%%%%%%%%%%%%%%%%%%%%%%%%%%%%%%%%%%

\begin{figure}
\begin{center}
  \includegraphics[width=14cm]{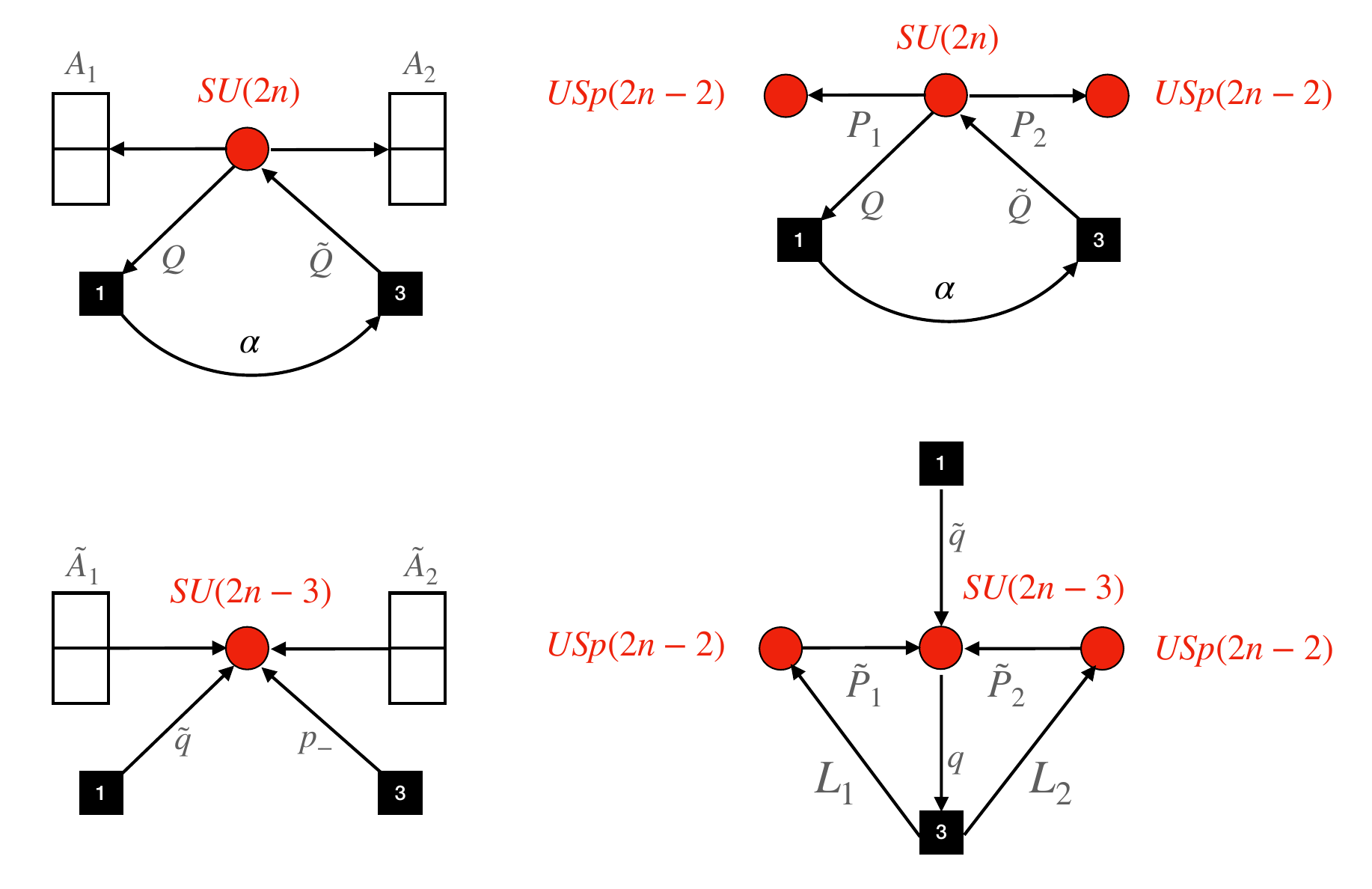}
  \end{center}
  \caption{In this figure we have summarized the various dual step implemented in the text of Subsection \ref{Sub_AA13ev}. On the top-left corner we provide the quiver for the starting $SU(2n)$ theory, with two antisymmetric tensors, 1 fundamental and 3 antifundamentals. On the top-right corner we show the auxiliary quiver where the antisymmetric tensors are traded with two $USp(2n-2)$ gauge group with new bifundamentals $P_{1,2}$. On the bottom-right corner we present the quiver of the theory after having dualized the central $SU(2n)$ gauge node to $SU(2n-3)$. In the last bottom-left quiver we present the theory after the confinement of the two $USp(2n-2)$ gauge nodes. We refer to the text for the details regarding the superpotential, the duality mappings and the singlets appearing at each step. }
    \label{2AS13ev}
\end{figure}

We then consider an $SU(2n)$ gauge group with two antisymmetric tensors $A_{1,2}$, one fundamental $Q$ and three antifundamentals $\tilde Q$.
This theory was studied in \cite{Nii:2019ebv}, where a confining duality was proposed in terms of the singlets
$M=Q \tilde Q$, 
$\tilde B = A \tilde Q^2$, 
$T^{(n)} = A^n$, 
$P^{(1)} = A^{n-1} (A \tilde Q) Q$,
 $P^{(2)} = A^{n-1} (A \tilde Q)^2$ 
 and 
$P^{(3)} = A^{n-2} (A \tilde Q)^3 Q$ .
  There is also a dressed monopole $Y^{dressed} = Y^{bare}_{SU(2n-2)}A^{2n-5}$
 and the superpotential of the dual WZ model in this case is
  \begin{equation}
 \label{WKeita3}
 W = Y^{dressed} \left( T^{(n)} \tilde B P^{(1)} + P^{(1)} P^{(2)} +T^{(n)} P^{(3)} +T^{(n)} P^{(2)} M \right).
 \end{equation}
Here we want to obtain a proof of such confining duality in terms of others (elementary) dualities along the lines of tensor deconfinement. 
Actually, instead of considering a vanishing superpotential, here we flip the operators Pf$\,A_{1,2}$, the operators $A_{1,2} \tilde Q^2$ and the meson $Q \tilde Q$. The flippers are denoted as $\rho_{1,2}$, $\beta_{1,2}$ and $\alpha$ respectively.
In this way the superpotential associated to the first quiver in Figure \ref{2AS13ev} is
\begin{equation}
\label{W2AS13evfirst}
W = \sum_{i=1}^2 (\rho_i \, \text{Pf} \, A_i+\beta_i A_i \tilde Q^2 )+ \alpha Q \tilde Q.
\end{equation}
We then deconfine the two tensors $A_{1,2}$ using two $USp(2n-2)$ gauge group. The two new bifundamentals emerging from the deconfinement are denoted as $P_{1,2}$ in Figure \ref{2AS13ev} and the
superpotential of this phase is
\begin{equation}
\label{W2AS13evsecond}
W =  \sum_{i=1}^2 \beta_i P_i^2 \tilde Q^2 + \alpha Q \tilde Q.
\end{equation}
The $SU(2n)$ gauge node has $4n-3$ fundamentals and $3$ antifundamentals.  The dual theory was proposed in 
 \cite{Nii:2018bgf} and it was reviewed in appendix {\bf D.1.1} of \cite{Amariti:2024gco}. It  is an $SU(2n-3)$ gauge theory  
and its superpotential is 
\begin{equation}
\label{W2AS13evthirdmassless}
W =  \sum_{i=1}^2 (L_i \tilde P_i q+\beta_i L_i^2),
\end{equation}
where we have integrated out the massive meson $M_{Q \tilde Q}$ and the flipper $\alpha$.

At this point we observe that the two $USp(2n-2)$ gauge groups are confining and each one gives rise to a
conjugate antisymmetric, $\tilde A_{1}$ and $\tilde A_{2}$ respectively and a meson in the antifundamental representation of $SU(2n-3)$,  $ \tilde p_1 = L_1 \tilde P_1$ and  $ \tilde p_2 = L_2 \tilde P_2$ respectively.
Integrating out the massive fields the final superpotential is
\begin{equation}
\label{W2AS13evfourth}
W = \sum_{i=1}^2 \tilde{\rho}_i \tilde{A}_i^{n-3} p_{-}^3,
\end{equation}
where $\tilde p_- = \tilde p_1 - \tilde p_2$.

Next we can use again the confinement of $SU(2n-3)$ with two antisymmetric tensors and four fundamentals.
Observe that in the case at hand here the $SU(4) \times SU(2)$ non abelian flavor symmetry is partially broken by the superpotential (\ref{W2AS13evfourth}). Furthermore the model described in the fourth quiver in Figure \ref{2AS13ev} differs from the one in appendix  \ref{app2AS4fund}  for an overall conjugation.
Keeping in mind these differences, here we split the indices of the $SU(4)$ flavor symmetry into the indices of a non-abelian $SU(3)_b$ flavor symmetries, in addition to an abelian $U(1)_a$.
Using these rules, the $SU(2n-3)$ model is confining in terms of the following singlets 
\begin{equation}
\label{singletsfirstrel13}
\begin{split}
&t^{n-2,a}_j = \Tilde{A}_1^{j} \Tilde{A}_2^{n-2-j} \tilde{q}, \quad j=0,...,n-2 \\
&t^{n-2,b}_j = \Tilde{A}_1^{j} \Tilde{A}_2^{n-2-j} p_{-}, \quad j=0,...,n-2 \\
& t^{n-3,abb}_j = \Tilde{A}_1^{j} \Tilde{A}_2^{n-3-j} \tilde{q}_1 p_{-}^2, \quad j=0,...,n-3 \\
& t^{n-3,bbb}_j = \Tilde{A}_1^{j} \Tilde{A}_2^{n-3-j} p_{-}^3, \quad j=0,...,n-3 \\
\end{split}
\end{equation}

These singlets  are mapped to the singlets of the original $SU(2n)$ theory with $1 \square$ and $3 \Bar{\square}$ as 
\begin{equation}
\label{singletssecondrel13}
    \begin{array}{lcllcllcl}
         & t^{n-2,a}_j \longrightarrow T^{n}_{n-1-j}, \quad t^{n-2,b}_j \longrightarrow P^{(1)}_{n-j-2}, \quad t^{n-3,abb}_j \longrightarrow P^{(2)}_{n-3-j}, \\
        & t^{n-3,bbb}_j \longrightarrow P^{(3)}_{n-4-j}, \quad y^{d}_{k} \longrightarrow Y^{d}_{2n-k-5}.
    \end{array}
\end{equation}
The final superpotential is given by (\ref{Wconf4e}) in addition to the deformation (\ref{W2AS13evfourth}). Using the singlets in formula (\ref{singletsfirstrel13}) this gives the following confining superpotential
\begin{equation}
\label{intermediofinale13}
    W = y^d_k (t^{n-2,a}_j t^{n-3, bbb}_{2n-k-j-5} + t^{n-2,b}_j t^{n-3, abb}_{2n-k-j-5}) + \Tilde{\rho}_1 t^{n-3,bbb}_{0} + \Tilde{\rho}_2 t^{n-3,bbb}_{n-3},
\end{equation}
where the sums over  $k=0,...,2n-5$ and $j= 0,..., n-2 $ are understood.
The deformation (\ref{W2AS13evfourth}) is mapped in the last two terms in (\ref{intermediofinale13})
where the monopoles $\Tilde{\rho}_i$ give mass to  $t^{(n-3,bbb)}_0$ and $t^{(n-3,bbb)}_{n-3}$, which then are not mapped to any of the $P$ fields. Using  the mapping (\ref{singletssecondrel13}), the final superpotential is 
\begin{equation}
\label{lastour13}
   W = \sum_{\{ k,j\}}  Y^d_{2n-k-5}  ( T^{n}_{n-1-j} P^{(3)}_{k+j+1-n} + P^{(1)}_{N-j-2} P^{(2)}_{k+j+2-n}).
\end{equation}
We conclude the analysis observing that  the superpotential (\ref{lastour13})  is exactly the one that is obtained by adding the deformation (\ref{W2AS13evfirst}) to the  superpotential (\ref{WKeita3}).

We can also reproduce the proof of the confining duality given above by studying the matching of the three sphere partition function.
The relation that we want to prove in this case is
\begin{eqnarray}
\label{1fund3antifund2ASeven}
&& Z_{SU(2n)} (\mu;\vec \nu;\cdot;\cdot;\vec \tau;\cdot ) = 
\prod_{\ell=1}^2 
 \prod_{1\leq a<b\leq 3}  \!\!\!\!
\Gamma_h (\tau_{\ell} + \nu_a + \nu_b)
\prod_{a=1}^3
\Gamma_h (\mu +\nu_a)
\nonumber \\
&&
\prod _{j=0}^{n-3} \prod_{1 \leq a <b \leq 3} 
\Gamma_h ( (n-1-j)\tau _1+(j+2) \tau _2+\nu _a+\nu _b )
\nonumber \\
&&
\prod _{j=0}^{n-5} \Gamma_h ( (n-j-2)\tau _1 +(j+3) \tau _2+\mu +
\sum_{a=1}^3 \nu _a)
\nonumber \\
&&
\prod _{j=0}^{2 n-5} \Gamma_h \left(2 \omega - (2 n-j-2)\tau _1  -(j+3) \tau _2-\mu -\sum_{a=1}^3 \nu_a  \right)
\nonumber \\
&&
\prod _{j=0}^{n} \Gamma_h \left((n-j)\tau _1+j \tau _2\right)
\nonumber \\
&&
\prod _{j=0}^{n-2} \prod_{a=1}^3 \Gamma_h \left(\tau _1 (n-1-j)+(j+1) \tau _2+\mu +\nu _a\right).
\end{eqnarray}

In order to prove such relation we follow the deconfinement and 
duality steps discussed from the field theory approach above.
We start by reading the mass parameters of the flippers  $\alpha,\beta$
and $\rho$ in the first quiver in Figure \ref{2AS13ev}. They are
\begin{equation}
m_{\rho_{\ell}} = 2\omega-n \tau_{\ell},\quad
m_{\beta_\ell^{(a,b)}} = 2\omega - \tau_{\ell} - \nu_a -\nu_b
,\quad
m_{\alpha^{(a)}} = 2\omega - \mu -\nu_a,
\end{equation}
with $\ell=1,2$, $a,b=1,2,3$ and $a<b$.
Then we deconfine the  two antisymmetric tensors and the mass parameter for the new bifundamentals $P_{1,2}$  are
\begin{equation}
m_{P_{\ell}} = \frac{\tau_{\ell}}{2}.
\end{equation}
The duality step requires to define an auxiliary quantity
\begin{equation}
X =\frac{(n-1) (\tau_1+\tau_2)+\mu}{2n-3},
\end{equation}
such that the masses of the fields in the third quiver in Figure \ref{2AS13odd} are
\begin{equation}
m_{\tilde P_{1,2}} = X -\frac{\tau_{1,2}}{2},\quad
m_{L_{1,2}^{(a)}} = \frac{\tau_{1,2}}{2}+\nu_{a},\quad
m_{q^{(a)}} = 2\omega - X - \nu_a,\quad
m_{\tilde q} = X - \mu.
\end{equation}
Then the confinement of the two symplectic nodes gives rise to the singlets
\begin{eqnarray}
m_{L_{\ell}^{(a)}L_{\ell}^{(b)} } = \tau_{\ell}+\nu_{a}+\nu_{b}, 
\quad m_{\tilde \rho_{1,2}} = 2\omega-(n-1)\tau_{1,2}-2\tau_{2,1}-\sum_{a=1}^{3}\nu_{a}-\mu,
\end{eqnarray}
while the two terms $m_{L_{1,2}^2} $  disappear together with the flippers and $\beta_{1,2}$, because of the inversion relation, the  singlets in the second term have to be considered.
The mass parameters of the charged matter fields are 
\begin{equation}
m_{\tilde A_{\ell}} = 2 X - \tau_{\ell}, \quad m_{\tilde p_-^{(a)}}= X+\nu_a
,\quad
m_{\tilde q} = X -\mu,
\end{equation}
where we can reorganize the four antifundamentals as
\begin{eqnarray}
\hat m = \{m_{\tilde q}, m_{\tilde p_-^{(1)}},m_{\tilde p_-^{(2)}},m_{\tilde p_-^{(3)}}\}.
\end{eqnarray}
In this way we have arrived to an identity between the partition function of the first and the last quivers in Figure \ref{2AS13odd}, that reads
\begin{eqnarray}
\label{id14even2n13}
&&
\left(
\prod_{\ell=1}^{2} \Gamma_h(m_{\rho_{\ell}})\prod_{1 \leq a<b \leq 3} \Gamma_h(m_{\beta_\ell^{(a,b)}})
\right)
\prod_{a=1}^{3}  \Gamma_h(m_{\alpha^{(a)}} )
Z_{SU(2n)} ( \mu; \vec \nu;\cdot;\cdot;\vec \tau;\cdot)
\nonumber \\
=
&& 
\prod_{\ell=1}^{2}
\Gamma_h(m_{\tilde \rho_\ell})
Z_{SU(2n-3)} (\cdot ;\vec {\hat m};\cdot;\cdot;\cdot ;2X-\vec \tau).
\end{eqnarray}
The integral $Z_{SU(2n-1)} (\cdot ;\vec {\hat m};\cdot;\cdot;\cdot ;2X-\vec \tau)$ can be evaluated using (\ref{4fund2ASodd}), up to an overall conjugation that does not modify the result.
Once such integral is plugged in (\ref{id14even2n13}) we apply the inversion relation to move the Gamma function of the flippers $\rho, \beta$ and $\alpha$ on the RHS and eliminate the singlets $\tilde \rho$, finally obtaining (\ref{1fund3antifund2ASeven}).
%
%
%
%
%
%
%
%
%%%%%%%%%%%%%%%%%%%%%%%%%%%%%%%%%%%
%%%%%%%%%%%%%%%%%%%%%%%%%%%%%%%%%%%
%%%%%%%%%%%%%%%%%%%%%%%%%%%%%%%%%%%
\subsection{$SU(2n+1)$ with 1 fundamental and 3 antifundamentals}
\label{Sub_AA13odd}

\begin{figure}
\begin{center}
  \includegraphics[width=14cm]{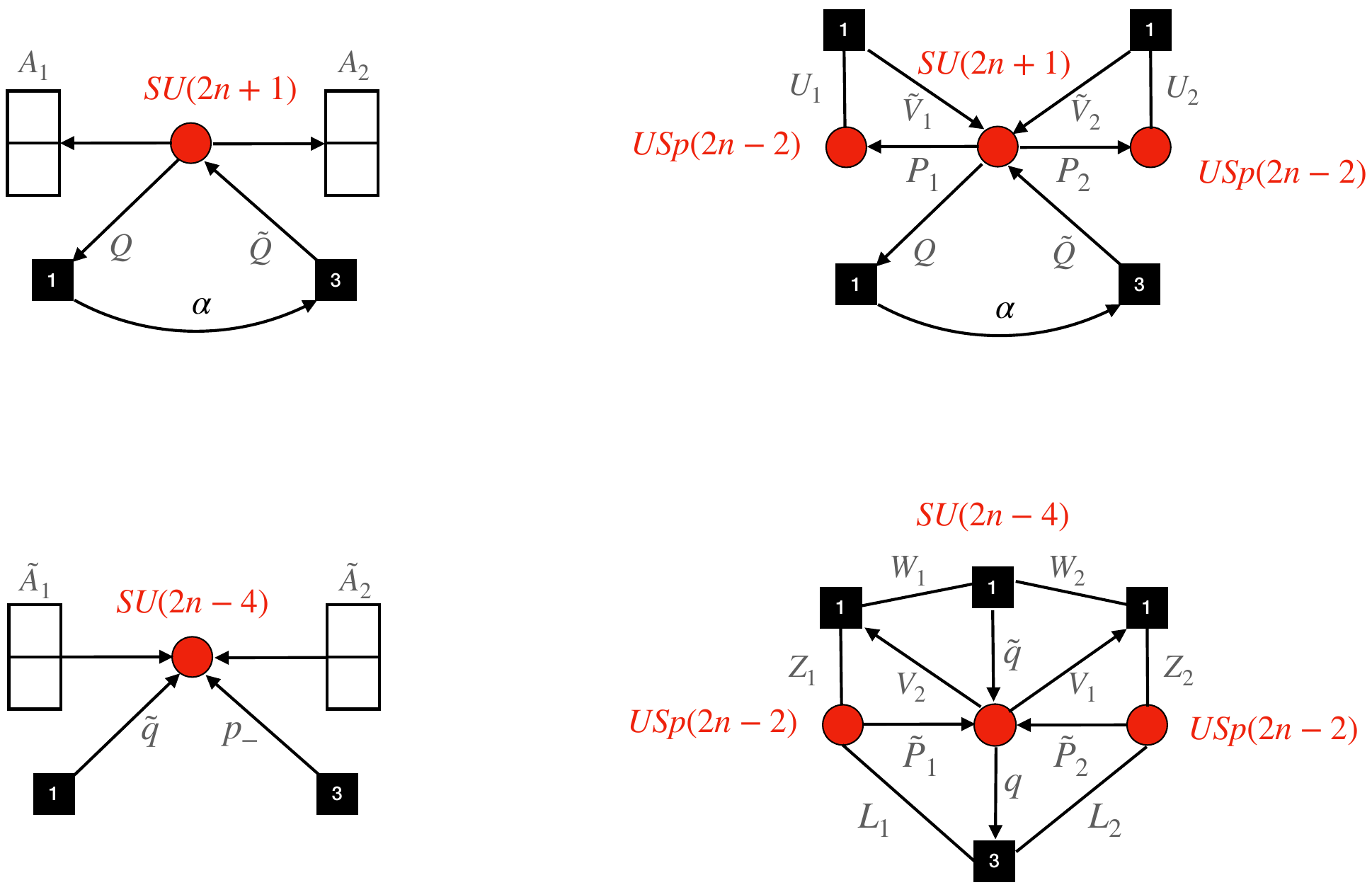}
  \end{center}
  \caption{In this figure we have summarized the various dual step implemented in the text of Subsection \ref{Sub_AA13odd}. On the top-left corner we provide the quiver for the starting $SU(2n+1)$ theory, with two antisymmetric tensors, 1 fundamental and 3 antifundamentals. On the top-right corner we show the auxiliary quiver where the antisymmetric tensors are traded with two $USp(2n-2)$ gauge group with new bifundamentals $P_{1,2}$ and new charged fields $U_{1,2}$. On the bottom-right corner we present the quiver of the theory after having dualized the central $SU(2n+1)$ gauge node to $SU(2n-4)$. In the last bottom-left quiver we present the theory after the confinement of the two $USp(2n-2)$ gauge nodes. We refer to the text for the details regarding the superpotential, the duality mappings and the singlets appearing at each step.  }
    \label{2AS13odd}
\end{figure}

Here we consider an $SU(2n+1)$ gauge group (with $n>1$) with two antisymmetric tensors $A_{1,2}$, one fundamental $Q$ and three antifundamentals $\tilde Q$.
This theory was studied in \cite{Nii:2019ebv}, where a confining duality was proposed in terms of the singlets
$M=Q \tilde Q$, 
$\tilde B = A \tilde Q^2$, 
$T^{(n)} = A^n Q $, 
$P^{(n)} = A^{n} (A \tilde Q) $,
 $P^{(n-1)} = A^{n-1} (A \tilde Q)^3$ 
 and 
$R = A^{n-1} (A \tilde Q)^2 Q$ .
  There is also a dressed monopole $Y^{dressed} = Y^{bare}_{SU(2n-1)}A^{2n-4}$
 and the superpotential of the dual WZ model in this case is
  \begin{equation}
 \label{WKeita4}
 W=Y^{dressed} \Big(M {P^{(n)}}^2+\tilde B P^{(n)}  T^{(n)} +T^{(n)} P^{(n-1)} +
 P^{(n)} R  \Big).
  \end{equation}

Here we want to prove  such confining duality in terms of others (elementary) dualities along the lines of tensor deconfinement. 
Actually, instead of considering a vanishing superpotential, here we flip the operators  the operators $A_{1,2} \tilde Q^2$ and the meson $Q \tilde Q$. The flippers are denoted as $\beta_{1,2}$ and $\alpha$ respectively.
In this way the superpotential associated to the first quiver in Figure \ref{2AS13odd} is
\begin{equation}
\label{W2AS13oddfirst}
W = \sum_{i=1}^2  \beta_i A_i \tilde Q^2+ \alpha Q \tilde Q.
\end{equation}

We then deconfine the two tensors $A_{1,2}$ using two $USp(2n-2)$ gauge groups. The two new bifundamentals emerging from the deconfinement are denoted as $P_{1,2}$ in Figure \ref{2AS13odd}. There are also two new $USp(2n-2)_{1,2}$ charged fields denoted as $U_{1,2}$, associated to the original fundamentals $Q_{1,2}$ through the relations $Q_{1,2} = U_{1,2} P_{1,2}$.
There are two new $SU(2n+1)$  antifundamentals $V_{1,2}$.
The superpotential of this phase is
\begin{equation}
\label{W2AS13oddsecond}
W = \sum_{i=1}^2  (Y_{USp(2n-2)}^{(i)}+P_i U_i \tilde V_i)+\alpha Q \tilde Q + \beta_i P_i^2 \tilde{Q}^2.
\end{equation}
Observe that in this case, differently from the cases studied above, we have deconfined the two tensors by  using symplectic gauge groups with linear monopole superpotentials. 
This duality was first discussed in \cite{Aharony:2013dha}, and it corresponds to the effective duality arising from the circle reduction of the s-confining limit of the Intriligator-Pouliot duality. The finite size effects are reflected in the presence of the linear monopole superpotential, corresponding in this case to the Kaluza-Klein monopole, associated to the affine root of the symplectic algebra when considering the compact Coulomb branch.

The $SU(2n+1)$ gauge node has has $4n-3$ fundamentals and $5$ antifundamentals.  The dual theory was proposed in 
 \cite{Nii:2018bgf} and it was reviewed in appendix {\bf D.1.1} of \cite{Amariti:2024gco}. It is  an $SU(2n-4)$ gauge theory and its superpotential is
\begin{equation}
W = ( L_1 \tilde P_1+L_2 \tilde P_2) q+\tilde{P}_1  V_2 Z_1+\tilde{P}_2  V_1 Z_2
+(V_2 W_1+V_1 W_2) \tilde q+L_1^2 \beta_1+L_2^2 \beta_2,
\end{equation}
where we have integrated the massive meson $M_{Q \tilde{Q}}$ and the flipper $\alpha$ and the $U_i$ fields with the mesons $M_{P_i \tilde{V}_i}$.

At this point we observe that the two $USp(2n-2)$ gauge groups are confining and each one gives rise to a conjugate antisymmetric tensor field, $\tilde{A}_1$ and $\tilde A_2$ respectively, and mesons in the antifundamental representation of $SU(2n-4)$, $\tilde{p}_1 = L_1 \tilde{P}_1$ and $\tilde{p}_2 = L_2 \tilde{P}_2$ respectively. Integrating out the massive fields the final superpotential is
\begin{equation}
\label{W2AS13oddfourth}
W= \sum_{i=1}^2 \tilde{\rho}_i (\tilde{A}_i^{n-3} p_{-}^2 M_{Z_i L_i} + \tilde{A}_i^{n-4} p_{-}^3 W_i \tilde{q}).
\end{equation}

Next we can use again the confinement of $SU(2n-4)$ with two antisymmetric and four fundamentals.
Observe that in the case at hand here the $SU(4) \times SU(2)$ non abelian flavor symmetry is partially broken by the superpotential (\ref{W2AS13oddfourth}). Furthermore the model described in the fourth quiver in Figure \ref{2AS13odd} differ from the one in appendix  \ref{app2AS4fund}  for an overall conjugation.
Keeping in mind these differences, here we split the indices of the $SU(4)$ flavor symmetry into the indices of a non-abelian $SU(3)_b$ flavor symmetries, in addition to an abelian $U(1)_a$.
Using these rules the $SU(2n-4)$ model is confining in terms of the following singlets 
\begin{equation}
\label{singletsfirstrel13odd}
\begin{split}
&t^{n-2,a}_j = \Tilde{A}_1^{j} \Tilde{A}_2^{n-2-j} \tilde{q}, \quad j=0,...,n-2 \\
&t^{n-2,b}_j = \Tilde{A}_1^{j} \Tilde{A}_2^{n-2-j} p_{-}, \quad j=0,...,n-2 \\
& t^{n-3,abb}_j = \Tilde{A}_1^{j} \Tilde{A}_2^{n-3-j} \tilde{q}_1 p_{-}^2, \quad j=0,...,n-3 \\
& t^{n-3,bbb}_j = \Tilde{A}_1^{j} \Tilde{A}_2^{n-3-j} p_{-}^3, \quad j=0,...,n-3 \\
& t^{n-4}_j = \Tilde{A}_1^{j} \Tilde{A}_2^{n-4-j} \tilde{q}_1 p_{-}^3, \quad j=0,...,n-4 \\
\end{split}
\end{equation}

These singlets  are mapped to the singlets of the original $SU(2n+1)$ theory with $1 \square$ and $3 \Bar{\square}$ as 
\begin{equation}
\label{singletssecondrel13odd}
    \begin{array}{lcllcllcl}
         & t^{n-2}_j \longrightarrow T^{n}_{n-2-j}, \quad t^{n-3,ab}_j \longrightarrow P^{(n)}_{n-3-j}, \quad t^{n-3,bb}_j \longrightarrow R_{n-3-j},\\
        & t^{n-4}_j \longrightarrow P^{(n-1)}_{n-4-j}, \quad y^d_{k} \longrightarrow Y^d_{2n-5-k}.
    \end{array}
\end{equation}
The final superpotential is given by (\ref{Wconf4e}) in addition to the deformation (\ref{W2AS13oddfourth}). Using the singlets in formula (\ref{singletsfirstrel13odd}) this gives the following confining superpotential
\begin{align}
\label{intermediofinale13odd}
     W = & y^d_k (t^{n-2}_j t^{n-4}_{2n-k-j-6} + t^{n-3,ab}_l t^{n-3, bb}_{2n-k-l-6}) + Y^d_{0} ( T^n_n P^{n-1}_{n-4} + P^{(n)}_{n-1} R_{n-3} ) + \notag \\ & Y^d_{2n-4} (T^n_0 P^{n-1}_0 + P^{(n)}_{0} R_{0}),
\end{align}
where the sums over  $k=0,...,2n-6$, $j= 1,..., n-1 $ and $l=0,...,n-1$ are understood.
The deformation (\ref{W2AS13oddfourth}) is mapped in the last two terms in (\ref{intermediofinale13odd})
where the monopoles $\Tilde{\rho}_i$ reconstruct the full symmetry in the monopole superpotential. Using  the mapping (\ref{singletssecondrel13odd}), the final superpotential is 
\begin{align}
\label{lastour13odd}
   W=&  Y^d_{2n-5-k}  ( T^{n}_{j} P^{(n-1)}_{2n-4-k-j} + P^{(n)}_{l} R_{2n-4-k-l}) + Y^d_{0} ( T^n_n P^{n-1}_{n-4} + P^{(n)}_{n-1} R_{n-3} ) + \notag \\ & Y^d_{2n-4} (T^n_0 P^{n-1}_0 + P^{(n)}_{0} R_{0}).
\end{align}
We conclude the analysis observing that  the superpotential (\ref{lastour13odd})  is exactly the one that is obtained by adding the deformation (\ref{W2AS13oddfirst}) to the  superpotential (\ref{WKeita4}).

We can also reproduce the proof of the confining duality given above by studying the matching of the three sphere partition function.
The relation that we want to prove in this case is
\begin{eqnarray}
\label{1fund3antifund2ASodd}
&& Z_{SU(2n+1)} (\mu;\vec \nu;\cdot;\cdot;\vec \tau;\cdot ) = 
\prod_{\ell=1}^2 
 \prod_{1\leq a<b\leq 3}  \!\!\!\!
\Gamma_h (\tau_{\ell} + \nu_a + \nu_b)
\prod_{a=1}^3
\Gamma_h (\mu +\nu_a)
\nonumber \\
&&
\prod _{j=0}^{n} \Gamma_h \left((n-j)\tau _1+j \tau _2+\mu\right)
\cdot
\prod _{j=0}^{n-1} \prod_{a=1}^3  \Gamma_h \left( (n-j)\tau _1 +(j+1) \tau _2+\nu _a\right)
\nonumber \\
&&
\prod _{j=0}^{n-3}  \prod_{1\leq a<b\leq 3}  \Gamma_h \left( (n-1-j)\tau _1 +(j+2) \tau _2+\mu +\nu _a+\nu _b\right)
\nonumber \\
&&
\prod _{j=0}^{2 n-4} \Gamma_h \left(2 \omega  -(2 n-j-1)\tau _1-(j+3) \tau _2-\mu -\sum_{a=1}^3\nu _a\right)
\nonumber \\
&&
\prod _{j=0}^{n-4} \Gamma_h\left( (n-1-j)\tau _1 +(j+3) \tau _2+\sum_{a=1}^3\nu _a\right).
\end{eqnarray}
In order to prove such relation we follow the deconfinement and 
duality steps discussed from the field theory approach above.
We start by reading the mass parameters of the flippers  $\alpha$ and $\beta$ in the first quiver in Figure \ref{2AS13odd}. They are
\begin{equation}
m_{\beta_{\ell}^{(a,b)}} = 2\omega - \tau_\ell - \nu_a -\nu_b
,\quad
m_{\alpha^{(a)} }= 2\omega - \mu -\nu_a,
\end{equation}
with $\ell=1,2$,  $a,b=1,2,3$ and $a<b$.
Then we deconfine the  two antisymmetric tensors and the mass parameter for the new bifundamentals $P_{1,2}$, $\tilde{V}_{1,2}$ and the new fields $U_{1,2}$  are
\begin{align}
& m_{P_{\ell}} = \frac{\tau_{\ell}}{2}, \quad  m_{\tilde{V}_{\ell}} = \frac{n \tau_{\ell}}{2}, \quad m_{U_{\ell}} = 2 \omega - \frac{2n+1 }{2} \tau_{\ell}.
\end{align}
The duality step requires to define an auxiliary quantity
\begin{equation}
X =\frac{(n-1) (\tau_1+\tau_2)+\mu}{2n-4},
\end{equation}
such that the masses of the fields in the third quiver in Figure \ref{2AS13odd} are
\begin{eqnarray}
&&
m_{\tilde P_{\ell}} = X -\frac{\tau_{\ell}}{2},\quad
m_{L_{\ell}^{(a)} }= \frac{\tau_{\ell}}{2}+\nu_{a}, \quad m_{Z_{1,2}} = \frac{\tau_{1,2}}{2}+\frac{n}{2} \tau_{2,1}, \quad m_{\tilde q} = X - \mu, \nonumber \\
&&
m_{V_{1,2}} = 2 \omega - m_{P_{2,1}} - m_{Z_{2,1}}, \quad m_{q^{(a)} }= 2\omega - X - \nu_a, \quad m_{W_{1,2}} = \frac{n}{2} \tau_{2,1} + \mu.
\end{eqnarray}
Then the confinement of the two symplectic nodes gives rise to the singlets 
\begin{eqnarray}
&&m_{L_{\ell}^{(a)} L_{\ell}^{(b)} } = \tau_{\ell}+\nu_{a}+\nu_{b}, \nonumber \\
&& m_{M_{Z_{1,2} L_{1,2}^{(a)}}} = \tau_{1,2}+ \frac{n}{2} \tau_{2,1} + \nu_{a},  \\
&&m_{\tilde \rho_{1,2}} = 2\omega-(n-1)\tau_{1,2}-2\tau_{2,1}-\sum_{a=1}^{3}\nu_{a}-\mu,\nonumber
\end{eqnarray}
while the masses $m_{L_{\ell}^{(a)} L_{\ell}^{(b)} } $   disappear together with $m_{\beta_{\ell}^{(a,b)}}$, due of the inversion relation. The other singlets in the last two lines have to be considered.
The mass parameters of the charged matter fields are 
\begin{equation}
m_{\tilde A_{\ell}} = 2 X - \tau_{\ell}, \quad m_{\tilde p_-^{(a)}}= X+\nu_a
,\quad
m_{\tilde q} = X -\mu,
\end{equation}
where we can reorganize the four antifundamentals as
\begin{eqnarray}
\hat m = \{m_{\tilde q}, m_{\tilde p_-^{(1)}},m_{\tilde p_-^{(2)}},m_{\tilde p_-^{(3)}} \}.
\end{eqnarray}
In this way we have arrived to an identity between the partition function of the first and the last quivers in Figure \ref{2AS13odd}, that reads

\begin{eqnarray}
\label{id14odd2n13}
&&
\prod_{\ell=1}^{2}\prod_{1 \leq a<b \leq 3} \Gamma_h(m_{\beta_{\ell}^{(a,b)}})
\prod_{a=1}^{3}  \Gamma_h(m_{\alpha^{(a)}} )
Z_{SU(2n+1)} ( \mu; \vec \nu;\cdot;\cdot;\vec \tau;\cdot)
\nonumber \\
=
&& 
\prod_{\ell=1}^{2} \left(
\Gamma_h(m_{\tilde \rho_\ell})\Gamma_h(m_{W_\ell})
\prod_{a=1}^{3} \Gamma_h(m_{M_{Z_{\ell} L_{\ell}^{(a)}}}) \right)
Z_{SU(2n-4)} (\cdot ;\vec {\hat m};\cdot;\cdot;\cdot ;2X-\vec \tau).
\end{eqnarray}
The integral $Z_{SU(2n-4)} (\cdot ;\vec {\hat m};\cdot;\cdot;\cdot ;2X-\vec \tau)$ can be evaluated using (\ref{4fund2ASeven}), up to an overall conjugation that does not modify the result.
Once such integral is plugged in (\ref{id14odd2n13}), we apply the inversion relation to move the Gamma function of the flippers $\beta$ and $\alpha$ on the RHS, finally obtaining (\ref{1fund3antifund2ASodd}).

\subsection{$SU(2n)$ with 4 antifundamentals}
\label{Sub_AA04ev}

This case corresponds to  an $SU(2n)$ gauge group with two antisymmetric tensors $A_{1,2}$, four  antifundamentals $\tilde Q$.
This theory was studied in \cite{Nii:2019ebv}, where a confining duality was proposed in terms of the singlets
$\tilde B = A \tilde Q^2$, 
$T^{(n)} = A^{n} $,
$T^{(n-1)} = A^{n-1} (A \tilde Q)^2 $ and
 $T^{(n-2)} = A^{n-2} (A \tilde Q)^4$.
  There is also a dressed monopole $Y^{dressed} = Y^{bare}_{SU(2n-2)}A^{2n-6}$
 and the superpotential of the dual WZ model in this case is
  \begin{equation}
 \label{WKeita5}
 W=Y^{dressed} \Big( T^{(n)} T^{(n-2)} +{T^{(n-1)} }^2+
 \tilde B^2 {T^{(n)} }^2 +  \tilde B T^{(n)}  T^{(n-1)} \Big).
  \end{equation}

\begin{figure}
\begin{center}
  \includegraphics[width=12cm]{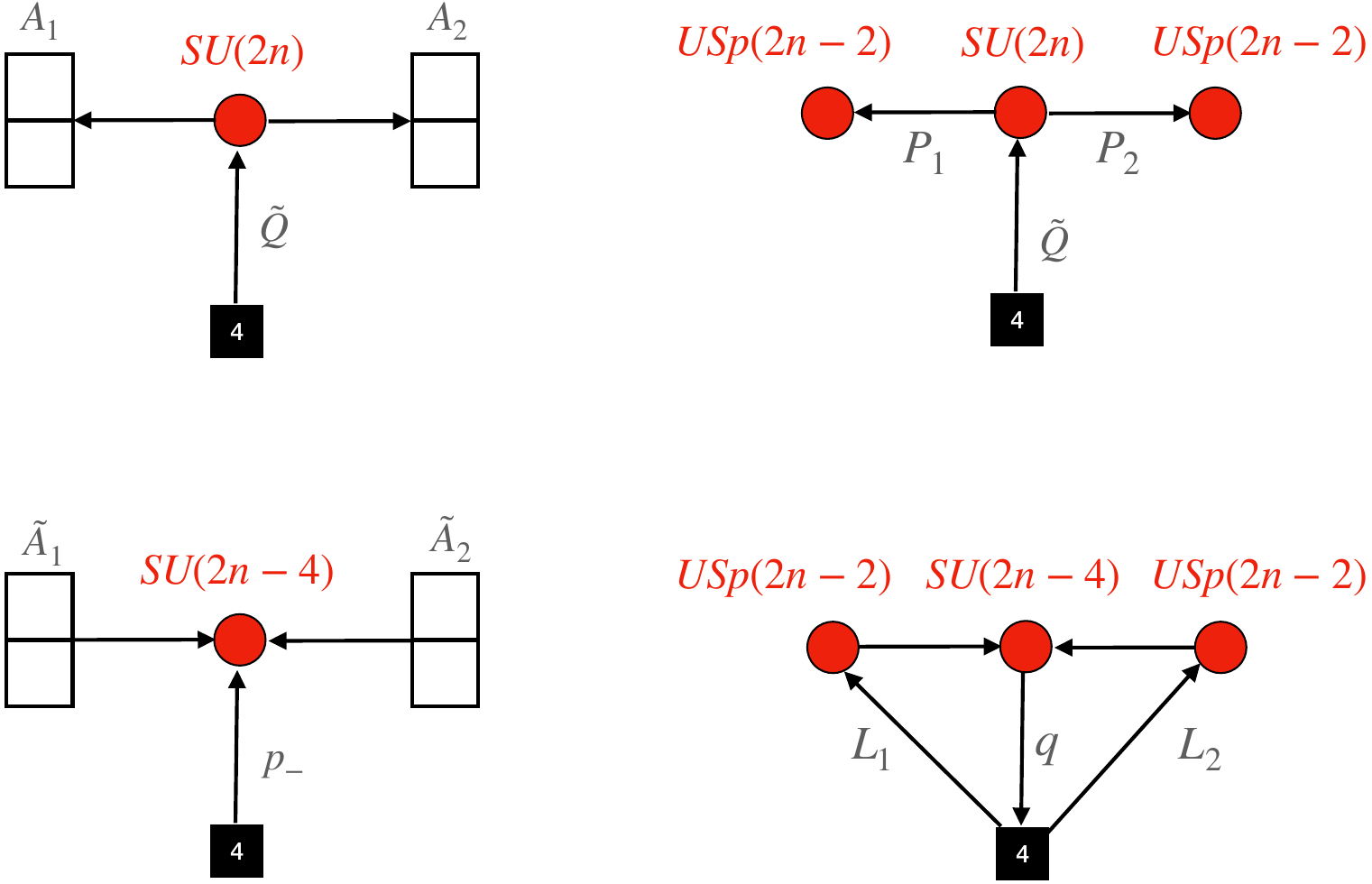}
  \end{center}
  \caption{In this figure we have summarized the various dual step implemented in the text of Subsection \ref{Sub_AA04ev}. On the top-left corner we provide the quiver for the starting $SU(2n)$ theory, with two antisymmetric tensors and 4 antifundamentals. On the top-right corner we show the auxiliary quiver where the antisymmetric tensors are traded with two $USp(2n-2)$ gauge group with new bifundamentals $P_{1,2}$. On the bottom-right corner we present the quiver of the theory after having dualized the central $SU(2n)$ gauge node to $SU(2n-4)$. In the last bottom-left quiver we present the theory after the confinement of the two $USp(2n-2)$ gauge nodes. We refer to the text for the details regarding the superpotential, the duality mappings and the singlets appearing at each step.  }
    \label{2AS04ev}
\end{figure}

Here we want to obtain proof of such confining duality in terms of others (elementary) dualities along the lines of tensor deconfinement. 
Actually, instead of considering a vanishing superpotential, here we flip the operators  the operators $A_{1,2} \tilde Q^2$ and $\text{Pf} A_{1,2}$. The flippers are denoted as $\beta_{1,2}$ and $\rho_{1,2}$ respectively.
In this way the superpotential associated to the first quiver in Figure \ref{2AS04ev} is
\begin{equation}
\label{W2AS04evfirst}
W = \sum_{i=1}^2 \rho_i \, \text{Pf} \, A_i+(\beta_1 A_1 +\beta_2 A_2) \tilde Q^2.
\end{equation}
We then deconfine the two tensors $A_{1,2}$ using two $USp(2n-2)$ gauge groups. The two new bifundamentals emerging from the deconfinement are denoted as $P_{1,2}$ in the second quiver in Figure \ref{2AS04ev} and the superpotential of this phase is
\begin{equation}
W =(\beta_1 P_1^2 +\beta_2 P_2^2) \tilde Q^2 .
\end{equation}
The $SU(2n)$ gauge node has $4n-4$ fundamentals and $4$ antifundamentals. The dual theory, reviewed in appendix {\bf D.1.1} of \cite{Amariti:2024gco}, is an $SU(2n-4)$ gauge theory and its superpotential is
\begin{equation}
W =\beta_1 L_1^2 +\beta_2 L_2^2 +(\tilde P_1 L_1 + \tilde P_2 L_2)q.
\end{equation}
At this point we observe that the two $USp(2n-2)$ gauge groups are confining and each one gives rise to a conjugate antisymmetric, $\tilde A_1$ and $\tilde A_2$ respectively, and a meson in the antifundamental representation of $SU(2n-4)$, $\tilde p_1 = L_1 \tilde P_1$ and $\tilde p_2 = L_2 \tilde P_2$ respectively. Integrating out the massive fields we obtain the final superpotential
\begin{equation}
\label{W2AS04evfourth}
W=\sum_{i=1}^2 \tilde{\rho}_i \tilde{A}_i^{N-4} p_{-}^4.
\end{equation}
Next we can use again the confinement of $SU(2n-4)$ with two antisymmetric and four fundamentals.
Observe that in the case at hand here the $SU(4) \times SU(2)$ non abelian flavor symmetry is partially broken by the superpotential (\ref{W2AS04evfourth}). Furthermore the model described in the fourth quiver in Figure \ref{2AS04ev} differ from the one in appendix  \ref{app2AS4fund}  for an overall conjugation.
Keeping in mind these differences the $SU(2n-4)$ model is confining in terms of the following singlets 
\begin{equation}
\label{singletsfirstrel04}
\begin{split}
&t^{n-2}_j = \Tilde{A}_1^{j} \Tilde{A}_2^{n-2-j}, \quad j=0,...,n-2 \\
& t^{n-3}_j = \Tilde{A}_1^{j} \Tilde{A}_2^{n-3-j} \tilde{q}^2, \quad j=0,...,n-3 \\
& t^{n-4}_j = \Tilde{A}_1^{j} \Tilde{A}_2^{n-4-j} \tilde{q}^4, \quad j=0,...,n-4 \\
\end{split}
\end{equation}

These singlets  are mapped to the singlets of the original $SU(2n)$ theory with $4 \Bar{\square}$ as 
\begin{equation}
\label{singletssecondrel04}
    \begin{array}{lcllcllcl}
         & t^{n-2}_j \longrightarrow T^{(n)}_{n-1-j}, \quad t^{n-3}_j \longrightarrow T^{(n-1)}_{n-3-j}, \quad t^{n-4}_j \longrightarrow T^{(n-2)}_{n-5-j} , \quad y^d_k \longrightarrow Y^d_{2n-6-k}.
    \end{array}
\end{equation}
The final superpotential is given by (\ref{Wconf4e}) in addition to the deformation (\ref{W2AS04evfourth}). Using the singlets in formula (\ref{singletsfirstrel04}) this gives the following confining superpotential
\begin{equation}
\label{intermediofinale04}
    W = y^d_k (t^{n-2}_j t^{n-4}_{2n-k-j-6} + t^{n-3}_l t^{n-3}_{2n-k-l-6}) + \rho_1 t^{n-4}_{n-4} + \rho_2 t^{n-4}_{0},
\end{equation}
where the sums over  $k=0,...,2n-6$, $j= 0,..., n-6 $ and $l=0,...,n-3$ are understood.
The deformation (\ref{W2AS04evfourth}) is mapped in the last two terms in (\ref{intermediofinale04})
where the monopoles $\Tilde{\rho}_i$ give mass to  $t^{n-4}_0$ and $t^{n-4}_{n-4}$, which then are not mapped to any of the $T^{(n-2)}$ fields. Using  the mapping (\ref{singletssecondrel04}), the final superpotential is 
\begin{equation}
\label{lastour04}
   W = Y^d_{2n-6-k} ( T^n_{n-1-j} T^{n-2}_{k+j+1-n} + T^{n-1}_{n-3-l} T^{n-1}_{k+l+3-n} ).
\end{equation}
We conclude the analysis observing that  the superpotential (\ref{lastour04})  is exactly the one that is obtained by adding the deformation (\ref{W2AS04evfirst}) to the  superpotential (\ref{WKeita5}).

We can also reproduce the proof of the confining duality given above by studying the matching of the three sphere partition function.
The relation that we want to prove in this case is
\begin{eqnarray}
\label{0fund4antifund2ASeven}
&& Z_{SU(2n)} (\cdot;\vec \nu;\cdot;\cdot;\vec \tau;\cdot ) \!=\! 
\prod_{\ell=1}^2 \!
 \prod_{1\leq a<b\leq 4}  \!\!\!\!
\Gamma_h (\tau_{\ell} + \nu_a + \nu_b)\! \cdot \!\!
\prod _{j=0}^{n-6} \Gamma_h \big( (n\!-\!j\!-\!2)\tau _1\! +\!(j+4) \tau _2 \!+\!\sum_{a=1}^4 \nu _a\big)
\nonumber \\
&&
\prod _{j=0}^{n} \Gamma_h \left((n-j)\tau _1+j \tau _2\right)
\cdot
\prod _{j=0}^{n-3} \prod_{1\leq a<b\leq 4} \Gamma_h \left( (n-j-1)\tau _1 +(j+2) \tau _2+\nu _a + \nu_b \right)
\nonumber \\
&&
\prod _{j=0}^{2 n-6} \Gamma_h \left(2 \omega  -(2 n-j-2)\tau _1-(j+4) \tau _2 -\sum_{a=1}^4\nu _a\right).
\end{eqnarray}

In order to prove such relation we follow the deconfinement and 
duality steps discussed from the field theory approach above.
We start by reading the mass parameters of the flippers  $\alpha$ and $\beta$ in the first quiver in Figure \ref{2AS04ev}. They are
\begin{equation}
m_{\beta_{\ell}^{(a,b)}} = 2\omega - \tau_{\ell} - \nu_a -\nu_b
,\quad
m_{\rho_{\ell}} = 2\omega - n \tau_{\ell},
\end{equation}
with $\ell=1,2$,  $a,b=1,\dots,4$ and $a<b$.
Then we deconfine the  two antisymmetric tensors and the mass parameter for the new bifundamentals $P_{1,2}$ are
\begin{align}
& m_{P_{1,2}} = \frac{\tau_{1,2}}{2}.
\end{align}
The duality step requires to define an auxiliary quantity
\begin{equation}
X =\frac{(n-1) (\tau_1+\tau_2)}{2n-4},
\end{equation}
such that the masses of the fields in the third quiver in Figure \ref{2AS04ev} are
\begin{eqnarray}
&&
m_{\tilde P_{1,2}} = X -\frac{\tau_{1,2}}{2},\quad
m_{L_{1,2}}^{(a)} = \frac{\tau_{1,2}}{2}+\nu_{a},  \quad m_{q}^{(a)} = 2\omega - X - \nu_a.
\end{eqnarray}
Then the confinement of the two symplectic nodes gives rise to the singlets 
\begin{eqnarray}
&&m_{L_{1,2}^{(a)} L_{1,2}^{(b)} } = \tau_{1,2}+\nu_{a}+\nu_{b}, \nonumber \\
&&m_{\tilde \rho_{1,2}} = 2\omega-(n-1)\tau_{1,2}-3\tau_{2,1}-\sum_{a=1}^{3}\nu_{a},
\end{eqnarray}
while the terms in the first line  disappear together with the flippers $\beta$, because of the inversion relation, the  singlets in the last lines have to be considered.
The mass parameters of the charged matter fields are 
\begin{equation}
m_{\tilde A_{1,2}} = 2 X - \tau_{1,2}, \quad m_{\tilde p_-^{(a)}}= X+\nu_a,
\end{equation}
where we can re-organize the four antifundamentals as
\begin{eqnarray}
\hat m = \{m_{\tilde p_-^{(1)}},m_{\tilde p_-^{(2)}} , m_{\tilde p_-^{(3)}} , m_{\tilde p_-^{(4)}} \}.
\end{eqnarray}
In this way we have arrived to an identity between the partition function of the first and the last quivers in Figure \ref{2AS04ev}, that reads

\begin{eqnarray}
\label{id14even2n04}
&&
\prod_{\ell=1}^{2} \Gamma_h(m_{\rho_{\ell}}) \prod_{1 \leq a<b \leq 4} \Gamma_h(m_{\beta_{\ell}^{(a,b)}}) Z_{SU(2n)} ( \cdot; \vec \nu;\cdot;\cdot;\vec \tau;\cdot)
\nonumber \\
=
&& 
\prod_{\ell=1}^{2}
\Gamma_h(m_{\tilde \rho_\ell}) 
{
%\color{red}
%\Gamma_h(m_{M_{Z_{\ell} L_{\ell}}}) \Gamma_h(m_{W_\ell})
} \cdot
Z_{SU(2n-4)} (\cdot ;\vec {\hat m};\cdot;\cdot;\cdot ;2X-\vec \tau).
\end{eqnarray}
The integral $Z_{SU(2n-4)} (\cdot ;\vec {\hat m};\cdot;\cdot;\cdot ;2X-\vec \tau)$ can be evaluated using (\ref{4fund2ASeven}), up to an overall conjugation that does not modify the result.
Once such integral is plugged in (\ref{id14even2n04}), we apply the inversion relation to move the Gamma function of the flippers $\beta$ and $\rho$ on the RHS and eliminate the monopole $\tilde{\rho}_{1,2}$, finally obtaining (\ref{0fund4antifund2ASeven}).
%
%
%
%
%
%
%
%
%
%%%%%%%%%%%%%%%%%%%%%%%%%
%%%%%%%%%%%%%%%%%%%%%%%%%
%%%%%%%%%%%%%%%%%%%%%%%%%
\subsection{$SU(2n+1)$ with 4 antifundamentals}
\label{Sub_AA04odd}

The last case corresponds to  an $SU(2n+1)$ gauge group (with $n>1$) with two antisymmetric tensors $A_{1,2}$ and four  antifundamentals $\tilde Q$.
This theory was studied in \cite{Nii:2019ebv}, where a confining duality was proposed in terms of the singlets
$\tilde B = A \tilde Q^2$, $P^{(n)} = A^{n} (A \tilde Q) $, $P^{(n-1)} = A^{n-1} (A \tilde Q)^3$.
There is also a dressed monopole $Y^{dressed} = Y^{bare}_{SU(2n-1)}A^{2n-5}$ and the superpotential of the dual WZ model in this case is
  \begin{equation}
 \label{WKeita6}
 W=Y^{dressed} \Big( P^{(n)} P^{(n-1)} +
 \tilde B (P^{(n)})^2  \Big).
  \end{equation}

\begin{figure}
\begin{center}
  \includegraphics[width=12cm]{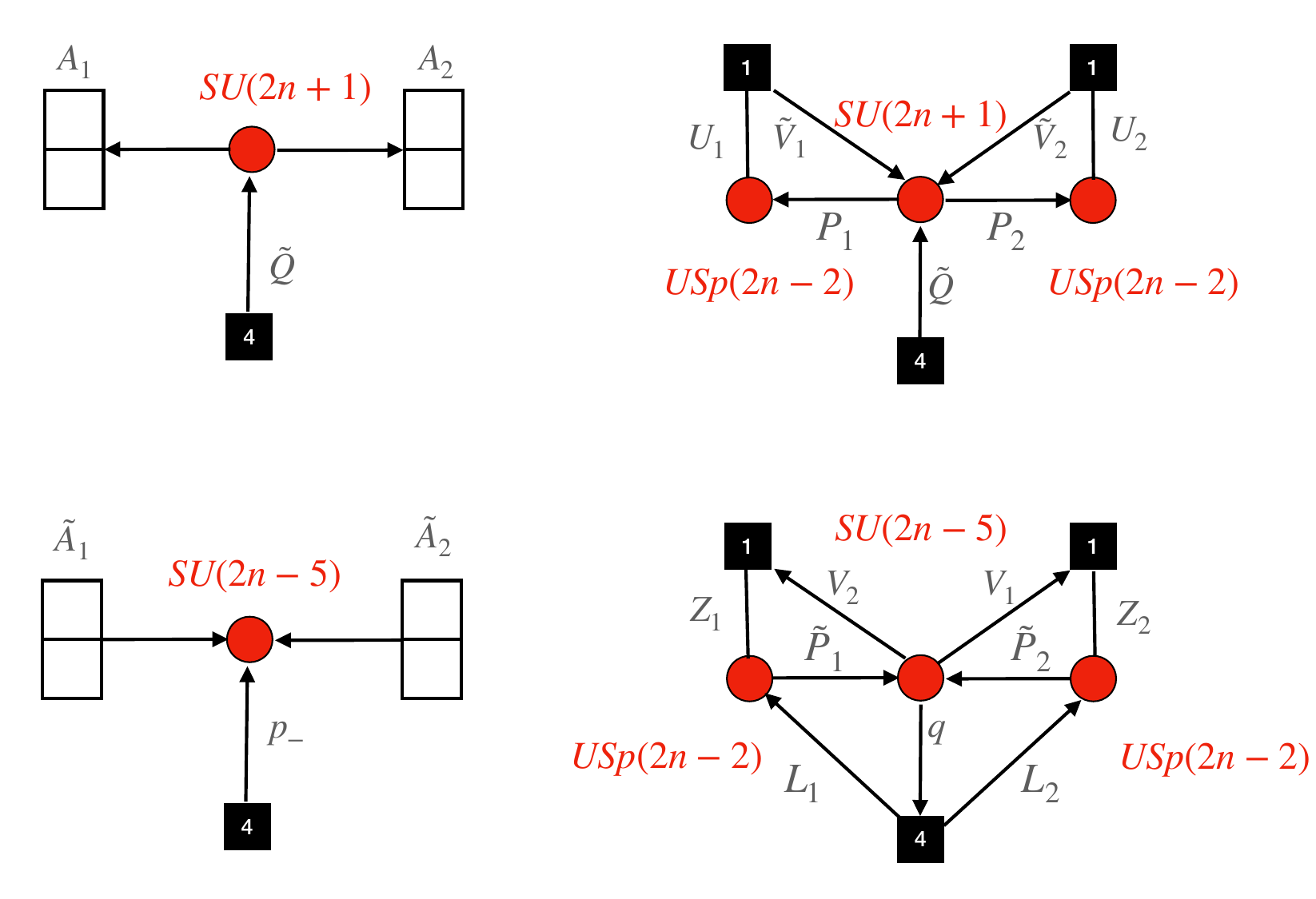}
  \end{center}
  \caption{In this figure we have summarized the various dual step implemented in the text of Subsection \ref{Sub_AA04odd}. On the top-left corner we provide the quiver for the starting $SU(2n+1)$ theory, with two antisymmetric tensors and 4 antifundamentals. On the top-right corner we show the auxiliary quiver where the antisymmetric tensors are traded with two $USp(2n-2)$ gauge group with new bifundamentals $P_{1,2}$ and new charged fields $U_{1,2}$. On the bottom-right corner we present the quiver of the theory after having dualized the central $SU(2n+1)$ gauge node to $SU(2n-5)$. In the last bottom-left quiver we present the theory after the confinement of the two $USp(2n-2)$ gauge nodes. We refer to the text for the details regarding the superpotential, the duality mappings and the singlets appearing at each step.  }
    \label{2AS04odd}
\end{figure}

Here we want to obtain proof of such confining duality in terms of others (elementary) dualities along the lines of tensor deconfinement. 
Actually, instead of considering a vanishing superpotential, here we flip the operators  the operators $A_{1,2} \tilde Q^2$. The flippers are denoted as $\beta_{1,2}$ respectively.
In this way the superpotential associated to the first quiver in Figure \ref{2AS04odd} is
\begin{equation}
\label{W2AS04oddfirst}
W =(\beta_1 A_1 +\beta_2 A_2) \tilde Q^2.
\end{equation}
We then deconfine the two tensors $A_{1,2}$ using two $USp(2n-2)$ gauge groups. The two new bifundamentals emerging from the deconfinement are denoted as $P_{1,2}$,  in the second quiver in Figure \ref{2AS04odd}. There are also two new auxiliary $USp(2n-2)_{1,2}$ charged fields denoted as $U_{1,2}$ and also two new auxiliary $SU(2n+1)$ charged fields $V_{1,2}$. The superpotential of this phase is
\begin{equation}
W = Y_{USp(2n-2)}^{(1)}+ Y_{USp(2n-2)}^{(2)}+P_1 U_1 \tilde V_1 +P_2 U_2 \tilde V_2+(\beta_1 P_1^2 +\beta_2 P_2^2) \tilde Q^2 .
\end{equation}
Where we have deconfined the two tensors by  using symplectic gauge groups with linear monopole superpotentials. 
The $SU(2n+1)$ gauge node has $4n-4$ fundamentals and $6$ antifundamentals. The dual theory, 
reviewed in appendix {\bf D.1.1} of \cite{Amariti:2024gco}, is an $SU(2n-5)$ gauge theory (see third quiver of Figure \ref{2AS04odd}) and its superpotential is
\begin{equation}
W=\beta_1 L_1^2 +\beta_2 L_2^2+(L_1 \tilde P_1+L_2 \tilde P_2) q + Z_2 \tilde P_2 V_1 +Z_1  \tilde P_1 V_2.
\end{equation}
At this point we observe that the two $USp(2n-2)$ gauge groups are confining and each one gives rise to a conjugate antisymmetric, $\tilde A_1$ and $\tilde A_2$ respectively, a meson in the antifundamental representation of $SU(2n-5)$, $\tilde p_1 = L_1 \tilde P_1$ and $\tilde p_2 = L_2 \tilde P_2$ respectively and two gauge singlets $\psi_1 = L_1 Z_1$, $\psi_2 = L_2 Z_2$.  Integrating out the massive fields we obtain the final superpotential
\begin{equation}
\label{W2AS04oddfourth}
W = (\rho_1 \tilde A_1^{n-4} \psi_1 +\rho_2 \tilde A_2^{n-4} \psi_2) p_-^3.
\end{equation}
Next we can use again the confinement of $SU(2n-5)$ with two antisymmetric and four fundamentals.
Observe that in the case at hand here the $SU(4) \times SU(2)$ non abelian flavor symmetry is partially broken by the superpotential (\ref{W2AS04oddfourth}). Furthermore the model described in the fourth quiver in Figure \ref{2AS04odd} differ from the one in appendix  \ref{app2AS4fund}  for an overall conjugation.
Keeping in mind these differences, the $SU(2n-5)$ model is confining in terms of the following singlets
\begin{equation}
\label{singletsfirstrel04odd}
\begin{split}
&t^{n-3}_j = \tilde{A}_1^{j} \tilde{A}_2^{n-3-j}, \quad j=0,...,n-3 \\
& t^{n-4}_j = \tilde{A}_1^{j} \tilde{A}_2^{n-4-j} \tilde{q}^2, \quad j=0,...,n-4 \\
\end{split}
\end{equation}
These singlets  are mapped to the singlets of the original $SU(2n+1)$ theory with $4 \Bar{\square}$ as 
\begin{equation}
\label{singletssecondrel04odd}
    \begin{array}{lcllcllcl}
         & t^{n-3}_j \longrightarrow P^{(n)}_{n-2-j}, \quad t^{n-4}_j \longrightarrow P^{(n-1)}_{n-4-j}, \quad y^d_k \longrightarrow Y^d_{2n-6-k}.
    \end{array}
\end{equation}
The final superpotential is given by (\ref{Wconf4o}) in addition to the deformation (\ref{W2AS04oddfourth}). Using the singlets in formula (\ref{singletsfirstrel04odd}) this gives the following confining superpotential
\begin{equation}
\label{intermediofinale04odd}
    W = y^d_k (t^{n-3}_j t^{(n-4)}_{2n-7-k-j}) + Y^d_{0} P^{(n)}_{n-1} P^{(n-1)}_{n-4} + Y^d_{2n-5} P^{(n)}_{0} P^{(n-1)}_{0},
\end{equation}
where the sums over  $k=0,...,2n-7$ and $j= 0,..., n-3 $ are understood. The deformation (\ref{W2AS04oddfourth}) is mapped in the last two terms in (\ref{intermediofinale04odd}) where the monopoles $\Tilde{\rho}_i$ reconstruct the full symmetry in the monopole superpotential. Using  the mapping (\ref{singletssecondrel04odd}), the final superpotential is 
\begin{equation}
\label{lastour04odd}
   W = Y^d_{2n-6-k} (P^{n}_{n-2-j} P^{(n-1)}_{k+j-n+3}) + Y^d_{0} P^{(n)}_{n-1} P^{(n-1)}_{n-4} + Y^d_{2n-5} P^{(n)}_{0} P^{(n-1)}_{0}.
\end{equation}
We conclude the analysis observing that  the superpotential (\ref{lastour04odd})  is exactly the one that is obtained by adding the deformation (\ref{W2AS04oddfirst}) to the  superpotential (\ref{WKeita6}).

We can also reproduce the proof of the confining duality given above by studying the matching of the three sphere partition function.
The relation that we want to prove in this case is
\begin{eqnarray}
\label{0fund4antifund2ASodd}
&& Z_{SU(2n+1)} (\cdot;\vec \nu;\cdot;\cdot;\vec \tau;\cdot ) = 
\prod_{\ell=1}^2 \!
 \prod_{1\leq a<b\leq 4}  \!\!\!\!
\Gamma_h (\tau_{\ell} + \nu_a + \nu_b)
\prod _{j=0}^{n-1} \prod_{a=1}^4 \!\Gamma_h \left( (n\!-\!j)\tau _1 \!+\!(j+1) \tau _2\!+\!\nu _a \right)
\nonumber \\
&&
\prod _{j=0}^{n-4} \prod_{a=1}^4 \Gamma_h \left( (n\!-\!j\!-\!1)\tau _1 +(j+3) \tau _2 \!-\! \nu _a\right)\!
\prod _{j=0}^{2 n-5} \! \Gamma_h \big(2 \omega  \!-\!(2 n\!-\!j\!-\!1)\tau _1\!-\!(j+4) \tau _2 \!-\!\sum_{a=1}
^4\nu _a\big)\nonumber \\.
\end{eqnarray}

In order to prove such relation we follow the deconfinement and 
duality steps discussed from the field theory approach above.
We start by reading the mass parameters of the flippers $\beta$ in the first quiver in Figure \ref{2AS04odd}. They are
\begin{equation}
m_{\beta_{\ell}^{(a,b)}} = 2\omega - \tau_{\ell} - \nu_a -\nu_b,
\end{equation}
with $\ell=1,2$,  $a,b=1,...,4$ and $a<b$.
Then we deconfine the  two antisymmetric tensors and the mass parameter for the new bifundamentals $P_{1,2}$ and for the new fields $\tilde{V}_{1,2}$, $U_{1,2} $ are
\begin{align}
& m_{P_{1,2}} = \frac{\tau_{1,2}}{2}, \quad  m_{\tilde{V}_{1,2}} = \frac{n \tau_{1,2}}{2}, \quad m_{U_{1,2}} = 2 \omega - \frac{2n+1 }{2} \tau_{1,2}.
\end{align}
The duality step requires to define an auxiliary quantity
\begin{equation}
X =\frac{(n-1) (\tau_1+\tau_2)}{2n-5},
\end{equation}
such that the masses of the fields in the third quiver in Figure \ref{2AS04ev} are
\begin{eqnarray}
&&
m_{\tilde P_{1,2}} = X -\frac{\tau_{1,2}}{2},\quad
m_{L_{1,2}}^{(a)} = \frac{\tau_{1,2}}{2}+\nu_{a},  \quad m_{q}^{(a)} = 2\omega - X - \nu_a, \nonumber \\
&& m_{Z_{1,2}} = \frac{\tau_{1,2}}{2}+\frac{n}{2} \tau_{2,1}, \quad m_{V_{1,2}} = 2 \omega - m_{P_{2,1}} - m_{Z_{2,1}}.
\end{eqnarray}
Then the confinement of the two symplectic nodes gives rise to the singlets 
\begin{eqnarray}
&&m_{L_{1,2}^{(a)}  L_{1,2}^{(b)}} = \tau_{1,2}+\nu_{a}+\nu_{b}, \nonumber \\
&&m_{\Psi_{1,2}^{(a)}} = \tau_{1,2} + \frac{n}{2} \tau_{2,1}+\nu_{a}, \nonumber \\
&&m_{\tilde \rho_{1,2}} = 2\omega-(2n-1)\tau_{1,2}-4\tau_{2,1}-\sum_{a=1}^{4}\nu_{a},
\end{eqnarray}
while the terms in the first line  disappear together with the flippers $\beta$, because of the inversion relation, the other singlets in the last two lines have to be considered.
The mass parameters of the charged matter fields are 
\begin{equation}
m_{\tilde A_{1,2}} = 2 X - \tau_{1,2}, \quad m_{\tilde p_-^{(a)}}= X+\nu_a,
\end{equation}
where we can reorganize the four antifundamentals as
\begin{eqnarray}
\hat m = \{m_{\tilde p_-^{(1)}} ,m_{\tilde p_-^{(2)}} ,m_{\tilde p_-^{(3)}} ,m_{\tilde p_-^{(4)}}  \}.
\end{eqnarray}
In this way we have arrived to an identity between the partition function of the first and the last quivers in Figure \ref{2AS04odd}, that reads

\begin{eqnarray}
\label{id14odd2n04}
&&
\prod_{\ell=1}^2
\prod_{1 \leq a<b \leq 4}\!\! \!\!\!\! \Gamma_h(m_{\beta_{\ell }^{(a,b)}}) Z_{SU(2n+1)} ( \cdot; \vec \nu;\cdot;\cdot;\vec \tau;\cdot)
 \nonumber\\
=&& 
\prod_{\ell=1}^2
\Gamma_h(m_{\tilde \rho_{\ell}}) \prod_{a=1}^4 \Gamma_h(m_{\Psi_{\ell}^{(a)}})
Z_{SU(2n-5)} (\cdot ;\vec {\hat m};\cdot;\cdot;\cdot ;2X-\vec \tau).
\end{eqnarray}

The integral $Z_{SU(2n-5)} (\cdot ;\vec {\hat m};\cdot;\cdot;\cdot ;2X-\vec \tau)$ can be evaluated using (\ref{4fund2ASodd}), up to an overall conjugation that does not modify the result.
Once such integral is plugged in (\ref{id14odd2n04}), we apply the inversion relation to move the Gamma function of the flippers $\beta$ to the RHS, finally obtaining (\ref{0fund4antifund2ASodd}).

\section{3d $SU(N)$ with symmetric tensors and $W_{monopole}$}
\label{sec3}
In this section we support the claim that two 3d $SU(N)$ gauge theories  with a two index symmetric tensor (in addition to an antisymmetric tensor and/or fundamentals) are confining in presence 
of a non trivial superpotential that includes a linear monopole deformations.
We indeed expect that such theories are confining because the three sphere partition function can be computed explicitly and it can be reorganized in terms of gauge singlets 
corresponding to the chiral ring operators of the gauge theory.
The evaluation of the partition function follows from the 4d/3d reduction of two parent 4d confining gauge theories, with fundamentals and antisymmetric matter field(s).
Once the associated integral identities relating the supersymmetric index of the 4d confining theories are reduced on the circle they give origin to integral identities 
for the squashed three sphere partition function.
Then we apply the duplication formula for the hyperbolic gamma functions by \emph{freezing} some of the mass parameters for the fundamentals. The resulting identities 
are compatible with the claim that the $SU(N)$ gauge theories with two index symmetric tensors and monopole superpotential are confining.
The claim is then supported by applying the technique of tensor deconfinement, where a two index tensor becomes equivalent to a new confining 
gauge node interacting with the original one through an auxiliary bifundamental field. 

We further dualize the original gauge node in such deconfined phase, obtained a new dual phase where the original gauge node either confines or it is dual to a model with a different (generically reduced)  amount of colors. In the models studied in this section we show that by deconfining the tensors and then, by sequentially confining the various gauge nodes, one recovers the dual WZ model expected from the duplication formula above.

The two dualities obtained in this way can further flow, through opportune real mass deformations, that remove the linear monopole deformations, to some of the dualities  proposed in \cite{Amariti:2024gco}.  Furthermore the UV picture discussed here, in presence of linear monopole
deformations, clarifies the origin of some terms that were claimed in \cite{Amariti:2024gco} to be dynamically generated, and this represents a check of validity of the results in that reference as well.
\subsection{$SU(N)$ with $S$, $Q$, $\tilde Q_S$ and $N$ $\tilde Q$}
\label{subsec3.1}
In this section we start by considering a 4d $SU(N)$ gauge theory with an antisymmetric, four fundamentals and $N$ antifundamentals. The model is s-confining for each parity of $N$, i.e. $N=2n$ and $N=2n+1$, nevertheless the details of such confinements are different and the two cases require a different analysis. We refer the reader to the original reference \cite{Csaki:1996zb} for further discussion on these models and to \cite{Benvenuti:2020wpc} for a proof of such s-confining dualities through tensor deconfinement. 
For completeness in appendix \ref{apptuttobello} we have reported a more detailed description of these models and on their circle reduction, by providing the explicit expressions relating the three sphere partition function for the circle reduction of the confining dualities, corresponding to the identities  (\ref{W2np14}) and (\ref{W2n4}) for the case $N=2n+1$ and $N=2n$ respectively.

At this point we \emph{freeze} three out of the four mass parameters for the fundamentals \footnote{In principe we could have also studied the case with an extra frozen parameter 
$\mu_1 = \frac{\tau_S}{2}$. We avoid the analysis for this case here because it does not give origin to an interacting model in the dual picture where only the fields $\Phi_1=\det S$ and $\Phi_2= S\tilde Q^2$ survive, and the combination $\Phi_1 \Phi_2^N$ is uncharged, without giving origin to a 3d superpotential.}. 
First we re-name the mass parameter $\tau_A$ for the antisymmetric tensor in (\ref{W2np14}) and (\ref{W2n4}) as $\tau_S$. Then we fix the values of such masses, in both cases, as
\begin{equation}
\mu_2 = \frac{\omega_1}{2}+\frac{\tau_S}{2},
\quad
\mu_3 = \frac{\omega_2}{2}+\frac{\tau_S}{2},
\quad
\mu_4 = \frac{\tau_S}{2}.
\end{equation}
Then we manipulate the expressions by using the duplication formula for the hyperbolic Gamma functions, that we report here for completeness
\begin{equation}
\label{dupform}
\Gamma_h(2z) = \Gamma_h(z)  \Gamma_h\left(z+\frac{\omega_1}{2} \right)
\Gamma_h\left(z+\frac{\omega_2}{2} \right) 
\Gamma_h(z+\omega).
\end{equation}
Even if we started with two different identities, (\ref{W2np14}) and (\ref{W2n4}), in presence of an antisymmetric, here, after the application of the duplication formula,
we have obtained an unified formula for both $N=2n$ and $N=2n+1$, corresponding to 
\begin{eqnarray}
\label{iddopodup}
&&
Z_{SU(N)} ( \mu;\vec \nu;\tau_S;-;-;-)
=
 \prod_{b=1}^{N} \Gamma_h(\mu +\nu_b)  \Gamma_h\left(\omega-\frac{\tau_S}{2} -\nu_b\right)
 \nonumber \\
 &&
 \Gamma_h(N \tau_S)
 \Gamma_h((N-1)\tau_S+ 2\mu)
 \Gamma_h\left(\sum_{b=1}^{N} \nu_b \right)
 \prod_{b\leq c}^{N} \Gamma_h(\tau_S +\nu_b +\nu_c),
 \end{eqnarray}
 where we re-named the parameter $\mu_1$ simply as $\mu$ above.
 Moreover this identity is valid if the mass parameters satisfy a constraint, often 
 referred to as balancing condition, 
 that follows from the relations (\ref{BalAS1}) and  (\ref{BalAS2}). Then the balancing condition here becomes
\begin{equation}
\label{balancingoddS}
\left(N-\frac{1}{2} \right)\tau_S + \mu +  \sum_{b=1}^{N} \nu_b =\omega,
\end{equation}
and corresponds to a constraint on the global symmetries, usually enforced in 3d by the presence of a monopole superpotential. Indeed such constraint was imposed by the KK monopole superpotential for the theory with an antisymmetric, $4$ fundamentals and $N$ antifundamentals before the application of the duplication formula.

At this point we provide an interpretation of the identity obtained by the application of the duplication formula in terms of a 3d effective confining duality, where the terminology refers to the fact that we expect a monopole superpotential appearing on the gauge theory side of the duality.
The gauge theory  corresponds to $SU(N)$ with a symmetric $S$, a fundamental $Q$, an antifundamental $\tilde Q_S$
and $N$ antifundamentals $\tilde Q$ and superpotential
\begin{equation}
W = S Q_S^2+Y_{SU(N-2)}^{bare},
\end{equation}
where we checked that the constraint (\ref{balancingoddS}) in exactly enforced by the linear 
monopole deformation $Y_{SU(N-2)}^{bare}$ in the superpotential.
Observe further that using the constraint (\ref{balancingoddS}) we can re-write the last term in the first line of (\ref{iddopodup}) as
\begin{equation}
\Gamma_h\left(\omega-\frac{\tau_S}{2} -\nu_b\right)
=
\Gamma_h\left( (N-1) \tau_S + \mu +  \sum_{c=1}^{N} \nu_c-\nu_b \right),
\end{equation}
implying that this term corresponds to the gauge invariant combination $Q (S \tilde Q)^{N-1}$.

Summarizing, the singlets of this confining duality appearing in the RHS of (\ref{iddopodup}) are
\begin{eqnarray}
&&
\Phi_1 \equiv \det S,\quad
\Phi_2 \equiv Q \tilde Q, \quad
\Phi_3 \equiv Q(S \tilde Q)^{N-1}\nonumber \\
&&
\Phi_4 \equiv \tilde Q^{N},\quad\,\,\,\,
\Phi_5 \equiv S \tilde Q^{2},\quad\!\!
\Phi_6 \equiv S^{N-1} Q^2.
\end{eqnarray}
The most general superpotential for the WZ dual description compatible with the constraints from the global symmetries is
\begin{equation}
\label{Ipredicted1}
W = \Phi_6 \det \Phi_5 + \Phi_5 \Phi_3^2 +\Phi_1 \Phi_4^2 \Phi_6+ \Phi_1 \Phi_4 \Phi_2\Phi_3+\Phi_5^{N-1} \Phi_2^2  \Phi_1.
\end{equation}
In the following we show that this confining duality can be derived from tensor deconfinement, reproducing the expected superpotential (\ref{Ipredicted1}).
\begin{figure}
\begin{center}
  \includegraphics[width=10cm]{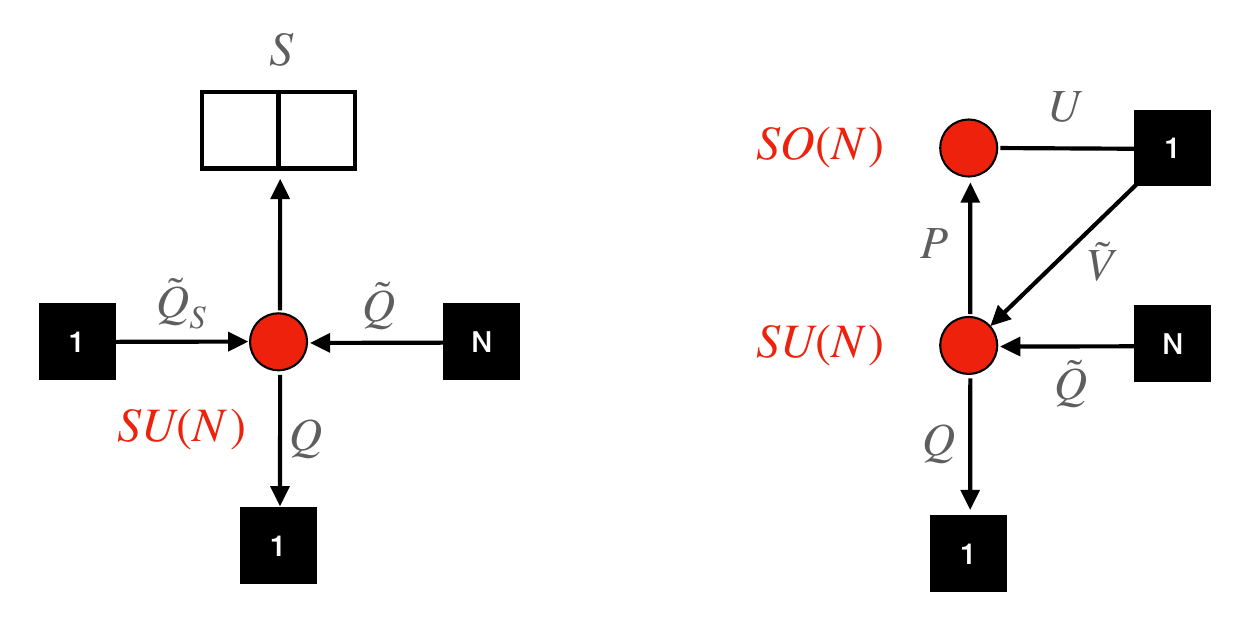}
  \end{center}
  \caption{In the first figure we provide the quiver description of the $SU(N)$ gauge theory with a symmetric, one fundamental $Q$, $N$ antifundamentals $\tilde Q$ and one antifundamental $\tilde Q_S$, with superpotential $W= S Q_S^2$. In the second figure we provide the quiver description after deconfining the symmetric tensor in terms of an $SO(N)$ gauge node.}
    \label{firstfigure}
\end{figure}
We start by deconfining the symmetric tensor using a confining duality for 3d $\mathcal{N}=2$ 
$SO(N)$ SQCD discussed in \cite{Benvenuti:2021nwt} (see also \cite{Benini:2011mf,Aharony:2011ci} for related discussions). The duality relates  
$SO(N)$ with $N + 1$ vectors and
linear monopole superpotential $W = Y_{SO(N)}^+$
and  a WZ model with superpotential
$W = S q^2 + \det S$, where $S$ is a symmetric meson of the electric description and $q$ correspond to the baryons. 

Here we use this duality in order to trade the symmetric tensors in favour of an $SO(N)$ gauge group, by further adding the flippers $\tilde V$, $\gamma$ and $\alpha$. In this way 
the model on the RHS of Figure \ref{firstfigure} has superpotential
\begin{equation}
W = Y_{SO(N)}^+ + Y_{SU(N)} + \alpha U^2 + PU\tilde V + \gamma \, \epsilon_N \cdot P^{N}.
\end{equation}

The $SU(N)$ gauge group is then confining having $N+1$ pairs of fundamentals and antifundamentals and 
a linear monopole superpotential. Its confinement can be read from 4d, by reducing the s-confining limiting case of $SU(N)$ SQCD Seiberg duality as shown in \cite{Aharony:2013dha}.

The mesons of the confining phase are 
$M_{\tilde Q P}$, $M_{\tilde Q Q}$, $M_{\tilde V Q}$ and $M_{\tilde V P}$
while the baryons\footnote{Where  the antisymmetric $\epsilon$ contractions are left implicit.} are
$B_1=P^{N}$,
$B_2 \equiv P^{N-1} Q$,
$\tilde B_1 \equiv \tilde Q^{N}$ and 
$\tilde B_2 \equiv \tilde Q^{N-1} \tilde V$.
The superpotential of the leftover $SO(N)$ gauge theory is 
\begin{eqnarray}
 W &=& Y_{SO(N)}^+ + \alpha U^2 + M_{\tilde V P} U + \gamma B_1
 +M_{\tilde Q P}B_2\tilde B_2+M_{\tilde Q Q}B_1 \tilde B_2+
  \nonumber\\
 &&+
 M_{\tilde V Q}B_1 \tilde B_1
 +M_{\tilde V P}B_2 \tilde B_1
 +
 \det \left(
 \begin{array}{cc}
 M_{\tilde Q P}&  M_{\tilde Q Q}\\
 M_{\tilde V P}&  M_{\tilde V Q}
\end{array}
  \right).
 \end{eqnarray}
 Such superpotential, after integrating out the massive fields, becomes
  \begin{equation}
  \label{WSOfinflip}
W = Y_{SO(N)}^+ + M_{\tilde Q P}B_2\tilde B_2+\alpha  (B_2 \tilde B_1+M_{\tilde Q P}^{N-1}M_{\tilde Q Q})^2+M_{\tilde V Q}\det  M_{\tilde Q P}.
\end{equation}

The leftover $SO(N)$ gauge theory has $N$ vectors $M_{\tilde Q P}$ and one vector $B_2$, in addition to the monopole superpotential $Y_{SO(N)}^+$. This node then confines, being a flipped version (i.e. with superpotential (\ref{WSOfinflip})) of the confining theory of  \cite{Benvenuti:2021nwt} discussed above.
The confined description is described by the baryons 
 \begin{equation}
 q_0=\epsilon_N \epsilon_N  (M_{\tilde Q P}^{N-1} B_2), \quad
 q_1 =\det(M_{\tilde Q P}),
 \end{equation}
 and by the symmetric meson
 \begin{equation}
 S=\left(
 \begin{array}{cc}
 S_{00}& S_{01}\\
 S_{01} & S_{11}
 \end{array}
 \right)
 =
 \left(
 \begin{array}{cc}
 M_{\tilde Q P}^2&  M_{\tilde Q P}B_2\\
  M_{\tilde Q P}B_2& B_2^2
 \end{array}
 \right).
  \end{equation}
 The superpotential of the confining theory is
 \begin{eqnarray}
 W &=& \det S + S_{00} q_0^2 +  S_{01} q_0 q_1+ S_{11} q_1^2+ S_{01}\tilde B_2+\alpha S_{11} \tilde B_1^2 \nonumber \\
 &+&M_{\tilde V Q} q_1+\alpha \tilde B_1 M_{\tilde Q Q} q_0+\alpha M_{\tilde Q Q }^2S_{00}^{N-1}.
 \end{eqnarray}
 After integrating out the massive fields, it becomes
 \begin{eqnarray}
 \label{WdecI}
W =  S_{11}\det S_{00} + S_{00} q_0^2 +\alpha S_{11} \tilde B_1^2+\alpha \tilde B_1 M_{\tilde Q Q} q_0+\alpha M_{\tilde Q Q }^2S_{00}^{N-1}.
 \end{eqnarray}
 At this point of the discussion we can read the duality map
\begin{eqnarray}
&&
\alpha = \det S = \Phi_1, \quad
M_{Q\tilde Q} = Q\tilde Q =\Phi_2 , \quad
q_0 = Q (S \tilde Q)^{N-1}=\Phi_3, \nonumber \\
&&
\tilde B_1 = \tilde Q^{N}=\Phi_4, \quad\,
S_{00} =S \tilde Q^2=\Phi_5, \quad \,\,
S_{11}=S^{N-1} Q^2=\Phi_6 .
\end{eqnarray} 
 Observing that (\ref{WdecI}) coincides with (\ref{Ipredicted1}).

 Once we have provided a proof of the duality in terms of other dualities we can also connect the results of this section with the ones of \cite{Amariti:2024gco} for $SU(N)$ with a symmetric, a lower amount of fundamentals or antifundamentals and without monopole superpotential.
 The two relevant models discussed in \cite{Amariti:2024gco} corresponds to the cases denoted there as I-A and I-B .
 
 The duality discussed here flows to those two cases by performing a real mass flow that removes the monopole superpotential. There are two possibilities:
 \begin{itemize}
 \item The flow to the model I-A requires to assign to two opposite real masses to $\tilde Q_{N-1}$ and $\tilde Q_{N}$. On the electric side we are left with $SU(N)$ with a symmetric $S$, one fundamental $Q$, one antifundamental $\tilde Q_S$ and $N-2$ antifundamentals $\tilde Q$.
The monopole superpotential is lift and the is not any constraint to impose on the global symmetries, or equivalently on the mass parameters on the on the partition function. 

On the WZ side the fields $\Phi_1$, $\Phi_4 $ and $\Phi_6$ are not modified by the real mass deformation, while only $N-2$ some components of the meson $\Phi_2$ and of the baryon $\Phi_3$ remain in the low energy spectrum.
On the other hand the field $\Phi_5$ splits into two massless fields\footnote{There are in addition $2(N-2)$ massive combinations arising from $\Phi_5$.}, one, denoted as $\hat \Phi_5$, corresponds to the flavor  symmetric combination $S \tilde Q_a \tilde Q_b $  for $a,b=1,\dots,N-2$, the other corresponds to  $S \tilde Q_{N-1} \tilde Q_{N} $ with respect of the original variables. The singlets $Y$ and $\Phi_4$ are massless in the dual side but they are not associated to any mesonic or baryonic gauge invariant combination of the massless elementary fields on the electric side. These fields are indeed  monopoles of the electric theory acting as a singlet in the dual side. These combinations are 
$Y_{SU(N-2)}^{bare} S$ and $Y_{SU(N-2)}^{bare} \tilde Q^{N-2}$, for $Y$ and $\Phi_4$ respectively.
The superpotential associated to this WZ model is then
\begin{equation}
\label{Ipredicted}
W = Y^2 \Phi_6 \det \hat \Phi_5 + \hat \Phi_5 \Phi_3^2 +\Phi_1 \Phi_4^2 \Phi_6+ \Phi_1 \Phi_4 \Phi_2\Phi_3.
\end{equation}
This corresponds to the superpotential expected for model I-A in \cite{Amariti:2024gco}.
 \item The flow to the model I-B requires to assign to two opposite real masses to $Q$ and $\tilde Q_{N}$. On the electric side we are left with $SU(N)$ with a symmetric $S$, one antifundamental $\tilde Q_S$ and $N-1$ antifundamentals $\tilde Q$. 
 The monopole superpotential is lift and there is not any constraint to impose on the global symmetries, or equivalently on the mass parameters on the on the partition function. 
 On the WZ side the fields $\Phi_4$ and $\Phi_6$ disappear from the low energy spectrum while
 the field $\Phi_1$ is not modified by the real mass deformation. $N (N-1)/2$ components of the symmetric meson $\Phi_5$ are massless and remain in the low energy spectrum as well. 
Only one  component of the meson $\Phi_2$ and $N-1$ components of the baryon $\Phi_3$ remain in the low energy spectrum. They correspond to the gauge invariant monopoles 
$Y_{SU(N-2)}^{bare}$ and $Y_{SU(N-2)}^{bare} \tilde Q^{N-2} S^{N-1}$  respectively.
The superpotential associated to this WZ model is then:
 \begin{equation} 
W =  \Phi_5 \Phi_3^2 + \Phi_1 \Phi_2^2\Phi_5^{N-1},
 \end{equation}
 where $\Phi_{2,3}$ are gauge invariant monopoles of the electric phase and $\Phi_{1,5}$ are gauge invariant combinations of the chiral fields of the electric phase.
 This corresponds to the superpotential expected for model I-C in \cite{Amariti:2024gco}.
 \end{itemize}

We conclude this section by deriving the identity \ref{iddopodup} applying the deconfinement techniques on the partition function.
The first step consists of deconfining the symmetric tensor $S$, by trading it with an $SO(N)$ 
gauge theory. The mass parameters for the fields in the  $SU(N)\times SO(N)$ quiver are
\begin{eqnarray}
m_P &=& \frac{\tau_S}{2},\quad
m_U = \omega-\frac{N}{2} \tau_S,\quad
m_{\tilde V} = \omega+\frac{N-1}{2} \tau_S,\quad
m_Q  = \nu_b, \nonumber \\
m_{\tilde Q} &=& \mu,\quad
m_\gamma= 2\omega- \frac{N}{2} \tau_S,\quad
m_\alpha = N \tau_S,\quad
\end{eqnarray}
where the constraint imposed by the linear monopole for the $SO(N)$ gauge groups are automatically satisfied by this parameterization. 
The singlets arising from confining the  $SU(N)$ gauge node are
\begin{equation}
\begin{array}{l}
m_{B_1}=\frac{N}{2} \tau_S\\
m_{B_2}=\frac{N-1}{2} \tau_S+\mu\\
m_{\tilde B_1 }=\sum_{c=1}^{N} \nu_c\\
m_{\tilde B_2} =\sum_{c=1}^{N} \nu_c-\nu_b + \omega+\frac{N-1}{2} \tau_S
\end{array}
\quad\quad
\begin{array}{l}
m_{M_{\tilde Q P}}=\nu_b +\frac{1}{2} \tau_S\\
m_{M_{\tilde Q Q }}=\mu + \nu_b
\\
m_{M_{\tilde V Q}}=\omega+\frac{N-1}{2} \tau_S+\mu
\\
m_{M_{\tilde V P}}=\omega+\frac{N}{2} \tau_S
\end{array}
\end{equation}
with $b=1,\dots,N$.
The massive combinations in the  superpotential are $M_{\tilde V P}U$ and $\gamma B_1$ and it reflects in the relations 
$ \Gamma_h(m_{M_{\tilde V P}}) \Gamma_h( m_U) =  \Gamma_h( m_\gamma) \Gamma_h( m_{B_1}) = 1$ on the one loop determinants, which follow from the inversion relation.

 We then confine the $SO(N)$ gauge node with $N+1$ vectors and linear monopole superpotential. The constraint imposed by this superpotential corresponds to the constraint on the masses of $M_{\tilde Q P}$ and $B_2$, that corresponds indeed to 
the original balancing condition (\ref{balancingoddS}). The baryons $q_{0}$ and $q_1$ and the mesons $S_{00}$, $S_{01}$ and $S_{11}$ of this confining duality have mass parameters
 \begin{eqnarray}
 m_{q_0}=\omega-\nu_b-\frac{\tau_S}{2}, \quad && \quad
 m_{q_1} =\omega-(N-1)\frac{\tau_S}{2}-\mu, \nonumber
 \\
 m_{S_{00}} = \tau_S +\nu_b +\nu_c, \quad
 m_{S_{01}} &=& \frac{N}{2} \tau_S+\mu+ \nu_b, \quad
  m_{S_{11}} = (N-1)\tau_S+ 2\mu.
\nonumber
  \end{eqnarray}
The massive combinations in the final superpotential are $M_{\tilde V Q} q_1$ and $S_{01} \tilde B_2$ and it reflects in the relations 
$ \Gamma_h(m_{M_{\tilde V Q}}) \Gamma_h(m_{q_1}) = 
  \Gamma_h(m_{S_{01}}) \Gamma_h(m_{\tilde B_2})=1$ on the one loop determinants.
We are left in the final WZ model with the  combinations of hyperbolic Gamma functions
corresponding to the RHS of (\ref{iddopodup}), i.e. we have confirmed the validity of the identity by using the tensor deconfinement technique.

 \subsection{$SU(2n+1)$ with $S,\tilde A$, $\tilde Q_S$ and $3$ $\tilde Q$}

Here we consider a 4d $SU(N)$ gauge theory with an antisymmetric and a conjugate antisymmetric (from now on an antisymmetric flavor), and three pairs of fundamentals and antifundamentals (i.e. three fundamental flavors). The model is s-confining for each parity of $N$, i.e. $N=2n$ and $N=2n+1$, and again the details of such confinements are different and the two cases require a different analysis. We refer the reader to the original reference \cite{Csaki:1996zb} for further discussion on these models and fo \cite{Benvenuti:2020wpc} for a proof of such s-confining dualities through tensor deconfinement. 
For completeness in appendix \ref{apptuttobello} we have reported a more detailed description of these models and on their circle reduction, by providing the explicit expressions relating the three sphere partition function for the circle reduction of the confining dualities, corresponding to the identities (\ref{WAAtodd}) and (\ref{Z2naat3fl}) for the case $N=2n+1$ and $N=2n$ respectively.

Here, differently from the case studied above, we need to distinguish the analysis also after 
\emph{freezing} the mass parameters and applying the duplication formula.
We start by applying the duplication formula to (\ref{WAAtodd}) by freezing the three masses for the fundamentals as\footnote{We could also have chosen to freeze three mass parameters for the three antifundamentals, in such case we would have obtained, up to an overall charge conjugation, an identical description.} 
\begin{equation}
\label{dupg}
\mu_1 = \frac{\omega_1}{2}+\frac{\tau_S}{2},
\quad
\mu_2 = \frac{\omega_2}{2}+\frac{\tau_S}{2},
\quad
\mu_3 = \frac{\tau_S}{2}.
\end{equation}
After applying the duplication formula and after some rearrangement we obtain the following identity 
 \begin{eqnarray}
 \label{II-Boddmum}
 &&
 Z_{SU(2n+1)}(-;\omega-\frac{\tau_S}{2},\vec \nu;\tau_S;-;-;\tau_{\tilde A})=
 \Gamma_h((2n+1) \tau_S)
 \prod_{\ell=1}^{n}\Gamma_h \left( 2\ell(\tau_S + \tau_{\tilde A}) \right)
 \nonumber \\
&& \Gamma_h((n-1) \tau_{\tilde A} + \nu_1+\nu_2+\nu_3)
  \prod_{1\leq a\leq b\leq 3} \prod_{j=0}^{n-1}\Gamma_h((2j+1)\tau_{S} + 2j \tau_{\tilde A} + \nu_a +\nu_b)
    \nonumber \\
&&  \prod_{a=1}^3\Gamma_h(n \tau_{\tilde A} + \nu_a)
  \prod_{1\leq a<b \leq 3} \prod_{j=0}^{n-1}\Gamma_h(2(j+1)\tau_{S} + (2j+1) \tau_{\tilde A} + \nu_a +\nu_b),
    \end{eqnarray}
  with the balancing condition 
\begin{equation}
\label{balancing3.2}
\left( 2n+\frac{1}{2} \right) \tau_S +(2n-1) \tau_{\tilde A} +  \sum_{b=1}^3  \nu_b =  \omega.
   \end{equation}

\begin{figure}
\begin{center}
  \includegraphics[width=12cm]{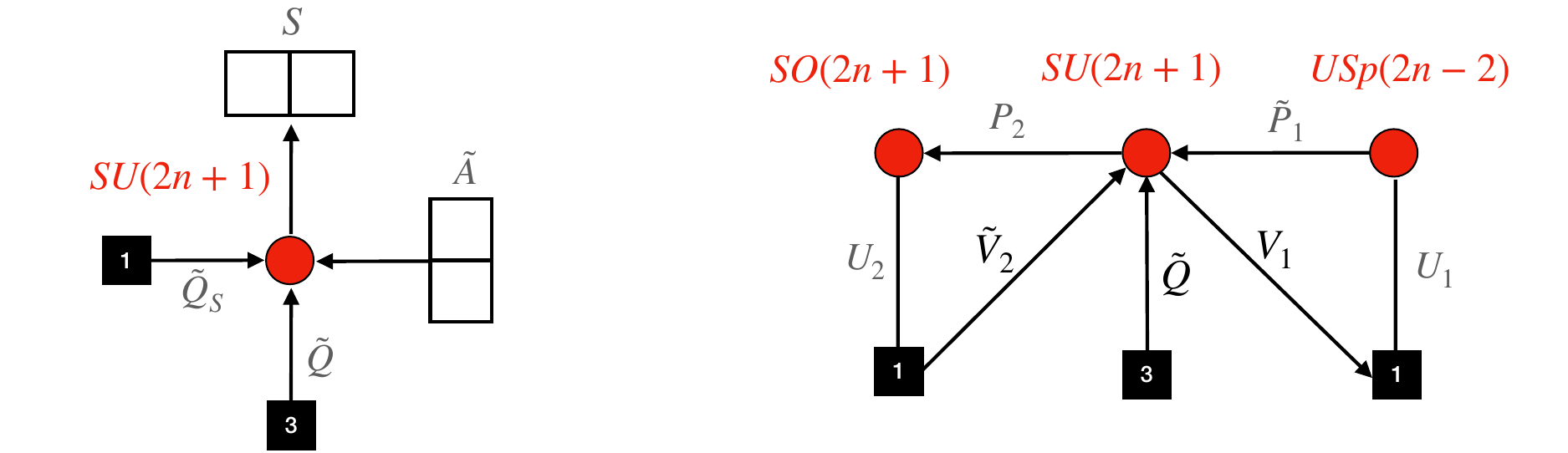}
  \end{center}
  \caption{In the first figure we provide the quiver description of the $SU(2n+1)$ gauge theory with a symmetric, a conjugate antisymmetric,  $3$ antifundamentals $\tilde Q$ and one antifundamental $\tilde Q_S$, with superpotential $W= S Q_S^2$. In the second figure we provide the quiver description after deconfining the symmetric and the conjugate antisymmetric tensors in terms of an $SO(2n+1)$ and an $USp(2n-2)$ gauge node respectively.}
    \label{CaseIIoddmon}
\end{figure}

We interpret this relation as the fact that a 3d $\mathcal{N}=2$ $SU(2n+1)$ gauge theory with a symmetric $S$, a conjugate antisymmetric $\tilde A$,
one antifundamental $\tilde Q_S$ and three antifundamentals $\tilde Q$ is confining in presence of the superpotential
\begin{equation}
W = S \tilde Q_S^2 + Y_{SU(2n-1)}^{bare}.
\end{equation}
We summarized the charged field content in the first quiver in Figure \ref{CaseIIoddmon}.
We then look at the RHS of the identity (\ref{II-Boddmum}), where we read the following gauge invariant combinations
\begin{eqnarray}      
\label{GIOs}
D &=&  \det S, \nonumber \\
T_{\ell} &=&  (S \tilde A)^{2\ell}, \quad \ell=1,\dots,N\nonumber \\
\tilde b_3 &=& \tilde A^{n-1} \tilde Q^3, \nonumber \\
\tilde P_{2j} &=&   \tilde Q_a S (S \tilde A)^{2j} \tilde Q_b,  \quad j=0,\dots,n-1\, \&\, a\leq b\nonumber \\
\tilde b_1 &=&  \tilde A^{n} \tilde Q, \nonumber \\
\tilde P_{2j+1} &=&  \tilde Q_a S (S \tilde A)^{2j+1}   \tilde Q_b, \quad j=0,\dots,n-1 \,\& \,a< b
    \end{eqnarray}
with $j=0,\dots,n-1$ and  $ \ell=1,\dots,n$.
We summarized  the various gauge invariant combinations in (\ref{GIOs}) with the same order in which they appear in the arguments of the hyperbolic Gamma function in the RHS of  (\ref{II-Boddmum}).
Observe that the singlets $\tilde P_{2j+1} $ are antisymmetric in the $SU(3)$ flavor indices while the 
singlets $\tilde P_{2j} $ are symmetric in the $SU(3)$ flavor indices.

 We then support the claim about such confining duality conjectured from the 
hyperbolic identity using tensor deconfinement.
We start by deconfining the two-index tensors $S$ and $\tilde A$.

We start by deconfining the symmetric tensor using the confining duality for 3d $\mathcal{N}=2$ 
$SO(2n+1)$ SQCD discussed in \cite{Benvenuti:2021nwt} already discussed in sub-section \ref{subsec3.1}, with linear monopole superpotential. 
The (conjugate) antisymmetric tensor on the other hand is deconfined by using the 
confining duality for 3d $\mathcal{N}=2$ 
$USp(2n-2)$ SQCD discussed in \cite{Aharony:2013dha} already discussed in sub-section \ref{Sub_AA13odd}, with linear monopole superpotential. 
The gauge and the field  obtained after deconfining the two tensors is represented  
in the second quiver in Figure \ref{CaseIIoddmon}. The superpotential for this model is
\begin{equation}
W = \gamma P_2^{2n+1}+\alpha U_2^2+U_2 \tilde V_2 P_2 +  \tilde P_1 V_1 U_1 +Y_{SU(2n+1)}+Y_{SO(2n+1)}^+ +Y_{USp(2n-2)}.
 \end{equation}
 Then we observe that the $SU(2n+1)$ gauge group has a linear monopole superpotential and $2n+2$ flavors. It then corresponds to the effective s-confining SQCD on $S^1$, where the 
IR degrees of freedom are the meson $\mathcal M$, the baryons $\mathcal{B}$ and the antibaryon $\tilde B$.
These fields correspond to the $SU(2n+1)$ invariant combinations
\begin{equation}
  \mathcal{M}
    \!=\! \left(\!
   \begin{array}{cc}
   M_1 & M_2 \\
   M_3 & M_4\\
   M_5 & M_6\!
   \end{array}
   \right)
     \!\!=\! \!\left(\! \begin{array}{cc}
    \tilde P_1 P_2&   \tilde P_1V_1\\
    \tilde V_2 P_2&  \tilde V_2 V_1\\
    \tilde Q P_2 & \tilde Q V_1
    \end{array}\!
    \right)\!\!,\,
  \mathcal{B}
    \!=\! \left( \begin{array}{c}
    \!B_2\! \\
    \!B_1\! 
    \end{array}\!
    \right)
     \!=\! \left( \begin{array}{c}
     P_2^{2n} V_1 \\
    P_2^{2n+1}
    \end{array}
    \right)\!\!, \,
 \tilde{\mathcal{B}}^T
    \!=\! \left(\! \begin{array}{c}
    \tilde B_3 \\  \tilde B_1 \\ \tilde B_2\\
    \end{array}\!
    \right)
     \!=\! \left( \!\begin{array}{c}
\tilde P_1^{2n-3}\tilde Q^3 \tilde V_2\\ \tilde P_1^{2n-2}\tilde Q^3 \\\tilde P_1^{2n-2}\tilde Q^2 \tilde V_2
    \end{array}\! 
    \right).
\end{equation}

The superpotential for the confining duality is
\begin{equation}
W=   \mathcal{M} \mathcal{B} \tilde{\mathcal{B}} +
  \det  \mathcal{M} 
   +
   \gamma B_1+\alpha U_2^2+U_2 M_3+   M_2 U_1 +Y_{SO(2n+1)}^+ +Y_{USp(2n-2)}.
\end{equation}    
 After integrating out the massive fields this superpotential becomes
\begin{eqnarray}
 W = Y_{SO(2n+1)}^+ +Y_{USp(2n-2)}+\alpha( B_2 \tilde B_1+M_6 M_5^2 M_1^{2n-2})^2+M_4 M_1^{2n-2} M_5^{3}. 
  \end{eqnarray}
 Then we observe that the $SO(2n+1)$ gauge group has $2n+2$ vectors,   corresponding to the fields $M_1$, $M_5$  and $B_2$ and linear monopole
 superpotential. 
 The $SO(2n+1)$ baryons are
\begin{eqnarray}
 q_1=M_1^{2n-3}M_5^3 B_2, \quad
 q_2=M_1^{2n-2}M_5^2 B_2 , \quad
 q_{3} = M_1^{2n-2}M_5^3,
   \end{eqnarray}
   while the components of the $SO(2n+1)$ symmetric meson are
   \begin{eqnarray}
&&
S_{11} = M_1^2, \quad
S_{12} = M_1 M_5,\quad
S_{13} = M_1 B_2, \nonumber \\
&&
S_{22} =  M_5^2, \quad
S_{23} = M_5  B_2,\quad
S_{33} =  B_2^2. 
\end{eqnarray}
After integrating out the massive fields, the superpotential of the leftover $USp(2n-2)$ gauge theory is 
   \begin{equation}
   \label{Wbefore}
   W = \alpha (S_{33} \tilde B_1^2+\tilde B_1 M_6 q_2+M_6^2 S_{22}^2 S_{11}^{2n-2} )+S_{11} q_1 q_1+S_{12} q_1 q_2+S_{22} q_2^2+Y_{USp(2n-2)}.
\end{equation}

The $USp(2n-2)$   gauge group has four fundamentals denoted as $S_{12}$ and $q_1$ and an adjoint $S_{11}$.
The field $S_{12}$ is in the fundamental representation of the $SU(3)$ flavor symmetry, while the fundamental $q_1$ interact with the adjoint through the coupling
$S_{11} q_1^2$. There is also a linear superpotential for the $USp(2n-2)$ fundamental monopole.
This model confines and its confinement was studied in \cite{Amariti:2022wae}.
Actually the analysis of the model is simplified  by flipping the tower of traces $Tr S_{11}^{2k}$, corresponding to add to (\ref{WUSpcontorre}) 
the deformation $\sum_{k=1}^{n-1} \beta_k Tr S_{11}^{2k}$ that corresponds in the original model to add the contribution
$\sum_{k=1}^{n-1} \beta_k Tr (S \tilde A)^{2k}$.

In this case the singlets of the dual WZ model are constructed from the symmetric and the antisymmetric contractions of two fundamentals $S_{12}$ with odd and even powers of the adjoint $S_{11}$, i.e.  $\mathcal{S}_{ab}^{(j)} = (S_{12})_a (S_{12})_b S_{11}^{2j+1}$  and $\mathcal{A}_{ab}^{(j)} =  (S_{12})_a (S_{12})_b S_{11}^{2j}$ with $j=0,\dots,n-2$. Furthermore the composite  operator $q_1 (S_{12})_a$ corresponds to   $\sum_{j=0}^{n-2} \epsilon^{bcd} \mathcal{S}_{ab}^{(j)} \mathcal{A}_{cd}^{(n-j-2)}$.
The superpotential of the WZ becomes 
\begin{eqnarray}
\label{spot2np132}
W&=&\epsilon^{r_1 r_2 r_3}\epsilon^{s_1 s_2 s_3}  
\big(
\mathcal{S}_{s_1,r_1}^{(\ell_1)}
\mathcal{S}_{s_2,r_2}^{(\ell_2)}
\mathcal{S}_{s_3,r_3}^{(\ell_3)}
\delta_{\ell_1 +\ell_2+\ell_3,2n-4}
+
\mathcal{A}_{s_1,r_1}^{(\ell_1)}
\mathcal{A}_{s_2,r_2}^{(\ell_2)}
\mathcal{S}_{s_3,r_3}^{(\ell_3)}
\delta_{\ell_1 +\ell_2+\ell_3,2n-3}
\big)
\nonumber  \\
&+&
\alpha (S_{33} \tilde B_1^2+\tilde B_1 M_6 q_2)+
S_{22} q_2^2+
\sum_{j=0}^{n-2} q_{2}^{a} \epsilon^{bcd} \mathcal{S}_{ab}^{(j)} \mathcal{A}_{cd}^{(n-j-2)},
\end{eqnarray}
where in the last term we have explicitly written the antifundamental $SU(3)$ index of the field $q_2$.
We can also, by following the duality map, associate the singlets here to the ones appearing in the original definition (\ref{GIOs})
\begin{itemize}
\item $\beta_{\ell}$ flip the combinations $T_{\ell} = (S \tilde A)^\ell$ for $\ell=1,\dots,n-2$. The only
singlet that has not been flipped corresponds to $S_{33}$ that indeed is identified with $T_{2n}$.
\item The singlets $\{\alpha,\tilde B_1,M_6\}$ correspond to the singlets $\{D,\tilde b_3,\tilde b_1 \}$
respectively.
\item The tower $P_{2j}$ is associated to the tower $\mathcal{S}^{(j-1)}$ for $j=1,\dots,n-1$, while $P_{0}$ corresponds to $S_{22}$.
\item The tower $P_{2j+1}$ is associated to the tower $\mathcal{A}^{(j)}$ for $j=0,\dots,n-2$, while $P_{2n-1}$ corresponds to $q_{2}$.
\end{itemize}
    
We can also flow from the duality obtained here to the one denoted as II-B
in  \cite{Amariti:2024gco}. Such a flow is triggered by two opposite real masses for two of the fundamentals $\tilde Q$.
The fate of the fields appearing in (\ref{GIOs}) after the real mass flow is
\begin{itemize}
\item The field $D$ survives  and it has been denoted as $\Psi_1$ in 
\cite{Amariti:2024gco}.
\item
The fields $T_\ell$ survive
and they have been denoted as $\Psi_4^{(j)}$ in 
\cite{Amariti:2024gco}.
\item The component of the field 
$\tilde b_3$ that survives becomes a monopole and it has been denoted as $\Psi_5$ 
 in 
\cite{Amariti:2024gco}
\item The components of the fields
$\tilde P_{2j+1}$ that survive reconstruct a part of the tower of monopoles, denoted as $\Psi_6^{(m)} $ in  \cite{Amariti:2024gco} .
\item 
From $\tilde b_1$ only one field survives, denoted as  $\Psi_2$ in 
\cite{Amariti:2024gco}.
\item The fields 
$\tilde P_{2j}$ that do not disappear split into a tower of monopoles and a tower of mesons, denoted respectively as 
$\Psi_6^{(m)}$ and $\Psi_3^{(j)}$ in \cite{Amariti:2024gco}. Observe that
the tower $\Psi_6^{(m)}$ is then fully reconstructed from $\tilde P_{2j}$
and $\tilde P_{2j+1}$.
\end{itemize}

We can also compare the  superpotential (\ref{spot2np132}) after the real mass flow with the one proposed in  {\bf Formula 5.23} of \cite{Amariti:2024gco}, after flipping the tower $\Psi_4^{(\ell)}$ for $\ell=0,\dots,n-2$. By inspection we can check that all most of the expected terms in are reproduced. For example the term
$\alpha S_{33} \tilde B_1^2$ becomes the term  $\Psi_5 \Psi_1 \Psi_4^{(n-1)}$ and the term 
$\alpha \tilde B_1 M_6 q_2$ becomes the term $\Psi_5 \Psi_6^{(0)} \Psi_2 \Psi_1$.
Also the various sums are reproduced through the real mass flow. The only term that is not obtained consists of the combination $\Psi_2^2 \Psi_1 \Psi_6^{(m)} \Psi_6^{(2n-m)}$. Such term, allowed from the symmetries,  is neither obtained here nor in \cite{Amariti:2024gco} from tensor deconfinement, and its existence and stability deserves more investigations.

We conclude this section by deriving the identity (\ref{II-Boddmum}) applying the deconfinement techniques on the partition function.
The first step consists of deconfining the symmetric tensor $S$, by trading it with an $SO(2n+1)$ 
gauge theory and the conjugate antisymmetric by trading it with an $USp(2n-2)$. The mass parameters for the fields in the  $SU(2n+1)\times SO(2n+1)\times USp(2n-2)$ quiver are
\begin{eqnarray}
m_{\tilde P_1} &=&\frac{\tau_{\tilde A}}{2}, \quad m_{P_2} =\frac{\tau_S}{2}
,\quad m_{\tilde V_2} =\omega + n\tau_S,\quad 
m_{V_1} =n \tau_{\tilde A},\quad 
m_{\gamma} =2\omega- \frac{2n+1}{2} \tau_S,\
\nonumber \\
m_{\tilde Q} &=&\nu_a,\quad 
m_{\alpha} =(2n+1) \tau_S,\quad 
m_{U_2} =\omega - \frac{2n+1}{2} \tau_S,\quad 
m_{U_1} =2\omega - \frac{2n+1}{2}\tau_{\tilde A}.
\nonumber
\end{eqnarray}

After confining the $SU(2n+1)$ node we obtain the partition function of the leftover $SO(2n+1)\times USp(2n-2)$ quiver gauge theory with the following mass parameters associated to the $SU(2n+1)$  singlets
\begin{equation}
\label{par3.2SUsinglets}
\begin{array}{l}
m_{M_1} =\frac{\tau_S+\tau_A}{2},\\
m_{M_2} = \left(n+\frac{1}{2}\right) \tau_{\tilde A},\\
m_{M_3} = \omega +\left(n+\frac{1}{2}\right) \tau_{\tilde S},\\
m_{M_4} = \omega+n(\tau_{\tilde A}+\tau_S),\\
m_{M_5} =\nu_b+\frac{\tau_S}{2},\\
m_{M_6} =\nu_b + n \tau_{\tilde A},
\end{array}
\quad
\begin{array}{l}
m_{B_1} =\frac{2n+1}{2} \tau_S,\\
m_{B_2} =n(\tau_S+\tau_{\tilde A}),\\
m_{\tilde B_1} = (n-1)\tau_{\tilde A} + \sum_{b=1}^{3} \nu_b,\\
m_{\tilde B_2} =\omega+n \tau_S + (n-1)\tau_{\tilde A} +\nu_a +\nu_b, \\
m_{\tilde B_3} =2\omega -\left(n+\frac{1}{2} \right) (\tau_S +\tau_{\tilde A}),
\end{array}
\end{equation}

The quadratic combinations in the  superpotential are $\gamma B_1$, $U_2 M_3$ and $M_2 U_1$ and these holomorphic masses  reflecting the relations 
$ \Gamma_h(m_\gamma)\Gamma_h(m_{B_1}) = \Gamma_h(m_{U_2})\Gamma_h(m_{M_3})=
\Gamma_h( m_{M_2})\Gamma_h(m_{U_1})=1$ on the one loop determinants.

We then confine the $SO(2n+1)$ gauge node with $N+1$ vectors and linear monopole superpotential. The constraint imposed by this superpotential corresponds to the constraint on the masses of $M_{\tilde Q P}$ and $B_2$, that corresponds indeed to 
the original balancing condition (\ref{balancingoddS}). The baryons and the mesons  of this confining duality have mass parameters
\begin{equation}
\begin{array}{l}
m_{q_1} =\omega-\frac{\tau_S+\tau_{\tilde A}}{2},
\\
m_{q_2} = 2n \tau_S + (2n-1)\tau_A + \nu_b + \nu_c \quad (b<c),
\\
m_{q_3} =\omega-n(\tau_{\tilde A}+\tau_S),
\end{array}
\end{equation}

\begin{equation}
\begin{array}{l}
m_{S_{11}}=\tau_S+\tau_{\tilde A},
\\
m_{S_{12}} = \nu_b+  \tau_S +\frac{\tau_A}{2},
\\
m_{S_{22}}=\tau_S+\nu_b +\nu_c \quad (b\leq c),
\end{array}
\quad
\begin{array}{l}
m_{S_{13}}=\left(n+\frac{1}{2}\right)(\tau_S+\tau_{\tilde A}),
\\
m_{S_{33}}=2n(\tau_S+\tau_{\tilde A}),
\\
m_{S_{23}}= \nu_b+\frac{\tau_S}{2}+n(\tau_S+\tau_{\tilde A}),
\end{array}
\end{equation}
The quadratic combinations in the  superpotential are $S_{23} \tilde B_2$, $S_{13} \tilde B_3$ and $q_3 M_4$ and these holomorphic masses  reflecting the relations 
$\Gamma_h(m_{S_{23}}) \Gamma_h(m_{\tilde B_2})
=\Gamma_h(m_{S_{13}}) \Gamma_h(m_{\tilde B_3})=
\Gamma_h(m_{q_{3}}) \Gamma_h(m_{M_4})
=1$ on the one loop determinants. Observe that the first relation hods provided we impose the balancing condition (\ref{balancing3.2}).

The last step consists of confining the $USp(2n-2)$ gauge node with an adjoint, four fundamentals, one of which interacts with the adjoint, and a monopole superpotential.
The necessary identity was given in {\bf Formula 4.1} of \cite{Amariti:2022wae}, and it was obtained by applying the duplication formula to the identities derived in \cite{Amariti:2018wht,Benvenuti:2018bav}.
There are three towers of singlets, one associated to the combinations $(S \tilde A)^{\ell}$, that have been flipped by $\beta_\ell$ for $\ell=1,\dots,n-2$. The only singlet that remains from this tower has mass $(n-1)(\tau_S + \tau_{\tilde A})$.
On the other hand the other two mesonic towers, of symmetric and antisymmetric contraction of the three fundamentals $S_{12}$ have masses
\begin{eqnarray}
\label{finaltowres}
(2j+3)\tau_S + (2j+2)\tau_{\tilde A}+\nu_a +\nu_b
\quad \text{with} \quad a \leq b \text{ and }  j=0,\dots,n-2,
\nonumber
\\
(2j+2)\tau_S + (2j+1)\tau_{\tilde A}+\nu_a +\nu_b
\quad  \text{with} \quad a<b  \text{ and }   j=0,\dots,n-2,
\end{eqnarray}
respectively.
The first line in (\ref{finaltowres}) reconstructs the final towers in last term in the second line of (\ref{II-Boddmum}) together with $\Gamma_h(m_{S_{22}})$.
The second line in (\ref{finaltowres}) reconstructs the final towers in last term in the third line of (\ref{II-Boddmum}) together with $\Gamma_h(m_{q_2})$.
The other contributions to the RHS of (\ref{II-Boddmum})  correspond to 
$\Gamma_h(m_{\tilde B_1},m_{S_{33}},m_{\alpha},m_{M_6})$
and once we consider their contribution we obtain the expected identity. 

 %
 %
 %
 %
 %
 %
 %
 %
 %%%%%%%%%%%%%%%%%%%%%%%%%%%%%%%%%%%%
 %%%%%%%%%%%%%%%%%%%%%%%%%%%%%%%%%%%%
 \subsection{$SU(2n)$ with $S,\tilde A$, $\tilde Q_S$ and $3$ $\tilde Q$}
 
 We conclude this section by focusing on the case of $SU(2n)$ with 
an antisymmetric and three fundamental flavors. 
In this case we apply the duplication formula to the identity (\ref{Z2naat3fl})  by freezing the three masses for the fundamentals as in formula (\ref{dupg}).
After some rearrangements we obtain the following identity 
   \begin{eqnarray}
   \label{II-Bevenmum}
 &&
 Z_{SU(2n)}(-;\omega-\frac{\tau_S}{2},\vec \nu;\tau_S;-;-;\tau_{\tilde A})= \Gamma_h(n \tau_{\tilde A} )
   \Gamma_h ( 2n \tau_S)     
    \\
&&   
   \prod_{\ell=1}^{n-1}\Gamma_h \left( 2\ell (\tau_S + \tau_{\tilde A}) \right)       \prod_{a<b} \prod_{j=0}^{n-2}\Gamma_h(2(j+1)\tau_{S} + (2j+1) \tau_{\tilde A} + \nu_a +\nu_b)
   \nonumber \\
&&   
   \prod_{a<b}\Gamma_h((n-1) \tau_{\tilde A} + \nu_a+\nu_b)  \prod_{a\leq b} \prod_{k=0}^{n-1}\Gamma_h((2k+1)\tau_{S} + 2k \tau_{\tilde A} + \nu_a +\nu_b),
      \nonumber
   \end{eqnarray}
with the balancing condition 
\begin{equation}
\label{balcingevenod33}
\left( 2n-\frac{1}{2} \right) \tau_S +(2n-2) \tau_{\tilde A} +  \sum_{b=1}^3  \nu_b =  \omega.
   \end{equation}

We interpret this relation as the fact that a 3d $\mathcal{N}=2$ $SU(2n)$ gauge theory with a symmetric $S$, an conjugate antisymmetric $\tilde A$ 
one antifundamental $\tilde Q_S$ and three antifundamentals $\tilde Q$ is confining in presence of the superpotential
\begin{equation}
W = S \tilde Q_S^2 + Y_{SU(2n-2)}^{bare}.
\end{equation}
From the RHS of (\ref{II-Bevenmum}) we read the gauge invariant combinations
\begin{eqnarray}
\label{GIOs2}
&&
\tilde P_{2j+1} =  \tilde Q_a S (S \tilde A)^{2j+1}   \tilde Q_b, \quad
\tilde P_{2k} =  \tilde Q_a S (S \tilde A)^{2k}   \tilde Q_b,
 \nonumber \\
 &&
 T_{\ell} = (S \tilde A)^{2\ell}, \quad
 \tilde b_1 =  \tilde A^{n},\quad
D = \det S,\quad
\tilde b_2=\tilde A^{n-1} \tilde Q^2,
\end{eqnarray}
with $j=0,\dots,n-2$, $k=0,\dots,n-1$ and  $ \ell=1,\dots,n-1$.
The singlets $\tilde P_{2j+1} $ are antisymmetric in the $SU(3)$ flavor indices while the 
singlets $\tilde P_{2k} $ are symmetric in the $SU(3)$ flavor indices.

\begin{figure}
\begin{center}
  \includegraphics[width=10cm]{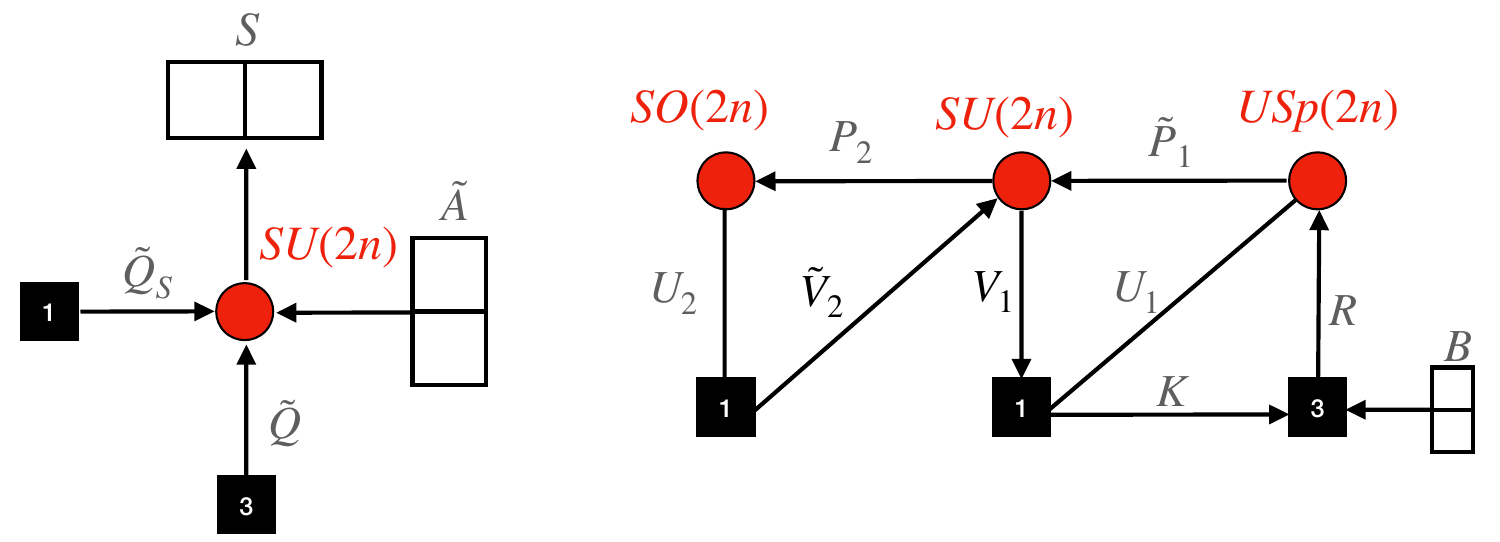}
  \end{center}
  \caption{In the first figure we provide the quiver description of the $SU(2n)$ gauge theory with a symmetric, a conjugate antisymmetric,  $3$ antifundamentals $\tilde Q$ and one antifundamental $\tilde Q_S$, with superpotential $W= S Q_S^2$. In the second figure we provide the quiver description after deconfining the symmetric and the conjugate antisymmetric tensors in terms of an $SO(2n)$ and an $USp(2n)$ gauge node respectively.}
    \label{CaseIIevenmon}
\end{figure}

We then support the claim about such confining duality conjectured from the 
hyperbolic identity using tensor deconfinement.
We start by deconfining the two-index tensors $S$ and $\tilde A$ using an $SO(2n)$ and an $USp(2n)$ gauge group
as in Figure \ref{CaseIIevenmon}. The superpotential for this model is
\begin{eqnarray}
 W &=& \gamma P_2^{2n}+\alpha U_2^2+U_2 \tilde V_2 P_2 +  \tilde P_1 V_1 U_1+ R U_1 K
 \nonumber \\
 &+& Y_{SU(2n)}+Y_{SO(2n)}^+ +Y_{USp(2n)}+B R^2.
\end{eqnarray}

The $SU(2n)$ gauge group has a linear monopole superpotential and $2n+1$ flavors. It then confines 
and the effective degrees of freedom are the meson $\mathcal M$, the baryons $\mathcal{B}$ and the antibaryon $\tilde{\mathcal{B}}$.
These fields correspond to the $SU(2n)$ invariant combinations
 
\begin{equation}
\mathcal{M} \!=\!
\left(
\begin{array}{cc}
M_1 &M_2 \\
M_3 & M_4
\end{array}\right)
\!=\!
\left(
\begin{array}{cc}
\tilde P_1  P_2 & \tilde P_1  V_1\\
 \tilde V_2  P_2& \tilde V_2  V_1
\end{array}\right)\!,
\,
\mathcal{B} \!=\!
\left(
\begin{array}{c}
B_2 \\
B_1
\end{array}\right)
\!=\!
\left(\!
\begin{array}{c}
P_2^{2n-1}  V_1 \\
P_2^{2n}
\end{array} \!\right)\!,
\,
\!\tilde {\mathcal{B} }^T\!=\!
\left(
\begin{array}{c}
\tilde B_2 \\\tilde B_1
\end{array}\right)
\!=\!
\left(\!
\begin{array}{c}
 \tilde P_2^{2n-1} \tilde V_2 \\ \tilde P_1^{2n}
\end{array}\!\right)\!\!.
\end{equation}
The superpotential for the confining duality is
   \begin{equation}
W = \mathcal{M}  \mathcal{B} \tilde{\mathcal{B}} 
+ \det \mathcal{M}+\gamma B_1+\alpha U_2^2+U_2 M_3 
+M_2 U_1 + R U_1 K+Y_{SO(2n)}^+ +Y_{USp(2n)}+B R^2
  \end{equation}
   After integrating out the massive fields this superpotential becomes
\begin{equation}
 W  =  Y_{SO(2n)}^+ +Y_{USp(2n)}+B_2 \tilde B_2 M_1+M_4 \det M_1 +\alpha (B_2 \tilde B_1+ R  K M_1^{2n-1})^2+B R^2.
  \end{equation}
 Then we observe that the $SO(2n)$ gauge group has $2n+1$ vectors,   corresponding to the fields $M_1$ and $B_2$ and linear monopole
 superpotential. This group confines and the confined degrees of freedom are the $SO(2n)$ baryons
$q_1=\det  M_1$ and $q_0=B_2 M_1^{2n-1}$ and the  $SO(2n)$ symmetric mesons
$S_{00} = M_1^2$, $S_{01} = M_1 B_2$ and $S_{11} =  B_2^2$.
The superpotential for the leftover $USp(2n)$ gauge theory is
\begin{eqnarray}     
W &=&   Y_{USp(2n)}+\alpha \tilde B_1^2 S_{11}+S_{01} \tilde B_2+M_4 q_1+\alpha \tilde B_1 q_0 R  K
\nonumber \\ &+&\alpha (R  K)^2 S_{00}^{2n-1}+ B R^2 S_{00} q_0^2 +S_{11} q_1^2 + S_{01} q_0q_1
+ \det 
\left(
\begin{array}{cc}
S_{00} & S_{01} \\
S_{01}^T& S_{11}
\end{array}
\right).
\end{eqnarray}
After integrating out the massive fields this superpotential becomes          
\begin{equation}  
\label{WUSpcontorre}        
W = Y_{USp(2n)}+\alpha \tilde B_1^2 S_{11}+\alpha \tilde B_1 q_0 R  K+\alpha (R  K)^2 S_{00}^{2n-1}+S_{00} q_0^2+B R^2 + S_{11} \det S_{00}.
\end{equation}
The $USp(2n)$   gauge group has four fundamentals denoted as $R$ and $q_0$ and an adjoint $S_{00}$.
The field $R$ is in the fundamental representation of the $SU(3)$ flavor symmetry, while the fundamental $q_0$ interact with the adjoint through the coupling
$S_{00} q_0^2$. There is also a linear superpotential for the $USp(2n)$ fundamental monopole.
This model confines and its confinement was studied in \cite{Amariti:2022wae}.
Actually the analysis of the model is simplified  by flipping the tower of traces $Tr S_{00}^{2k}$, corresponding to add to (\ref{WUSpcontorre}) 
the deformation $\sum_{k=1}^{n-1} \beta_k Tr S_{00}^{2k}$ that corresponds in the original model to add the contribution
$\sum_{k=1}^{n-1} \beta_k Tr (S \tilde A)^{2k}$. Observe that the superpotential (\ref{WUSpcontorre})
complete such tower of flippers $\beta_{\ell}$ by including the term \footnote{See {\bf Footnote 2} of \cite{Amariti:2022wae} for a more detailed discussion.}
$S_{00} \det S_{11} \simeq S_{00} Tr S_{11}^n$.

In this case the singlets of the dual WZ model are constructed from the symmetric and the antisymmetric contractions of two fundamentals $R$ with odd and even powers of the adjoint $S_{00}$, i.e.  $\mathcal{S}_{ab}^{(j)} = R_a R_b S_{00}^{2j+1}$  and $\mathcal{A}_{ab}^{(j)} = R_a R_b S_{00}^{2j}$ with $j=0,\dots,n-1$. Furthermore the composite  operator $q_0 R_a$ corresponds to   $\sum_{j=0}^{n-1} \epsilon^{bcd} \mathcal{S}_{ab}^{(j)} \mathcal{A}_{cd}^{(n-j-1)}$.
The superpotential of the WZ becomes 
\begin{eqnarray}
\label{spot2np133}
W&=&\epsilon^{r_1 r_2 r_3}\epsilon^{s_1 s_2 s_3}  
\big(\mathcal{S}_{s_1,r_1}^{(\ell_1)}
\mathcal{S}_{s_2,r_2}^{(\ell_2)}
\mathcal{S}_{s_3,r_3}^{(\ell_3)}
\delta_{\ell_1 +\ell_2+\ell_3,2n-2}
+
\mathcal{A}_{s_1,r_1}^{(\ell_1)}
\mathcal{A}_{s_2,r_2}^{(\ell_2)}
\mathcal{S}_{s_3,r_3}^{(\ell_3)}
\delta_{\ell_1 +\ell_2+\ell_3,2n-1}
\big)
\nonumber  \\
&+&
\alpha  K^2 \mathcal{S}^{(n)}
+B \mathcal{A}^{(0)}
+
\sum_{j=0}^{n-1} \alpha \tilde B_1 K^{a} \epsilon^{bcd} \mathcal{S}_{ab}^{(j)} \mathcal{A}_{cd}^{(n-1-j)}, 
\end{eqnarray}
where in the last term we have explicitly written the antifundamental $SU(3)$ index of the singlet $K$.

We can also, by following the duality map, associate the singlets here to the ones appearing in the original definition (\ref{GIOs2}):
\begin{itemize}
\item The fields $\beta_{\ell}$ flip the combinations $T_{\ell} = (S \tilde A)^\ell$ for $\ell=1,\dots,n-1$.
\item The singlets $\{\alpha,\tilde B_1,K\}$ correspond to the singlets $\{D,\tilde b_1,\tilde b_2 \}$
respectively.
\item The tower $\tilde P_{2j}$ is associated to the tower $\mathcal{S}^{(j)}$ for $j=0,\dots,n-1$.
\item The tower $\tilde P_{2j-1}$  is associated to the tower $\mathcal{A}^{(j)}$ for $j=1,\dots,n-1$, while $\mathcal{A}^{(0)}$ is massive because of the quadratic superpotential interaction with the field $B$.
\end{itemize}

We can also flow from the duality obtained here to the one denoted as II-B
in  \cite{Amariti:2024gco}. Such a flow is triggered by two opposite real masses for two of the fundamentals $\tilde Q$.
The fate of the fields appearing in (\ref{GIOs2}) after the real mass flow is
\begin{itemize}
\item The fields $D$ and  $\tilde b_1$ survive  and they have been denoted as $\Psi_1$ and $\Psi_2$ respectively in 
\cite{Amariti:2024gco}.
\item
The fields $T_\ell$ survive
and they have been denoted as $\Psi_4^{(j)}$ in 
\cite{Amariti:2024gco}.
\item The component of the field 
$\tilde b_2$ that survives becomes a monopole and it has been denoted as $\Psi_5$ 
 in 
\cite{Amariti:2024gco}.
\item The components of the fields
$\tilde P_{2j+1}$ that survive reconstruct a part of the tower of monopoles, denoted as $\Psi_6^{(m)} $ in  \cite{Amariti:2024gco} .
\item The fields 
$\tilde P_{2j}$ that do not disappear split into a tower of monopoles and a tower of mesons, denoted respectively as 
$\Psi_6^{(m)}$ and $\Psi_3^{(j)}$ in \cite{Amariti:2024gco}. Observe that
the tower $\Psi_6^{(m)}$ is then fully reconstructed from $\tilde P_{2j}$
and $\tilde P_{2j+1}$.
\end{itemize}
We can also compare the  superpotential (\ref{spot2np132}) after the real mass flow with the one proposed in  {\bf Formula 5.22} of \cite{Amariti:2024gco}, after flipping the tower $\Psi_4^{(\ell)}$ for $\ell=0,\dots,n-1$. 
By inspection we observe that the expected superpotential is fully reproduced, including the terms that have not been reconstructed from tensor deconfinement in  \cite{Amariti:2024gco}.

We conclude this section by deriving the identity (\ref{II-Bevenmum}) applying the deconfinement techniques on the partition function.
We start on the electric side to consider the addition of the flippers $\beta_\ell$, with mass parameters $m_{\beta_\ell}$ for $\ell=1,\dots,n-1$.
The second step consists of deconfining the symmetric tensor $S$, by trading it with an $SO(2n)$ 
gauge theory and the conjugate antisymmetric by trading it with an $USp(2n)$. The mass parameters for the fields in the  $SU(2n)\times SO(2n)\times USp(2n)$ quiver are
 \begin{eqnarray}
m_{R} &=&\nu_b - \frac{\tau_{\tilde A}}{2}
, \quad
m_{P_2} =\frac{\tau_S}{2}
, \quad
m_{\tilde P_1} =\frac{\tau_{\tilde A}}{2}
, \quad
m_{B} =2\omega+\tau_{\tilde A}-\nu_b-\nu_c
, \quad
m_{\alpha} =2 n \tau_S,
\nonumber \\
m_{V_1} &=&%\left(n-2\right)\tau_{\tilde A}+\sum_{b=1}^{3} \nu_b =
\omega-\left(2n-\frac{1}{2}\right) \tau_S -n \tau_{\tilde A} 
, \quad
m_{\tilde V_2} =\omega + \left(n-\frac{1}{2}\right) \tau_S
, \quad
m_{\gamma} =2\omega- n \tau_S,
 \\
m_{U_1} &=&%2\omega - \left(n-\frac{3}{2}\right)\tau_{\tilde A}-\sum_{b=1}^{3} \nu_b =
\omega +\left(n-\frac{1}{2}\right)\tau_{\tilde A}+\left(2n-\frac{1}{2}\right) \tau_S
, \quad
m_{U_2} =\omega - n \tau_S
, 
\\
m_{K} &=&\nu_b +\nu_c +\left(n-1\right)\tau_{\tilde A}.\nonumber 
\end{eqnarray}
where in the fourth and in the last term $1\leq b<c \leq 3$.

After confining the $SU(2n)$ node we obtain the partition function of the leftover $SO(2n)\times USp(2n)$ quiver gauge theory with the following mass parameters associated to the $SU(2n)$  singlets
\begin{equation}
\label{parsu2nduali}
\begin{array}{l}
m_{B_1} = n \tau_S,
\\
m_{B_2}  =\omega-n(\tau_{\tilde A} +\tau_S),
\\
m_{\tilde B_1} = n \tau_A,
\\
m_{\tilde B_2} = \omega + \left(n-\frac{1}{2}\right) (\tau_{\tilde A} + \tau_S),
\end{array}
\quad
\begin{array}{l}
m_{M_1} = \frac{\tau_S+\tau_A}{2},
\\
m_{M_2} =\omega-\left(2n-\frac{1}{2}\right)\tau_S-\left(n-\frac{1}{2}\right)\tau_{\tilde A},
\\
m_{M_3} = \omega + n \tau_S,
\\
m_{M_4} =2\omega-n(\tau_S + \tau_{\tilde A}).
\end{array}
\end{equation}
The quadratic combinations in the  superpotential are $\gamma B_1$, $U_2 M_3$ and $M_2 U_1$ and these holomorphic masses  reflecting the relations 
$ \Gamma_h(m_\gamma)\Gamma_h(m_{B_1}) = \Gamma_h(m_{U_2})\Gamma_h(m_{M_3})=
\Gamma_h( m_{M_2})\Gamma_h(m_{U_1})=1$ on the one loop determinants.

We then confine the $SO(2n)$ gauge node with $2n+1$ vectors and linear monopole superpotential. The constraint imposed by this superpotential corresponds to the constraint on the masses of $M_1$ and $B_2$, that is imposed by the parametrization (\ref{parsu2nduali}). The baryons $q_{0}$ and $q_1$ and the mesons $S_{00}$, $S_{01}$ and $S_{11}$ of this confining duality have mass parameters
\begin{eqnarray}
&&
m_{S_{11}}=2\omega-2n(\tau_{\tilde A} +\tau_S),\quad
m_{S_{01}}=\omega-\left(n-\frac{1}{2}\right)(\tau_S+\tau_A),
\nonumber \\
&&m_{S_{00}}=\tau_S + \tau_{\tilde A},\quad
m_{q_0}=\omega-\frac{\tau_S + \tau_{\tilde A}}{2},\quad
m_{q_1}=n(\tau_S + \tau_{\tilde A}).
\end{eqnarray}
The quadratic combinations in the  superpotential are $S_{01}\tilde B_2 $ and $q_1 M_4$ and these holomorphic masses  reflecting the relations 
$\Gamma_h(m_{S_{01}}) \Gamma_h(m_{\tilde B_2})
=
\Gamma_h(m_{q_{1}}) \Gamma_h(m_{M_4})
=1$ on the one loop determinants. 

The last step consists of confining the $USp(2n)$ gauge node with an adjoint, four fundamentals, one of which interacts with the adjoint, and a monopole superpotential.
The necessary identity was given in {\bf Formula 4.1} of \cite{Amariti:2022wae} and it was obtained by applying the duplication formula to the identities derived in \cite{Amariti:2018wht,Benvenuti:2018bav}.
There are three towers of singlets, one associated to the combinations $(S \tilde A)^{\ell}$, that have been flipped by $\beta_\ell$ for $\ell=1,\dots,n$. 
On the other hand the other two mesonic towers, of symmetric and antisymmetric contraction of the three fundamentals $R$ have masses
\begin{equation}
\label{finaltowres2}
\begin{array}{lcc}
(2j+1)\tau_S +2j \tau_{\tilde A}+\nu_a +\nu_b,\quad
&
\text{with}
&
1\leq a \leq b \leq 3 
\\
2j\tau_S +(2j-1) \tau_{\tilde A}+\nu_a +\nu_b,\quad
&
\text{with}
&
1\leq a < b \leq 3 
\end{array}
\quad
\text{and} 
\quad
{j=0,\dots,n-1} 
\end{equation}
respectively.
The first line in (\ref{finaltowres2}) reconstructs the final towers in last term in the third line of (\ref{II-Bevenmum}).
The second line in (\ref{finaltowres2}) reconstructs the final towers in last term in the second line of (\ref{II-Bevenmum}), except the term with $j=n-1$, that simplifies together with the  term  $\Gamma_h(m_{B})$.
The other contributions to the RHS of (\ref{II-Bevenmum})  correspond to 
$\Gamma_h(m_{\alpha},m_{\tilde B_1} ,m_{K})$. On the other hand the singlet $S_{11}$ simplifies in the last step, because it can be identified with a flipper of  tower $\beta_\ell$, i.e. the one that still misses with $\ell=n$, that flips the n-th power trace of the  $USp(2n)$ adjoint.
In this way we have reconstructed the identity (\ref{II-Bevenmum}) by applying tensor deconfinement and dualities on the squashed three sphere partition function.

%
%
%
%
%
%
%
%%%%%%%%%%%
%%%%%%%%%%%
\section{Conclusions}

In this paper we have studied 3d $\mathcal{N}=2$ confining gauge theories with $SU(N)$ gauge groups and 
tensor matter, antisymmetric and/or symmetric.
He have focused on two main classes of models, in section \ref{sec2} and \ref{sec3}.
The first class that we have studied has two antisymmetric tensors and $n_f+n_a=4$ fundamentals and antifundamentals. We have corroborated the fact that such models are confining by providing a derivation in terms of other dualities, by deconfining the tensors and by dualizing the  gauge groups. In this way we have completed the analysis, started in \cite{Nii:2019ebv}, for all the models with two antisymmetric proposed in \cite{Amariti:2024gco}.
The second class consists of model with symmetric tensors and linear monopole superpotential. These models generalize some of models studied in \cite{Amariti:2024gco} and they can be obtained by applying the duplication formula for the hyperbolic Gamma function on integral identities. These lasts correspond to the matching of the three sphere partition function for 4d s-confining gauge theories compactified on a circle. The identities obtained in this way correspond to confining gauge theories with a symmetric tensor and with (anti)-fundamental and/or a conjugate antisymmetric tensor. In each case there is a constraint on the masses of such fields that can is interpreted as a linear monopole superpotential, that needs to be imposed on the superpotential on the gauge theory side of the duality.
Also in these cases we have corroborated the validity of such dualities by deconfining the tensor(s) and dualizing the gauge groups, obtaining at the end of the process the expected WZ models.

We conclude  by pointing out some possible follow up and open questions.
A first comments is deserved by the models denoted as family II-A in \cite{Amariti:2024gco}, that in this paper did not have any counterpart with a monopole superpotential turned on. The reason is that it has not been possible to find any parent 4d duality with an antisymmetric flavor and at least four fundamental flavors, that would have been necessary in order to apply the duplication formula once the identity between the elliptic integrals is effectively reduced on $S^1$ as an identity between the hyperbolic hypergeometric integrals.
In principle the existence of such a 4d parent is just a sufficient condition to find the candidate 3d identity that we are looking for and one could try to construct the 3d model independently. However, by inspection we have not found any confining duality with a monopole superpotential turned on that, upon a real mass deformation, flows to the models of the II-A family. 
A more promising approach regards the 4d models recently discussed in \cite{Amariti:2024gco}, for $SU(2n)$ with one antisymmetric flavor $(A,\tilde A)$ and four fundamental flavors $(Q,\tilde Q)$, with a further superpotential term $W=\tilde A^{n-1} \tilde Q_1 \tilde Q_2$. This model is dual in 4d to $USp(2n)$ with eight fundamentals and a series of flippers. Once it is reduced to 3d 
we can further apply the duplication formula and obtain a 3d duality involving $SU(n)$ with a symmetric and a conjugate antisymmetric and an $USp(2n)$ model with an adjoint. In both cases we further have (anti)-fundamentals 
and monopole superpotential.
Then, through a real mass flow it is possible to flow to the II-A family discussed in \cite{Amariti:2024gco}, by observing that the two dual models are also dual to the expected confining gauge theory.
We are currently investigating in this direction.

The discussion is in principle generalizable to the cases with two antisymmetric tensors studied in Section \ref{sec2}. 
In such case finding a 4d non anomalous gauge theory requires to add a large amount of (anti)-fundamentals, possibly interacting with the antisymmetric(s), such to find a duality that becomes confining only after considering a real mass flow that eliminates the 2d monopole superpotential.
The discussion can then be potentially extended to the  cases denoted as family III in \cite{Amariti:2024gco} by applying the duplication formula for the hyperbolic gamma functions.

A further comment regards the unitarity of the dualities discussed here. In general it is always possible to flip some of the operators in the electric theories, such to leave only cubic interactions in the dual WZ models. In this way all the models discussed in the paper are unitary and there is not any  operator in the chiral ring hitting the unitary bound. 

It is also interesting to observe that many of the dualities studied here and in \cite{Nii:2019ebv,Amariti:2024gco} share some similarities with $\mathcal{N}=(0,2)$ dualities recently studied in \cite{Jiang:2024ifv,Amariti:2024usp,Amariti:2025jvi} (see also \cite{Sacchi:2020pet} for similar dualities with $USp(2n)$ and an antisymmetric chiral).
The relation between the 3d and the 2d dualities is in principle related to the bulk/boundary construction of \cite{Dimofte:2017tpi}. 
Along these lines it would be interesting to extend the analysis of \cite{Okazaki:2023hiv} to the cases with two antisymmetric  (without conjugation) and with symmetric tensors.

A last comments regards the possibility of finding similar dualities in 2d $\mathcal{N}=(0,2)$. There are indeed  dualities with $SU(N)$ gauge group and a symmetric tensor discussed in \cite{Jiang:2024ifv} that share many similarities with the dualities discussed in \cite{Amariti:2024gco} and here and it should be interesting to derive them by using the 4d/2d prescription of \cite{Gadde:2015wta} and the duplication formula for the Jacobi theta functions.

\section*{Acknowledgments}
 The work of A.A and S.R. has been supported in part by the Italian Ministero dell'Istruzione, Universit\'a e Ricerca (MIUR), in part by Istituto Nazionale di Fisica Nucleare (INFN) through the “Gauge Theories, Strings, Supergravity” (GSS) research project. The work of F.M. is funded by the Deutsche Forschungsgemeinschaft (DFG, German Research Foundation) – SFB 1624 – "Higher structures, moduli spaces and integrability" –506632645. The work of S.R. has been partially supported by the MUR-PRIN grant No. 2022NY2MXY.

\appendix

\section{$SU(M)$ with two antisymmetric tensors  and four fundamentals}
\label{app2AS4fund}

In this appendix we review the 3d $\mathcal{N}=2$ duality for $SU(N)$ with two antisymmetric tensors and four fundamentals, first proposed in \cite{Nii:2019ebv} and then further studied in \cite{Amariti:2024gco}, where the identity between the three sphere partition functions was proved.
This duality has played indeed a crucial role in our analysis, because all the models studied in Section \ref{sec2}
have been reduced, after deconfining the antisymmetric tensors, and dualizing the gauge nodes, to the case of $SU(N)$ with two antisymmetric and four fundamental fields. Then the fact that the dualities of Section \ref{sec2} directly follows from the confining duality reviewed here.

In the following we will be review only the basic aspects of this confining duality, referring the reader to the original references for an extended discussion.
The model requires to separate the case of even and odd gauge rank. 

For $M=2m$ the confining superpotential is
\begin{equation} 
\label{Wconf4e}
W = y^{dressed} (t_m t_{m-2} + t_{m-1}^2)
\end{equation}
where $t_m = A^m$, $t_{m-1} = A^{m-1} Q^2$ and $t_{m-2} = A^{m-2} Q^4$.
Furthermore the dressed monopole is defined as
\begin{equation} 
y^{dressed} = y_{SU(2m-2)}^{bare} A^{2m-2}
\end{equation}
The identity between the gauge and the confining phase at the level of the three sphere partition function is
\begin{eqnarray}
&&\label{4fund2ASeven}
Z_{SU(2m)} (\vec \mu;\cdot;\cdot;\cdot;\vec \tau;\cdot)
\!=\!
\prod_{j=0}^{m} \Gamma_h(j \tau_1 + (m\!-\!j) \tau_2)
\prod_{j=0}^{m-1}  \! \prod_{1\leq a <b \leq 4}\!\!\!\!\Gamma_h(j \tau_1 \!+\! (m\!-\!j\!-\!1) \tau_2 \!+\! \mu_a \!+\!\mu_b)
\nonumber \\
&&
\prod_{j=0}^{m-2} \Gamma_h \left(j \tau_1 \!+\! (n\!-\!j\!-\!2) \tau_2 + \sum_{a=1}^{4} \mu_a \right)
 \prod_{j=0}^{2m-2} \Gamma_h \left(\!
2\omega-j \tau_1 \! -(2m-j-2) \tau_2 -\! \sum_{a=1}^{4} \mu_a \! \right).
\nonumber \\
\end{eqnarray}

For $N=2m+1$ the confining superpotential is
\begin{equation} 
\label{Wconf4o}
W =y^{dressed} t_m t_{m-1}
\end{equation}
where $t_m = A^m$ and  $t_{m-1} = A^{m-1} Q^3$.
Furthermore the dressed monopole is defined as
\begin{equation} 
y^{dressed} = y_{SU(2m-1)}^{bare} A^{2m-1}
\end{equation}
The identity between the gauge and the confining phase at the level of the three sphere partition function is
\begin{eqnarray}
\label{4fund2ASodd}
Z_{SU(2m+1)} (\mu;\cdot;\cdot;\cdot;\vec \tau;\cdot)
&
=&
\prod_{j=0}^{m}  \prod_{a=1}^{4} \Gamma_h(j \tau_1 + (m-j) \tau_2 + \mu_a)
\nonumber \\
&\times&
\prod_{j=0}^{m-1}  \prod_{1\leq a <b<c \leq 4}\Gamma_h(j \tau_1 + (m-j-1) \tau_2 + \mu_a +\mu_b +\mu_c)
 \nonumber \\
&\times&
\prod_{j=0}^{2m-1}  \Gamma_h \left( 2\omega-j \tau_1 + (2m-j-1) \tau_2 -\sum_{a=1}^{4} \mu_a \right). 
\end{eqnarray}

\section{$SU(N)$ and antisymmetric}
\label{apptuttobello}
Here we review  two 4d confining dualities studied in \cite{Csaki:1996sm,Csaki:1996zb}, corresponding to 
\begin{itemize}
\item $SU(N)$ with an antisymmetric, $4$ fundamentals and $N$ antifundamentals
\item $SU(N)$ with three fundamental flavors and one antisymmetric flavor.
\end{itemize}
We will also discuss the reduction of the identity between the 4d superconformal indices (see \cite{Spiridonov:2009za} for a survey of these 4d identities) to the  3d partition function by using the prescription of \cite{Aharony:2013dha}.
The details of the chiral rings operator describing the confining dynamics are different in the case of $N=2n$ and $N=2n+1$, for this reason we must treat the four cases separately.

\subsubsection*{$SU(2n+1)$ with $A$, 4 $Q$ and $2n+1$ $\tilde Q$}

We begin by considering the 4d confining duality involving an $SU(2n+1)$ gauge group with an antisymmetric tensor, four fundamental flavors, $2n+1$ antifundamental flavors and vanishing superpotential.
We denote the antisymmetric tensor as $A $ and the fundamentals as $Q$ and $\tilde Q$. 
The singlet of the 4d confining phase are
\begin{align}
&M =Q \tilde Q, \quad \tilde B_2= A \tilde Q^2, \quad B_1=A^n Q, \quad  B_3=A^{n-1} Q^3, \quad \tilde B_{2n+1}=\tilde Q^{2n+1}.
\end{align}
The superpotential is a complicated function of these singlets.
Reducing this theory on $S^1$, the KK monopole forces the constraint on the global symmetry, which is forced by the requirement of anomaly freedom on the 4d R-symmetry in turn.
At the level of the superconformal index the reduction gives rise to the following hyperbolic identity for the squashed three sphere partition function  
\begin{eqnarray}
\label{W2np14}
&&
Z_{SU(2n+1)} (\vec \mu; \vec \nu;-;-;\tau_A;-)
=
 \prod_{a=1}^4 \prod_{b=1}^{2n+1} \Gamma_h(\mu_a +\nu_b)
 \prod_{b<c}^{2n+1} \Gamma_h(\tau_A +\nu_b +\nu_c)
\nonumber \\
&&
  \prod_{a=1}^{4} \Gamma_h(n \tau_A +\mu_a)
  \prod_{a<b<c}^{4} \Gamma_h((n-1) \tau_A +\mu_a+\mu_b+\mu_c)
   \Gamma_h\left(\sum_{b=1}^{2n+1} \nu_b \right),
\end{eqnarray}
where the KK monopole forces the constraint
\begin{equation}
\label{BalAS1}
(2n-1)\tau_A + \sum_{a=1}^4 \mu_a +  \sum_{b=1}^{2n+1} \nu_b =2\omega.
\end{equation}
This identity follows from the corresponding balancing condition of the elliptic case.

\subsubsection*{$SU(2n)$ with $A$, 4 $Q$ and $2n$ $\tilde Q$}

In this section we consider the 4d confining duality involving an $SU(2n)$ gauge group with an antisymmetric tensor, four fundamental flavors, $2n$ antifundamental flavors and vanishing superpotential.
We denote the antisymmetric tensor as $A $ and the fundamentals as $Q$ and $\tilde Q$. 
The singlet of the 4d confining phase are
\begin{align}
& M =Q \tilde Q, \quad  \tilde B_2= A \tilde Q^2, \quad 
B_0=A^n , \notag \\
& B_2=A^{n-1} Q^2, \quad  B_4=A^{n-2} Q^4, \quad \tilde B_{2n}=\tilde Q^{2n}. 
\end{align}
The superpotential is a complicated function of these singlets.
Here we are interested in reducing this confining duality on $S^1$, by turning on a KK monopole for the $SU(2n+1)$ gauge group.
At the level of the superconformal index the reduction gives rise to the following hyperbolic identity for the squashed three sphere partition function  
\begin{eqnarray}
\label{W2n4}
&&
Z_{SU(2n)} (\vec \mu; \vec \nu;-;-;\tau_A;-)
=
 \prod_{a=1}^4 \prod_{b=1}^{2n} \Gamma_h(\mu_a +\nu_b)
 \prod_{b<c}^{2n} \Gamma_h(\tau_A +\nu_b +\nu_c)
\nonumber \\
&&
\Gamma_h(n \tau_A)
  \prod_{a<b}^{4} \Gamma_h((n-1) \tau_A +\mu_a+\mu_b)
  \Gamma_h \left((n-2) \tau_A +\sum_{a=1}^{4} \mu_a \right)
   \Gamma_h\left(\sum_{b=1}^{2n} \nu_b \right).
\end{eqnarray}
The constraint forced by the KK monopole on the mass parameters is
\begin{equation}
\label{BalAS2}
(2n-2)\tau_A + \sum_{a=1}^4 \mu_a +  \sum_{b=1}^{2n} \nu_b =2\omega.
\end{equation}
This identity follows from the corresponding balancing condition of the elliptic case.

 \subsubsection*{$SU(2n+1)$ with $A,\tilde A$, $3$  $Q$ and $3$ $\tilde Q$}

Here we consider the 4d confining duality involving an $SU(2n+1)$ gauge group with an antisymmetric flavor and three fundamental flavors and vanishing superpotential.
 We denote the antisymmetric tensors as $A $ and $\tilde A$ and the fundamentals as $Q$ and $\tilde Q$. 
 The singlets of the 4d confining phase are
   \begin{equation}
  \begin{array}{llll}
    M_k = Q (A \tilde A)^k \tilde Q, &\quad H_k = \tilde A (A \tilde A)^k  Q^2 ,\quad           & \quad B_1 = A^n Q ,\quad                            &  \quad B_3 = A^{n-1} Q^3, \\
   T_m = (A \tilde A)^m,                 &\quad \tilde H_k =  A (A \tilde A)^k  \tilde Q^2 ,\quad & \quad \tilde B_1 = \tilde A^n \tilde Q,\quad   & \quad \tilde B_3 = \tilde A^{n-1} \tilde Q^3 ,
  \end{array}
     \end{equation} 
  with $k=0,\dots,n-1$ and $m=1,\dots,n$.
    The superpotential is a complicated function of these singlets, but its expression is simplified by flipping 
 the fields $B_3$, $\tilde B_3$ and $T_{m}$ as shown in \cite{Bajeot:2022kwt}.
  We refer the reader to the original references for details.
 Here we are interested in reducing this confining duality on $S^1$, by turning on a KK monopole for the $SU(2n+1)$ gauge group.
At the level of the superconformal index the reduction gives rise to the following hyperbolic identity for the squashed three sphere partition function  
 \begin{eqnarray}
 \label{WAAtodd}
 &&
 Z_{SU(2n+1)}(\vec \mu;\vec \nu;-;-;\tau_A;\tau_{\tilde A})=
 \prod_{a,b=1}^3 \prod_{k=0}^{n-1}\Gamma_h(k(\tau_A + \tau_{\tilde A}) + \mu_a +\nu_b)
\nonumber \\
&&
 \prod_{a<b} \prod_{k=0}^{n-1}\Gamma_h(\tau_{\tilde A} + k(\tau_A + \tau_{\tilde A}) + \mu_a +\mu_b,\tau_{A} + k(\tau_A + \tau_{\tilde A}) + \nu_a +\nu_b)
\nonumber \\
&&
    \prod_{a=1}^3\Gamma_h(n \tau_A + \mu_a,n \tau_{\tilde A} + \nu_a)
   \prod_{m=1}^{n}\Gamma_h(m(\tau_A + \tau_{\tilde A}))
   \nonumber \\
&&    \Gamma_h((n-1) \tau_A + \mu_1+\mu_2+\mu_3, (n-1) \tau_{\tilde A} + \nu_1+\nu_2+\nu_3),
  \end{eqnarray}
 where the KK monopole forces the constraint
\begin{equation}
\label{bc2np1aat3fl}
(2n-1)(\tau_A +\tau_{\tilde A}) + \sum_{b=1}^3 (\mu_b + \nu_b) = 2 \omega.
\end{equation}
This identity follows from the corresponding balancing condition studied of the elliptic case.

 \subsubsection*{$SU(2n)$ with $A,\tilde A$, $3$  $Q$  and $3$ $\tilde Q$}
    
 In this section we consider the 4d confining duality involving an $SU(2n)$ gauge group with an antisymmetric flavor, three fundamental flavors and vanishing superpotential.
 We denote the antisymmetric tensors as $A $ and $\tilde A$ and the fundamentals as $Q$ and $\tilde Q$. 
 The singlets of the 4d confining phase are
 \begin{equation}
 \label{singletssnaas}
 \begin{array}{llll}
 M_k = Q (A \tilde A)^k \tilde Q,  &\quad
H_m = \tilde A (A \tilde A)^m  Q^2 ,  &\quad
  B_0 = A^n , & \quad
  B_2 = A^{n-1} Q^2   \\
 T_\ell = (A \tilde A)^\ell , &\quad
\tilde H_m =  A (A \tilde A)^m  \tilde Q^2 , & \quad
 \tilde B_0 = \tilde A^n , & \quad
 \tilde B_2 = \tilde A^{n-1} \tilde Q^2,
 \end{array}
\end{equation}
  with $k=0,\dots,n-1$, $m=0,\dots,n-2$ and $\ell=1,\dots,n-2$.
 The superpotential is a complicated function of these singlets, but its expression is simplified by flipping 
 the fields $B_0$, $\tilde B_0$ and $T_{\ell}$ as shown in \cite{Bajeot:2022kwt}.
  We refer the reader to the original references for details.
 Here we are interested in reducing this confining duality on $S^1$, by turning on a KK monopole for the $SU(2n)$ gauge group.
At the level of the superconformal index the reduction gives rise to the following hyperbolic identity for the squashed three sphere partition function  
\begin{eqnarray}
\label{Z2naat3fl}
 &&
 Z_{SU(2n)}(\vec \mu;\vec \nu;-;-;\tau_A;\tau_{\tilde A})=
    \prod_{a,b=1}^3 \prod_{k=0}^{n-1}\Gamma_h(k(\tau_A + \tau_{\tilde A}) + \mu_a +\nu_b)
  \nonumber \\
&&    
    \prod_{a<b} \prod_{m=0}^{n-2}\Gamma_h(\tau_{\tilde A} + m(\tau_A + \tau_{\tilde A}) + \mu_a +\mu_b,\tau_{A} + m(\tau_A + \tau_{\tilde A}) + \nu_a +\nu_b)
     \Gamma_h(n \tau_A ,n \tau_{\tilde A} )
      \nonumber \\
&& 
     \prod_{a<b}\Gamma_h((n-1) \tau_A + \mu_a+\mu_b,(n-1) \tau_{\tilde A} + \nu_a+\nu_b)
     \prod_{\ell=1}^{n-1}\Gamma_h(\ell(\tau_A + \tau_{\tilde A})),
\end{eqnarray}
where the KK monopole forces the constraint
\begin{equation}
\label{bc2naat3fl}
(2n-2)(\tau_A +\tau_{\tilde A}) + \sum_{b=1}^3 (\mu_b + \nu_b) = 2 \omega.
\end{equation}
This identity follows from the corresponding balancing condition of the elliptic case.

\bibliographystyle{JHEP}
\bibliography{ref.bib}

\end{document}